\documentclass[a4paper]{article}
\usepackage[margin=1in]{geometry}

\newif\ifformat
\formatfalse 
\newcommand{\selectformat}[2]{%
  \ifformat%
  {#1}
  \else
  {#2}
  \fi
}

\makeatletter
\@ifundefined{lhs2tex.lhs2tex.sty.read}%
  {\@namedef{lhs2tex.lhs2tex.sty.read}{}%
   \newcommand\SkipToFmtEnd{}%
   \newcommand\EndFmtInput{}%
   \long\def\SkipToFmtEnd#1\EndFmtInput{}%
  }\SkipToFmtEnd

\newcommand\ReadOnlyOnce[1]{\@ifundefined{#1}{\@namedef{#1}{}}\SkipToFmtEnd}
\usepackage{amstext}
\usepackage{amssymb}
\usepackage{stmaryrd}
\DeclareFontFamily{OT1}{cmtex}{}
\DeclareFontShape{OT1}{cmtex}{m}{n}
  {<5><6><7><8>cmtex8
   <9>cmtex9
   <10><10.95><12><14.4><17.28><20.74><24.88>cmtex10}{}
\DeclareFontShape{OT1}{cmtex}{m}{it}
  {<-> ssub * cmtt/m/it}{}

\DeclareFontShape{OT1}{cmtt}{bx}{n}
  {<5><6><7><8>cmtt8
   <9>cmbtt9
   <10><10.95><12><14.4><17.28><20.74><24.88>cmbtt10}{}
\DeclareFontShape{OT1}{cmtex}{bx}{n}
  {<-> ssub * cmtt/bx/n}{}

\newcommand{\Conid}[1]{\mathit{#1}}
\newcommand{\Varid}[1]{\mathit{#1}}
\newcommand{\anonymous}{\kern0.06em \vbox{\hrule\@width.5em}}
\newcommand{\plus}{\mathbin{+\!\!\!+}}
\newcommand{\bind}{\mathbin{>\!\!\!>\mkern-6.7mu=}}

\renewcommand{\leq}{\leqslant}
\renewcommand{\geq}{\geqslant}
\usepackage{polytable}

\@ifundefined{mathindent}%
  {\newdimen\mathindent\mathindent\leftmargini}%
  {}%

\def\resethooks{%
  \global\let\SaveRestoreHook\empty
  \global\let\ColumnHook\empty}
\newcommand*{\savecolumns}[1][default]%
  {\g@addto@macro\SaveRestoreHook{\savecolumns[#1]}}
\newcommand*{\restorecolumns}[1][default]%
  {\g@addto@macro\SaveRestoreHook{\restorecolumns[#1]}}
\newcommand*{\aligncolumn}[2]%
  {\g@addto@macro\ColumnHook{\column{#1}{#2}}}

\resethooks

\newcommand{\onelinecommentchars}{\quad-{}- }
\newcommand{\commentbeginchars}{\enskip\{-}
\newcommand{\commentendchars}{-\}\enskip}

\newcommand{\visiblecomments}{%
  \let\onelinecomment=\onelinecommentchars
  \let\commentbegin=\commentbeginchars
  \let\commentend=\commentendchars}

\newcommand{\invisiblecomments}{%
  \let\onelinecomment=\empty
  \let\commentbegin=\empty
  \let\commentend=\empty}

\visiblecomments

\newlength{\blanklineskip}
\newcommand{\hsindent}[1]{\quad}
\let\hspre\empty
\let\hspost\empty

\EndFmtInput
\makeatother
\ReadOnlyOnce{polycode.fmt}%
\makeatletter

\newcommand{\hsnewpar}[1]%
  {{\parskip=0pt\parindent=0pt\par\vskip #1\noindent}}

\newcommand{\hscodestyle}{}

\newcommand{\sethscode}[1]%
  {\expandafter\let\expandafter\hscode\csname #1\endcsname
   \expandafter\let\expandafter\endhscode\csname end#1\endcsname}

  {\par\noindent
   \advance\leftskip\mathindent
   \hscodestyle
   \let\\=\@normalcr
   \let\hspre\(\let\hspost\)%
   \pboxed}%
  {\endpboxed\)%
   \par\noindent
   \ignorespacesafterend}

  {\hsnewpar\abovedisplayskip
   \advance\leftskip\mathindent
   \hscodestyle
   \let\hspre\(\let\hspost\)%
   \pboxed}%
  {\endpboxed%
   \hsnewpar\belowdisplayskip
   \ignorespacesafterend}

  {\hsnewpar\abovedisplayskip
   \advance\leftskip\mathindent
   \hscodestyle
   \let\\=\@normalcr
   \(\pboxed}%
  {\endpboxed\)%
   \hsnewpar\belowdisplayskip
   \ignorespacesafterend}

\newcommand{\plainhs}{\sethscode{plainhscode}}

\plainhs

  {\hsnewpar\abovedisplayskip
   \advance\leftskip\mathindent
   \hscodestyle
   \let\\=\@normalcr
   \(\parray}%
  {\endparray\)%
   \hsnewpar\belowdisplayskip
   \ignorespacesafterend}

  {\parray}{\endparray}

  {\(\parray}{\endparray\)}

\def\codeframewidth{\arrayrulewidth}
\RequirePackage{calc}

  {\parskip=\abovedisplayskip\par\noindent
   \hscodestyle
   \arrayrulewidth=\codeframewidth
   \tabular{@{}|p{\linewidth-2\arraycolsep-2\arrayrulewidth-2pt}|@{}}%
   \hline\framedhslinecorrect\\{-1.5ex}%
   \let\endoflinesave=\\
   \let\\=\@normalcr
   \(\pboxed}%
  {\endpboxed\)%
   \framedhslinecorrect\endoflinesave{.5ex}\hline
   \endtabular
   \parskip=\belowdisplayskip\par\noindent
   \ignorespacesafterend}

\newcommand{\framedhslinecorrect}[2]%
  {#1[#2]}

  {\(\def\column##1##2{}%
   \let\>\undefined\let\<\undefined\let\\\undefined
   \newcommand\>[1][]{}\newcommand\<[1][]{}\newcommand\\[1][]{}%
   \def\fromto##1##2##3{##3}%
   }{\) }%

  {\let\orighscode=\hscode
   \let\origendhscode=\endhscode
   \def\endhscode{\def\hscode{\endgroup\def\@currenvir{hscode}\\}\begingroup}
   \orighscode\def\hscode{\endgroup\def\@currenvir{hscode}}}%
  {\origendhscode
   \global\let\hscode=\orighscode
   \global\let\endhscode=\origendhscode}%

\makeatother
\EndFmtInput
\ReadOnlyOnce{lineno.fmt}%
\makeatletter

\newcounter{linenum}
\newcounter{linestep}
\newlength{\linetemp}

\newcommand{\linenumsetupleft}{\column{line}{@{}l@{}}\column{lineend}{}\column{B}{}}
\newcommand{\linenumsetupright}{\setlength{\linetemp}{\linewidth-\mathindent}\column[\linetemp]{line}{@{}l@{}}\column{lineend}{}\column{B}{}}
\newcommand{\linenumfromtoleft}{\refstepcounter{linenum}\fromto{line}{lineend}}
\newcommand{\linenumfromtoright}{\refstepcounter{linenum}\fromto{line}{lineend}}
\newcommand{\numbersright}{\let\linenumsetup\linenumsetupright\let\printlinebegin\empty\def\printlineend{\printline}\let\linenumfromto\linenumfromtoright\def\linenumalign{l}}
\newcommand{\numbersleft}{\let\linenumsetup\linenumsetupleft\def\printlinebegin{\printline}\let\printlineend\empty\let\linenumfromto\linenumfromtoleft\def\linenumalign{r}}
\numbersleft
\newcommand{\printlineyes}{%
  \linenumfromto{%
    \stepcounter{linestep}%
    \ifthenelse{\value{linestep}<\linenumstep}{}{%
      \formatlinenum{\thelinenum}%
      \setcounter{linestep}{0}}}}
\newcommand*{\formatlinenum}[1]{\makebox[\linenumwidth][\linenumalign]{\ \,{\footnotesize #1}\ \,}}
\newcommand*{\numberwidth}[1]{\def\linenumwidth{#1}}
\numberwidth{0pt}
\newcommand*{\numberstep}[1]{\def\linenumstep{#1}}
\numberstep{1}
\newcommand{\numberson}{\global\let\printline\printlineyes}
\newcommand{\numbersoff}{\global\let\printline\empty}
\newcommand{\numbersreset}{\setcounter{linenum}{0}}
\numberson

\usepackage{framed}
\usepackage{hyperref}
\usepackage{microtype} 
\usepackage{todonotes}
\usepackage{graphicx}
\usepackage[skip=2pt]{caption} 
\usepackage[aboveskip=2pt]{subcaption}
\usepackage{tikz-cd}
\usepackage{tikz}
\usetikzlibrary{shapes,arrows,positioning,calc,decorations.pathmorphing}
\tikzstyle{BLOCK}=[rectangle,rounded corners,thick,inner sep=7pt,minimum size=4mm]
\usepackage{pgfplots}
\usepackage{amsmath}
\usepackage{amsthm}
\usepackage{mathtools}
\usepackage{adjustbox}
\usepackage{multirow}
\usepackage[numbers]{natbib}
\usepackage{paralist}
\usepackage{blkarray}
\usepackage{epstopdf}

\pgfplotsset{scaled y ticks=false, scaled x ticks=false, compat=1.5}

\newtheorem{theorem}{Theorem}[section]
\newtheorem{definition}{Definition}[section]

\newif\ifappendix
\appendixfalse
\newcommand{\showappendix}[1]{%
  \ifappendix
  {#1}
  \fi
}

\definecolor{darkred}{HTML}{600018}
\definecolor{darkblue}{HTML}{1D2F73}
\definecolor{darkgreen}{HTML}{417505}
\definecolor{lightblue}{HTML}{76E9BB}

\renewcommand{\Varid}[1]{{\texttt{#1}}}
\renewcommand{\Conid}[1]{{\texttt{#1}}}

\DeclarePairedDelimiter{\abs}{\lvert}{\rvert}

\newcommand*\sfrac[2]{{}{#1}\!/{#2}}
\newcommand{\cmark}{$\color{darkgreen}\checkmark$}
\newcommand{\email}[1]{\href{mailto:#1}{\small\tt{#1}}}

  {\begin{list}{$\; \; \; \ \  \triangleright$}%
   {\leftmargin=0pt \itemsep=2pt \topsep=2pt
     \parsep=5pt \partopsep=0pt}}%
  {\end{list}}

\newenvironment{CompactItemizeee}%
  {\begin{list}{$\ \ \ \  \blacktriangleright$}%
   {\leftmargin=0pt \itemsep=2pt \topsep=2pt
     \parsep=0pt \partopsep=0pt}}%
 {\end{list}}

\begin{document}

\selectformat{\IEEEpeerreviewmaketitle}{}%
\title{A Programming Framework for Differential Privacy with Accuracy
  Concentration Bounds}

\selectformat{}{
\author{Elisabet Lobo-Vesga\\ \email{elilob@chalmers.se}  \and
        Alejandro Russo\\ \email{russo@chalmers.se} \and
        Marco Gaboardi\\ \email{gaboardi@bu.edu}}
}


\newcommand{\gb}[1]{\todo[color=red!40,inline]{#1}}
\newcommand{\gbb}[1]{\todo[color=red!40]{#1}}
\newcommand{\marco}[1]{\todo[color=green!40,inline]{#1}}
\newcommand{\marcoborder}[1]{\todo[color=green!40]{#1}}
\newcommand{\ale}[1]{\todo[color=blue!40,inline]{#1}}
\newcommand{\aleborder}[1]{\todo[color=blue!40]{#1}}
\newcommand{\eli}[1]{\todo[color=yellow!40,inline]{#1}}
\newcommand{\eliborder}[1]{\todo[color=yellow!40]{#1}}

\maketitle

\begin{abstract}
  Differential privacy offers a formal framework for reasoning about privacy and
  accuracy of computations on private data.
  It also offers a rich set of building blocks for constructing data analyses.
  When carefully calibrated, these analyses simultaneously guarantee privacy of
  the individuals contributing their data, and accuracy of their results for
  inferring useful properties about the population.
  %
  %
  The compositional nature of differential privacy has motivated the design and
  implementation of several programming languages aimed at helping a data
  analyst in programming differentially private analyses.
  However, most of the programming languages for differential privacy proposed
  so far provide support for reasoning about privacy but not for reasoning about
  accuracy of data analyses.
  To overcome this limitation, in this work we present DPella, a programming
  framework providing data analysts with support for reasoning about privacy,
  accuracy and their trade-offs.
  %
  %
  The distinguished feature of DPella is a novel component which statically
  tracks the accuracy of different data analyses.
  In order to make tighter accuracy estimations, this component leverages taint
  analysis for automatically inferring \emph{statistical independence} of the
  different noise quantities added for guaranteeing privacy.
  We show the flexibility of our approach by not only implementing classical counting
  queries (e.g., CDFs) but also by analyzing hierarchical
  counting queries (like those done by Census Bureaus), where accuracy have
  different constrains per level and data analysts should figure out the best
  manner to calibrate privacy to meet the accuracy requirements.

\end{abstract}


\section{Introduction}
\label{sec:intro}
Large amounts of individuals data are collected and stored every day
for research or statistical purposes.
Privacy concerns about the individuals contributing their data restrict the way
information can be used and released.
%
Differential privacy (DP) \cite{DMNS06} is emerging as a viable solution to
release statistical information about the population without compromising
data subjects' privacy.
A standard way to achieve DP is adding some statistical noise
to the result of a data analysis.
If the noise is carefully calibrated, it provides a \emph{privacy}
protection for the individuals contributing their data, and at the
same time it provides \emph{accurate} information about the population
from which the data are drawn.
Thanks to its quantitative formulation quantifying privacy by means of the
parameter $\epsilon$, DP provides a mathematical framework for rigorously
reasoning about the \emph{privacy-accuracy trade-offs}.
To be more precise, the accuracy requirement is not baked in the definition of
DP, rather it is a constraint that is made explicit for a
specific task at hand when a differentially private data analysis is designed.
%
%

An important property of DP is \emph{composeability}: multiple differentially
private data analyses can be composed with a graceful degradation of the privacy
parameter $\epsilon$.
This property allow one to reason about privacy as a \emph{budget}: a data
analyst can decide how much privacy budget (the $\epsilon$ parameter)to assign
to each of her analyses.
The compositionality aspects of DP motivated the design of several programming
frameworks~~\cite{PINQ09,RoySKSW10,ReedP10,HaeberlenPN11,GaboardiHHNP13,BartheGAHRS15,BartheFGAGHS16,ZhangK17,DBLP:journals/pacmpl/Winograd-CortHR17,JohnsonNS18,DBLP:conf/sigmod/ZhangMKHMM18,zhang-2019-fuzzi}
and
tools~\cite{MachanavajjhalaKAGV08,MohanTSSC12,MirICMW13,DBLP:journals/corr/GaboardiHKNUV16}
with built-in basic data analyses to help data analysts to design their own
differentially private data analysis.
At an high level, most of these programming frameworks and tools are based on a
similar idea for reasoning about privacy: use some primitives for basic tasks in
DP as building blocks, and use composition properties to combine these building
blocks making sure that the privacy cost of each data analysis sum up and that
the total cost does not exceed the privacy budget.
Programming frameworks such
as~\cite{PINQ09,RoySKSW10,ReedP10,HaeberlenPN11,GaboardiHHNP13,BartheGAHRS15,BartheFGAGHS16,ZhangK17,DBLP:journals/pacmpl/Winograd-CortHR17,JohnsonNS18,DBLP:conf/sigmod/ZhangMKHMM18,zhang-2019-fuzzi}
in addition usually also provides general support to further combine, through
programming techniques, the different building blocks and the results of the
different data analyses.
Differently, tools such
as~\cite{MachanavajjhalaKAGV08,MohanTSSC12,MirICMW13,DBLP:journals/corr/GaboardiHKNUV16}
usually are optimized for specific tasks at the price of restricting the kinds
of data analyses they can support.

Unfortunately, this simple approach for privacy cannot be directly applied to
accuracy.
Reasoning about accuracy is less compositional than reasoning about privacy, and
it depends both on the specific task at hand and on the specific accuracy
measure that one is interested in offering to data analysts.
Despite this, when restricted to specific mechanisms and specific
forms of data analyses, one can measure accuracy through estimates
given as \emph{confidence intervals}, or error bounds. 
%
As an example, most of the standard mechanisms from the differential privacy
literature come with theoretical confidence intervals or error bounds that can
be exposed to data analysts in order to allow them to take informed decisions
about the analyses that they want to run.
This approach has been integrated in tools such as GUPT~\cite{MohanTSSC12},
PSI~\cite{DBLP:journals/corr/GaboardiHKNUV16}, and Apex~\cite{Apex}.
Users of these tools, can specify the target confidence interval they want to
achieve, and the tools adjust accordingly the privacy parameters, when
sufficient budget is available\footnote{Apex actually goes beyond this by also
  helping user by selecting the right differentially private mechanism to
  achieve the required accuracy. Besides, it uses a combination off theoretical
  error bounds and statistical estimation, to provide confidence intervals for a
  large set of queries.}.

In contrast, all the programming frameworks proposed thus
far~\cite{PINQ09, RoySKSW10, ReedP10, HaeberlenPN11, GaboardiHHNP13,
  BartheGAHRS15,BartheFGAGHS16, ZhangK17,
  DBLP:journals/pacmpl/Winograd-CortHR17,JohnsonNS18,
  DBLP:conf/sigmod/ZhangMKHMM18, zhang-2019-fuzzi} do not offer any
support to programmers or data analysts for tracking, and reasoning
about, the accuracy of their data analyses.
This phenomenon is in large part due to the non-compositional nature
of accuracy when providing confidence intervals for arbitrary queries
that users of these frameworks may want to program and run.

In this paper we address this limitation by building a programming
framework for designing differentially private data analysis which
also supports a compositional form of reasoning about accuracy.
We achieve this by internalizing the use of \emph{probabilistic
  bounds}~\cite{dubhashi2009concentration} describing how to compose
different confidence intervals or error bounds.
Probabilistic bounds are part of the classical toolbox for the
analysis of randomized algorithms~\cite{dubhashi2009concentration},
and are the tools that differential privacy algorithms designers
usually employ for the accuracy analysis of classical
mechanisms~\cite{DworkR14, DworkRV10}.
Two important probabilistic  bounds are the \emph{union bound}, that
can be used to compose errors with no assumption on the way the random noise is
generated, and \emph{Chernoff bound}, which applies to the sum of random noise
when the different random variables characterizing noise generation are
statistically independent.
When applicable, and when the number of random variables grows,
Chernoff bound usually gives a much ``tighter'' error estimation than
the union bound.
We give a detailed formulation of these bounds in Section~\ref{sec:acc}.

Our main contributtion is showing that probabilistic bounds can be smoothly
integrated in a programming framework for differential privacy by using
techniques from information-flow control \cite{Sabelfeld:Myers:JSAC} (in the
form of taint analysis \cite{DBLP:conf/eurosp/SchoepeBPS16}).
While these probabilistic bounds are not enough to express every accuracy
guarantees one wants to express for arbitrary data analyses, they allow the
analysis of a large class of user-designed programs, complementing in this way
the approach followed by tools such as GUPT~\cite{MohanTSSC12},
PSI~\cite{DBLP:journals/corr/GaboardiHKNUV16}, and Apex~\cite{Apex}.

%

%
\subsection*{Our Contribution}
We present DPella, acronym for \underline{D}ifferential \underline{P}rivacy in
Hask\underline{ell} with \underline{a}ccuracy, a programming framework where
programmers and data analysts can explore the privacy-accuracy trade-off while
writing their differentially private data analyses.
The analyses that can be expressed in DPella are \emph{data independent} and can
be configured to used different norms for the accuracy of vectors.
They consist on counting queries (the bread and butter of statistical analysis),
average, and noisy max as well as any aggregation of their results.
DPella is an API implemented as a library in the general purpose language
Haskell\footnote{\href{https://www.haskell.org/}{https://www.haskell.org/}}; a
programming language that is well-known to support information-flow analyses as
libraries \cite{Li+:2010:arrows,Russo+:Haskell08}---rather than creating
interpreters or compilers from scratch as it is usually the case.
The API, and our accuracy calculations, are designed to be \emph{extensible}
through the addition of new primitives and new error measures.
In that manner, we expect DPella to expand the classes of supported analyses in
the future.

One of the main contribution of DPella is the compositional approach to
reasoning about accuracy when combining queries' results.
%
%
%
DPella extracts this information from a program implementing a data analysis
through a symbolic interpretation of the program and type-level reasoning.
More specifically, DPella builds an abstract-syntax tree corresponding to the
query that the programmer or data analyst is writing.
The tree carries information about the sensitivity\footnote{A
  quantitative measure of how much a query might amplify differences
  in the output, the formal definition is in
  Section~\ref{sec:teaser}.} of the query which is collected at the
type-level and needed to both guarantee privacy and accuracy.
DPella provides to the data analysts two distinct ways to symbolically interpret
such abstract-syntax tree: one for privacy and one for accuracy.
%
%
DPella's interpretation for privacy consists on decreasing the privacy
budget of a query by deducing the required budget of its sub-parts.
%
On the other hand, the accuracy interpretation uses as abstraction the
\emph{inverse Cumulative Distribution Function} (iCDF) representing an upper
bound to the (theoretical) error that the program incurs when guaranteeing DP.
The iCDF of a query is build out of the iCDFs of the different components, by
using as a basic composition principle the \emph{union bound}.
These interpretations provide overestimates of the corresponding quantities that
they track.
In order to make these estimates as precise as possible, DPella uses \emph{taint
  analysis} to track the use of noise to identify which variables are
\emph{statistically independent}.
This information is used by DPella to \emph{soundly} replace, when
needed, the union bounds with \emph{Chernoff bounds}, something that
to the best of our knowledge also program analyses focusing on
accuracy, such as \cite{DBLP:journals/pacmpl/SmithHA19}, do not
consider.



%
We envision these two ways to symbolically interpret queries by data analysts
as a mean to reason about privacy and accuracy.
In our experiments, we used them to implement two kind of programs in
DPella.
First, we wrote programs that repetitively call the interpreters on queries with
different values of $\epsilon$ to visualize the errors incurred by data
analyses---this allows data analyst to choose different values of $\epsilon$
depending on the accuracy that she/he wants to achieve.
Second, we used the interpretations to write optimization functions
aiming at finding the minimal $\epsilon$ for a given level of accuracy
in data analyses, this is, minimizing the \emph{privacy loss} of an
analysis.
These use cases align with the classical uses of tools such as GUPT, PSI, and
Apex.


\smallskip
In summary, our contributions are:

\selectformat{
\begin{CompactItemizeee}
\item We present DPella, a programming framework that allows data analysts to
  explore the privacy-accuracy trade-off.
  DPella combines symbolic interpretation and type-level analyses to support
  reasoning in a uniform way about privacy and accuracy.
  DPella codebase consists of just 540 lines of Haskell code.

\item We show how to use taint analysis to detect statistical independence of
  the noise that different primitives add, and how to use this information to
  achieve better error estimates.

\item We show on several examples how DPella can help data analysts to explore
  the trade-offs between the competing properties of privacy and accuracy.

\item We showcase DPella's error estimations by implementing PINQ-like
  queries from previous
  work~\cite{Mcsherry2011network,PINQ09,BartheGAHKS14} as well as some
  workloads from the matrix
  mechanism~\cite{Chao15MatMech,HayRMS10Boosting,XiaoWG11Privlet}.
\end{CompactItemizeee}
}{
\begin{itemize}[$\blacktriangleright$]
\item We present DPella, a programming framework that allows data analysts to
  explore the privacy-accuracy trade-off.
  DPella combines symbolic interpretation and type-level analyses to support
  reasoning in a uniform way about privacy and accuracy.
  DPella codebase consists of just 540 lines of Haskell code.

\item We show how to use taint analysis to detect statistical independence of
  the noise that different primitives add, and how to use this information to
  achieve better error estimates.

\item We show on several examples how DPella can help data analysts to explore
  the trade-offs between the competing properties of privacy and accuracy.

\item We showcase DPella's error estimations by implementing PINQ-like
  queries from previous
  work~\cite{Mcsherry2011network,PINQ09,BartheGAHKS14} as well as some
  workloads from the matrix
  mechanism~\cite{Chao15MatMech,HayRMS10Boosting,XiaoWG11Privlet}.
\end{itemize}
}
%




\section{DPella by example}\label{sec:teaser}




In this section, we present an overview of DPella, showcasing each of its
components.
We start by providing a brief background on the notions of privacy and
accuracy DPella considers.

\subsection{Background}

%
Differential privacy~\cite{DMNS06} is a quantitative notion of privacy that
bounds how much a single individual's private data can affect the
result of a data analysis. More formally, we can define differential
privacy as a property of a randomized query $\tilde{Q}(\cdot)$ representing
the data analysis, as follow.

\begin{definition}[{Differential Privacy (DP)}\cite{DMNS06}]
  A randomized query $\tilde{Q}(\cdot):\mathrm{db}\to R$ satisfies
  \emph{$\epsilon$-differential privacy} if and only if for any two
  datasets $D_1$ and $D_2$ in $\mathrm{db}$, which differ in one row,
  and for every output set $S\subseteq R$ we have
  \begin{equation}{\label{eq:DP}}
    \Pr[\tilde{Q}(D_1) \in S] \leq e ^{\epsilon} \Pr[\tilde{Q}(D_2) \in S]
  \end{equation}
\end{definition}

In the definition above, the parameter $\epsilon$ determines a bound
on the distance between the distributions induced by
$\tilde{Q}(\cdot)$ when adding or removing an individual from the
dataset---the farther away they are, the more at risk the privacy of
an individual is, and vice versa.
In other words, $\epsilon$ imposes a limit on the
\emph{privacy loss} that an individual can incur in, as a result of
running a data analysis.

A standard way to achieve $\epsilon$-differential privacy is adding
some carefully calibrated noise to the result of a query. To protect all the
different ways in which an individual's data can affect the result of a query,
the noise needs to be calibrated to the maximal change that the result
of the query can have when changing an individual's
data. This is formalized through the notion of \emph{sensitivity}.

\begin{definition}[\cite{DMNS06}]\label{def:sensitivity}
The \emph{(global) sensitivity} of a query $Q(\cdot):\mathrm{db}\to R$
is the quantity:
\selectformat{
\begin{align*}
\Delta_Q = \max \{ \abs{Q(D_1)-Q(D_2)} & \text{ for } D_1, D_2
             \text{ differing in} \\ & \text{~one row} \}
\end{align*}
}
{
\begin{equation*}
\Delta_Q = \max \{ \abs{Q(D_1)-Q(D_2)} \text{ for } D_1, D_2
             \text{ differing in one row} \}
\end{equation*}
}
\end{definition}

The sensitivity gives a measure of the amount of noise needed to
protect one individual's data. Besides, in order to achieve
differential privacy it is also important the choice of the kind of
noise that one adds. A standard approach is based on the addition of
noise sampled from the Laplace distribution.

\begin{theorem}[Laplace Mechanism~\cite{DMNS06}]
\label{lap:mech}
Let $Q(\cdot):\mathrm{db}\to R$ be a deterministic query with
sensitivity $\Delta_Q$. Let $\tilde{Q}(\cdot):\mathrm{db}\to R$ be a
randomized query defined as $\tilde{Q}(D)=Q(D)+\mathcal{N}$, where
$\mathcal{N}$ is sample from the Laplace distribution with mean
$\mu=0$ and scale $b=\Delta_Q/\epsilon$. Then $\tilde{Q}$ is
$\epsilon$-differentially private.
\end{theorem}

Notice that in the theorem above, for a given query, the smaller the
$\epsilon$ is, the more noise $\tilde{Q}(\cdot)$ needs to inject in
order to hide the contribution of one individual's data to the
result---this protects privacy but degrades how meaningful the result
of the query is.
In contrast, the bigger the $\epsilon$, the less noise $Q(\cdot)$
needs to inject---this increases the accuracy of the result but degrades privacy.

DPella enables data analysts and programmers to explore the trade-offs
between privacy and accuracy by giving bound estimates on the errors
caused by noise addition mechanisms, such as the Laplace mechanism
introduced above.
In general, the notion of \emph{accuracy} can be defined more formally as follows.
\begin{definition}[{Accuracy}, see e.g.\cite{DworkR14}] Given an
  $\epsilon$-differentiallly private query $\tilde{Q}(\cdot)$, a
  target query $Q(\cdot)$, a distance function $d(\cdot)$, a bound
  $\alpha$, and the probability $\beta$, we say that
  $\tilde{Q}(\cdot)$ is \emph{$(d(\cdot),\alpha,\beta)$-accurate} with
  respect to $Q(\cdot)$ if and only if for all dataset $D$:
\begin{equation}
  \Pr [d(\tilde{Q}(D) - Q(D)) > \alpha] \leq \beta
\end{equation}
\end{definition}

This definition allows one to express data independent error
statements such as: with probability at least $1-\beta$ the query
$\tilde{Q}(D)$ diverge from $Q(D)$, in terms of the distance
$d(\cdot)$, for less than $\alpha$.
Then, we will refer to $\alpha$ as the \emph{error} and $1-\beta$ as
the \emph{confidence probability} or simply \emph{confidence}.
In general, the lower the $\beta$ is, i.e., the higher the confidence
probability is, the higher the error $\alpha$ is.
On the other hand, the higher the $\beta$ is, i.e., the lower the
confidence probability is, the lower the error is. In this case, the
prediction will be valid less often than with a lower $\beta$.

As discussed in the previous section, an important property of
differential privacy is composeability.
\begin{theorem}[Sequential Composition~\cite{DMNS06}]
  Let $\tilde{Q}_1(\cdot)$ and $\tilde{Q}_2(\cdot)$ be two queries
  which are $\epsilon_1$- and $\epsilon_2$-differentially private,
  respectively.  Then, their sequential composition
  $\tilde{Q}(\cdot)=(\tilde{Q}_1(\cdot), \tilde{Q}_2(\cdot))$ is
  $(\epsilon_1 + \epsilon_2)$-differentially private.
\end{theorem}

\begin{theorem}[Parallel Composition~\cite{PINQ09}]
  \label{theo:parallel}
  Let $\tilde{Q}(\cdot)$ be a $\epsilon$-differentially private query.
  and $\mathrm{data}_1,\mathrm{data}_2$ be a partition of the set of
  data.  Then, the query
  $\tilde{Q}_1(D)=(\tilde{Q}(D\cap\mathrm{data_1}), \tilde{Q}(D\cap \mathrm{data}_2))$ is
  $\epsilon$-differentially private.
\end{theorem}
Thanks to the composition properties of differential privacy, we can
think about $\epsilon$ as a \emph{privacy budget} that one can spend
on a given data before compromising the privacy of individuals'
contributions to that data.
The \emph{global} $\epsilon$ for a given program can be seen as the
privacy budget for the entire data. This budget can be consumed by selecting the
\emph{local} $\epsilon$ to ``spend'' in each intermediate query.
Thanks to the composition properties, by tracking the local $\epsilon$
that are consumed, one can guarantee that a data analysis will not
consume more than the allocated privacy budget.

%

Given an $\epsilon$, DPella gives data analysts the possibility to
explore how to spend it on different queries and analyze the impact on
accuracy.
For instance, data analysts might decide to spend ``more'' epsilon on
sub-queries which results are required to be more accurate, while
spending ``less'' on the others.
The next examples (inspired by the use of DP in network trace
analyses~\cite{Mcsherry2011network}) show how DPella helps to quantify
what ``more'' and ``less'' means.

\subsection{Example: CDF }

Suppose we have a \emph{tcpdump} trace of packets which yields a table
where each row is represented as list of \ensuremath{\Conid{String}} values containing
the following information:

\numbersoff
\selectformat{
{\small
\begin{hscode}\linenumsetup\printlinebegin\SaveRestoreHook
\column{B}{@{}>{\hspre}c<{\hspost}@{}}%
\column{BE}{@{}l@{}}%
\column{4}{@{}>{\hspre}l<{\hspost}@{}}%
\column{E}{@{}>{\hspre}l<{\hspost}@{}}%
\>[B]{}[\mskip1.5mu {}\<[BE]%
\>[4]{}{\color{darkblue}  \texttt{<}}\Varid{id}{\color{darkblue}  \texttt{>}},{\color{darkblue}  \texttt{<}}\Varid{timestamp}{\color{darkblue}  \texttt{>}},{\color{darkblue}  \texttt{<}}\Varid{src}{\color{darkblue}  \texttt{>}},{\color{darkblue}  \texttt{<}}\Varid{dest}{\color{darkblue}  \texttt{>}},{\color{darkblue}  \texttt{<}}\Varid{protocol}{\color{darkblue}  \texttt{>}},{}\<[E]%
\printlineend\\
\printlinebegin\>[4]{}{\color{darkblue}  \texttt{<}}\Varid{length}{\color{darkblue}  \texttt{>}},{\color{darkblue}  \texttt{<}}\Varid{payload}{\color{darkblue}  \texttt{>}}\mskip1.5mu]{}\<[E]%
\printlineend\ColumnHook
\end{hscode}\resethooks
}}{
\begin{hscode}\linenumsetup\printlinebegin\SaveRestoreHook
\column{B}{@{}>{\hspre}l<{\hspost}@{}}%
\column{E}{@{}>{\hspre}l<{\hspost}@{}}%
\>[B]{}[\mskip1.5mu {\color{darkblue}  \texttt{<}}\Varid{id}{\color{darkblue}  \texttt{>}},{\color{darkblue}  \texttt{<}}\Varid{timestamp}{\color{darkblue}  \texttt{>}},{\color{darkblue}  \texttt{<}}\Varid{src}{\color{darkblue}  \texttt{>}},{\color{darkblue}  \texttt{<}}\Varid{dest}{\color{darkblue}  \texttt{>}},{\color{darkblue}  \texttt{<}}\Varid{protocol}{\color{darkblue}  \texttt{>}},{\color{darkblue}  \texttt{<}}\Varid{length}{\color{darkblue}  \texttt{>}},{\color{darkblue}  \texttt{<}}\Varid{payload}{\color{darkblue}  \texttt{>}}\mskip1.5mu]{}\<[E]%
\printlineend\ColumnHook
\end{hscode}\resethooks
}
From this table, we would like to inspect---in a differentially
private manner---the packet's length distribution by computing its
Cumulative Distribution function (CDF), defined as
$\text{CDF}(x) =$ number of records with value~$\leq x$.
Hence, we are just interested in the values of the attribute \ensuremath{{\color{darkblue}  \texttt{<}}\Varid{length}{\color{darkblue}  \texttt{>}}}.

In order to guarantee differential privacy,  we need to add some
randomness to our computation. So, the result will be an approximated version
of the original CDF.
How to best approximate a data analysis often depends on several
properties of the data analysis itself.
For the case of CDF,~\citet{Mcsherry2011network} proposed three
different ways to approximate it, and they argued for their different
levels of accuracy.
We revise two of these approximations here (and the third one can be found in
the extended version of the paper) to show how DPella can assist in showing the
accuracy of these analyses.





\begin{figure}[t]
\centering
\begin{subfigure}[b,tight]{1\columnwidth}
\centering
\numberson
{\small
\begin{hscode}\linenumsetup\printlinebegin\SaveRestoreHook
\column{B}{@{}>{\hspre}l<{\hspost}@{}}%
\column{3}{@{}>{\hspre}l<{\hspost}@{}}%
\column{5}{@{}>{\hspre}l<{\hspost}@{}}%
\column{11}{@{}>{\hspre}c<{\hspost}@{}}%
\column{11E}{@{}l@{}}%
\column{12}{@{}>{\hspre}l<{\hspost}@{}}%
\column{15}{@{}>{\hspre}l<{\hspost}@{}}%
\column{25}{@{}>{\hspre}c<{\hspost}@{}}%
\column{25E}{@{}l@{}}%
\column{28}{@{}>{\hspre}l<{\hspost}@{}}%
\column{32}{@{}>{\hspre}l<{\hspost}@{}}%
\column{50}{@{}>{\hspre}l<{\hspost}@{}}%
\column{E}{@{}>{\hspre}l<{\hspost}@{}}%
\>[B]{}\tt{cdf}_{1}\;\Varid{bins}\;\Varid{eps}\;\Varid{dataset}\mathrel{=}\mathbf{do}{}\<[E]%
\printlineend\\
\printlinebegin\>[B]{}\hsindent{3}{}\<[3]%
\>[3]{}\Varid{sizes}{}\<[11]%
\>[11]{}\leftarrow {}\<[11E]%
\>[15]{}{\color{darkred} \tt{dpSelect} }\;\Varid{getPktLen}\;\Varid{dataset}{}\<[E]%
\printlineend\\
\printlinebegin\>[B]{}\hsindent{3}{}\<[3]%
\>[3]{}\Varid{counts}{}\<[11]%
\>[11]{}\leftarrow {}\<[11E]%
\>[15]{}\Varid{sequence}\;{}\<[25]%
\>[25]{}[\mskip1.5mu {}\<[25E]%
\>[28]{}\mathbf{do}\;{}\<[32]%
\>[32]{}\Varid{elems}\leftarrow {\color{darkred} \tt{dpWhere} }\;{}\<[50]%
\>[50]{}(\leq \Varid{bin})\;{}\<[E]%
\printlineend\\
\printlinebegin\>[50]{}\Varid{sizes}{}\<[E]%
\printlineend\\
\printlinebegin\>[32]{}{\color{darkblue} \tt{dpCount} }\;\Varid{localEps}\;\Varid{elems}{}\<[E]%
\printlineend\\
\printlinebegin\>[25]{}\mid {}\<[25E]%
\>[28]{}\Varid{bin}\leftarrow \Varid{bins}\mskip1.5mu]{}\<[E]%
\printlineend\\
\printlinebegin\>[B]{}\hsindent{3}{}\<[3]%
\>[3]{}\Varid{return}\;({\color{darkgreen} \tt{norm}_{\infty} }\;\Varid{counts}){}\<[E]%
\printlineend\\
\printlinebegin\>[3]{}\hsindent{2}{}\<[5]%
\>[5]{}\mathbf{where}\;{}\<[12]%
\>[12]{}\Varid{localEps}\mathrel{=}\Varid{eps}\mathbin{/}(\Varid{length}\;\Varid{bins}){}\<[E]%
\printlineend\ColumnHook
\end{hscode}\resethooks
}
\caption{Sequential approach \label{fig:cdf1}}
\end{subfigure}
\begin{subfigure}[b,tight]{1\columnwidth}
\centering
{\small
\begin{hscode}\linenumsetup\printlinebegin\SaveRestoreHook
\column{B}{@{}>{\hspre}l<{\hspost}@{}}%
\column{3}{@{}>{\hspre}l<{\hspost}@{}}%
\column{8}{@{}>{\hspre}l<{\hspost}@{}}%
\column{21}{@{}>{\hspre}c<{\hspost}@{}}%
\column{21E}{@{}l@{}}%
\column{24}{@{}>{\hspre}l<{\hspost}@{}}%
\column{26}{@{}>{\hspre}l<{\hspost}@{}}%
\column{E}{@{}>{\hspre}l<{\hspost}@{}}%
\>[B]{}\tt{cdf}_{2}\;\Varid{bins}\;\Varid{eps}\;\Varid{dataset}\mathrel{=}\mathbf{do}{}\<[E]%
\printlineend\\
\printlinebegin\>[B]{}\hsindent{3}{}\<[3]%
\>[3]{}\Varid{sizes}\leftarrow {\color{darkred} \tt{dpSelect} }\;((\leq \Varid{max}\;\Varid{bins})\mathbin{\circ}\Varid{getPktLen})\;\Varid{dataset}{}\<[E]%
\printlineend\\
\printlinebegin\>[B]{}\hsindent{3}{}\<[3]%
\>[3]{}\mbox{\onelinecomment  \ensuremath{\Varid{parts}\mathbin{::}\Conid{Map}\;\Conid{Integer}\;({\color{darkgreen} \tt{Value}}\;\Conid{Double})}}{}\<[E]%
\printlineend\\
\printlinebegin\>[B]{}\hsindent{3}{}\<[3]%
\>[3]{}\Varid{parts}\leftarrow {\color{darkred} \tt{dpPartRepeat} }\;{}\<[26]%
\>[26]{}({\color{darkblue} \tt{dpCount} }\;\Varid{eps})\;\Varid{bins}\;\Varid{assignBin}\;{}\<[E]%
\printlineend\\
\printlinebegin\>[26]{}\Varid{sizes}{}\<[E]%
\printlineend\\
\printlinebegin\>[B]{}\hsindent{3}{}\<[3]%
\>[3]{}\mathbf{let}\;{}\<[8]%
\>[8]{}\Varid{counts}{}\<[21]%
\>[21]{}\mathrel{=}{}\<[21E]%
\>[24]{}\Varid{\Conid{Map}.elems}\;\Varid{parts}{}\<[E]%
\printlineend\\
\printlinebegin\>[8]{}\Varid{cumulCounts}{}\<[21]%
\>[21]{}\mathrel{=}{}\<[21E]%
\>[24]{}[\mskip1.5mu {\color{darkgreen} \tt{add} }\;(\Varid{take}\;\Varid{i}\;\Varid{counts}){}\<[E]%
\printlineend\\
\printlinebegin\>[24]{}\mid \Varid{i}\leftarrow [\mskip1.5mu {\color{darkgreen} \texttt{1}}\mathinner{\ldotp\ldotp}\Varid{length}\;\Varid{counts}\mskip1.5mu]\mskip1.5mu]{}\<[E]%
\printlineend\\
\printlinebegin\>[B]{}\hsindent{3}{}\<[3]%
\>[3]{}\Varid{return}\;({\color{darkgreen} \tt{norm}_{\infty} }\;\Varid{cumulCounts}){}\<[E]%
\printlineend\ColumnHook
\end{hscode}\resethooks
}\numbersoff
\caption{Parallel approach \label{fig:cdf2}}
\end{subfigure}
\caption{CDF's implementations \label{fig:cdfs_imp}}
\end{figure}

\subsubsection{Sequential \ensuremath{\tt{CDF}}}
  A simple approach to compute the CDF consists in splitting the
  range of lengths into \ensuremath{\Varid{bins}} and, for each \ensuremath{\Varid{bin}}, count the number
  of records that are \ensuremath{\leq \Varid{bin}}. A natural way to make this computation
  differentially private is to add independent Laplace noise to each
  count.
  %

  We show how to do this using DPella in Figure~\ref{fig:cdf1}.
  We define a function \ensuremath{\tt{cdf}_{1}} which takes as input the list of \ensuremath{\Varid{bins}}
  describing length ranges, the amount of budget \ensuremath{\Varid{eps}} to be spent by
  the entire query, and the \ensuremath{\Varid{dataset}} where it will be computed.
  For now, we assume that we have a fixed list of bins for packets' length.
%
  \ensuremath{\tt{cdf}_{1}} uses the primitive transformation\footnote{Anticipating on
    Section~\ref{sec:pinq}, in our code we will usually use the red
    color for transformations, the blue color for aggregate
    operations, and the green color for combinators for privacy and
    accuracy.} \ensuremath{{\color{darkred} \tt{dpSelect} }} provided by DPella to obtain from the
  dataset the length of each packet via a selector function
  \ensuremath{\Varid{getPktLen}\mathbin{::}\Conid{String}\to \Conid{Integer}}. This computation results in a new
  dataset \ensuremath{\Varid{sizes}}.
  Then,  we create a counting query for each bin using the primitive
  \ensuremath{{\color{darkred} \tt{dpWhere} }} provided by DPella. This filters all records that are less
  than the \ensuremath{\Varid{bin}} under consideration  (\ensuremath{\leq \Varid{bin}}). Finally, we perform
  a noisy count using the DPella primitive \ensuremath{{\color{darkblue} \tt{dpCount} }}.
  The noise injected by the primitive \ensuremath{{\color{darkblue} \tt{dpCount} }} is calibrated so that
  the execution of \ensuremath{{\color{darkblue} \tt{dpCount} }} is \ensuremath{\Varid{localEps}}-DP (line 8~\footnote{The
    casting operation \ensuremath{\Varid{fromIntegral}} is omitted for clarity}).
  The function \ensuremath{\Varid{sequence}} then takes the list of queries and compute them
  sequentially collecting their results in a list---to create a list of noisy
  counts.
  We then return this list. The combinator \ensuremath{{\color{darkgreen} \tt{norm}_{\infty} }} in line 7 is used
  to mark where we want the accuracy information to be collected, but
  it does not have any impact on the actual result of the cdf.


  On the privacy side, DPella provides primitives to statically explore the
  privacy budget. For instance, to ensure that \ensuremath{\tt{cdf}_{1}} is \ensuremath{\Varid{eps}}-differential
  privacy, we distributed the given budget \ensuremath{\Varid{eps}} evenly among the sub-queries
  (this is done in lines 5 and 8).
  However, a data analyst may forget to do so, e.g., she can define \ensuremath{\Varid{localEps}\mathrel{=}\Varid{eps}}, and in this case the final query is \ensuremath{(\Varid{length}\;\Varid{bins}){\color{darkblue}  \texttt{*}}\Varid{eps}}-DP, which is a
  significant change in the query's privacy price.
  To prevent such budget miscalculations or unintended expenditure of privacy
  budget, DPella provides the analyst with the function \ensuremath{{\color{darkblue} \tt{budget} }} (see
  Section~\ref{sec:pinq}) that, given a query, statically computes an upper
  bound on how much budget it will spend.  To see how to use this function,
  consider the function \ensuremath{\tt{cdf}_{1}} and a its modified version \ensuremath{\tt{cdf}_{1}'} with \ensuremath{\Varid{localEps}\mathrel{=}\Varid{eps}}. Suppose that we want to compute how much budget will be consumed by
  running it on a list of bins of size 10 (identified as \ensuremath{{\tt{bins}_{10}}}) and on a
  dataset \ensuremath{\Varid{networkTraffic}}. Then, the data analyst can ask this as follow:
  \numbersoff
  \begin{hscode}\linenumsetup\printlinebegin\SaveRestoreHook
\column{B}{@{}>{\hspre}l<{\hspost}@{}}%
\column{5}{@{}>{\hspre}l<{\hspost}@{}}%
\column{8}{@{}>{\hspre}l<{\hspost}@{}}%
\column{E}{@{}>{\hspre}l<{\hspost}@{}}%
\>[5]{}{\color{darkblue}  \texttt{>}}{}\<[8]%
\>[8]{}{\color{darkblue} \tt{budget} }\;(\tt{cdf}_{1}\;{\tt{bins}_{10}}\;{\color{darkgreen} \texttt{1}}\;\Varid{networkTraffic}){}\<[E]%
\printlineend\\
\printlinebegin\>[5]{}\epsilon\mathrel{=}{\color{darkgreen} \texttt{1}}{}\<[E]%
\printlineend\\[\blanklineskip]%
\printlinebegin\>[5]{}{\color{darkblue}  \texttt{>}}{}\<[8]%
\>[8]{}{\color{darkblue} \tt{budget} }\;(\tt{cdf}_{1}'\;{\tt{bins}_{10}}\;{\color{darkgreen} \texttt{1}}\;\Varid{networkTraffic}){}\<[E]%
\printlineend\\
\printlinebegin\>[5]{}\epsilon\mathrel{=}{\color{darkgreen} \texttt{10}}{}\<[E]%
\printlineend\ColumnHook
\end{hscode}\resethooks
  %
  The function \ensuremath{{\color{darkblue} \tt{budget} }} \emph{will not} execute the query, it simply performs an
  static analysis on the code of the query by symbolically interpreting it.
  The static analysis uses information encoded by the \emph{type} of the
  database \ensuremath{\Varid{networkTraffic}} (explained in Section \ref{sec:pinq}).
  For \ensuremath{{\color{darkblue} \tt{budget} }} to work, it is not necessary to provide the real dataset, just a
  dataset with the same static information.

  DPella also provides primitives to statically explore the accuracy of a
  query. The function \ensuremath{{\color{darkblue} \tt{accuracy} }} takes a query $Q(\cdot)$ and a probability
  $\beta$ and returns an estimate of the (theoretical) error that can be
  achieved with confidence probability $1-\beta$.
  Suppose that we want to estimate the error we will incur in by running \ensuremath{\tt{cdf}_{1}}
  with a budget of $\epsilon=1$ on with the same list of bins and dataset as
  above, and we want to have this estimate for $\beta=0.05$ and $\beta=0.2$,
  respectively.
  Then, the data analyst can ask this as follow:
  \numbersoff
  \begin{hscode}\linenumsetup\printlinebegin\SaveRestoreHook
\column{B}{@{}>{\hspre}l<{\hspost}@{}}%
\column{5}{@{}>{\hspre}l<{\hspost}@{}}%
\column{8}{@{}>{\hspre}l<{\hspost}@{}}%
\column{E}{@{}>{\hspre}l<{\hspost}@{}}%
\>[5]{}{\color{darkblue}  \texttt{>}}{}\<[8]%
\>[8]{}{\color{darkblue} \tt{accuracy} }\;(\tt{cdf}_{1}\;{\tt{bins}_{10}}\;{\color{darkgreen} \texttt{1}}\;\Varid{networkTraffic})\;{\color{darkgreen} \texttt{0.05}}{}\<[E]%
\printlineend\\
\printlinebegin\>[5]{}\alpha\mathrel{=}{\color{darkgreen} \texttt{53}}{}\<[E]%
\printlineend\\[\blanklineskip]%
\printlinebegin\>[5]{}{\color{darkblue}  \texttt{>}}{}\<[8]%
\>[8]{}{\color{darkblue} \tt{accuracy} }\;(\tt{cdf}_{1}\;{\tt{bins}_{10}}\;{\color{darkgreen} \texttt{1}}\;\Varid{networkTraffic})\;{\color{darkgreen} \texttt{0.2}}{}\<[E]%
\printlineend\\
\printlinebegin\>[5]{}\alpha\mathrel{=}{\color{darkgreen} \texttt{40}}{}\<[E]%
\printlineend\ColumnHook
\end{hscode}\resethooks

  Since the result of the query is a vector of counts, we measure the error
  $\alpha$ in terms of $\ell_\infty$ distance with respect to the CDF without
  noise.
  This is the max difference that we can have in a bin due to the noise.
  The way to read the information provided by DPella is that with confidence
  $95\%$ and $80\%$, we have errors $53$ and $40$, respectively.
  %
These error bounds can be used by a data analyst to figure out the
exact set of parameters that would be useful for her task.

\subsubsection{Parallel \ensuremath{\tt{CDF}}}
Another way to compute a CDF is by first generating an histogram of the
data according to the bins, and then building a cumulative sum for each
bin. To make this function private, an approach could be to add noise
at the different bins of the histogram, rather than to the cumulative sums
themself, so that we could use the parallel composition, rather than
the sequential one~\cite{Mcsherry2011network},
which we show how to implement in DPella in Figure~\ref{fig:cdf2}.
The symbol \ensuremath{\mathbin{::}} is used to describe the type of terms in Haskell and double-dashes
are used to introduce single-line comments.

In \ensuremath{\tt{cdf}_{2}}, we first select all the packages whose length is smaller
than the maximum bin, and then we partition the data accordingly to
the given list of bins.
To do this, we use \ensuremath{{\color{darkred} \tt{dpPartRepeat} }} operator to create as many (disjoint)
datasets as given \ensuremath{\Varid{bins}}, where each record in each partition belongs to the
range determined by an specific \ensuremath{\Varid{bin}}---where the record belongs is determined
by the function \ensuremath{\Varid{assignBin}\mathbin{::}\Conid{Integer}\to \Conid{Integer}}.
%
%
After creating all partitions, the primitive \ensuremath{{\color{darkred} \tt{dpPartRepeat} }} computes the given
query \ensuremath{{\color{darkblue} \tt{dpCount} }\;\Varid{eps}} in \emph{each partition}---the name \ensuremath{{\color{darkred} \tt{dpPartRepeat} }} comes
from repetitively calling \ensuremath{{\color{darkblue} \tt{dpCount} }\;\Varid{eps}} as many times as partitions we have.
As a result, \ensuremath{{\color{darkred} \tt{dpPartRepeat} }} returns a finite map where the keys are the \ensuremath{\Varid{bins}}
and the elements are the noisy count of the records per partition---i.e., the
histogram.
In what follows (lines 15--17), we compute the cumulative sums of the
noisy counts using the DPella primitive \ensuremath{{\color{darkgreen} \tt{add} }}, and finally we build
and return the list of values denoting the CDF.
%

The privacy analysis of \ensuremath{\tt{cdf}_{2}} is similar to the one of \ensuremath{\tt{cdf}_{1}}, except
that one can use the parallel composition, rather then the sequential
one.
\begin{hscode}\linenumsetup\printlinebegin\SaveRestoreHook
\column{B}{@{}>{\hspre}c<{\hspost}@{}}%
\column{BE}{@{}l@{}}%
\column{4}{@{}>{\hspre}l<{\hspost}@{}}%
\column{E}{@{}>{\hspre}l<{\hspost}@{}}%
\>[B]{}{\color{darkblue}  \texttt{>}}{}\<[BE]%
\>[4]{}{\color{darkblue} \tt{budget} }\;(\tt{cdf}_{2}\;{\tt{bins}_{10}}\;{\color{darkgreen} \texttt{1}}\;\Varid{networkTraffic}){}\<[E]%
\printlineend\\
\printlinebegin\>[4]{}\epsilon\mathrel{=}{\color{darkgreen} \texttt{1}}{}\<[E]%
\printlineend\ColumnHook
\end{hscode}\resethooks

%
The accuracy analysis of \ensuremath{\tt{cdf}_{2}} is more interesting: first it gets error
bounds for each cumulative sum, then these are used to give an error bound on
the maximum error of the vector.
The symbolic analysis of DPella implements this combination in an effective
way.
For the error bounds on the cumulative sums DPella uses either the union bound
or the Chernoff bound, depending on which one gives the lowest error.
For the maximum error of the vector, DPella uses the union bound,
similarly to what happens in \ensuremath{\tt{cdf}_{1}}.
Using DPella, also in this case, a data analyst can explore the accuracy of
\ensuremath{\tt{cdf}_{2}}.

\begin{hscode}\linenumsetup\printlinebegin\SaveRestoreHook
\column{B}{@{}>{\hspre}l<{\hspost}@{}}%
\column{5}{@{}>{\hspre}l<{\hspost}@{}}%
\column{8}{@{}>{\hspre}l<{\hspost}@{}}%
\column{E}{@{}>{\hspre}l<{\hspost}@{}}%
\>[5]{}{\color{darkblue}  \texttt{>}}{}\<[8]%
\>[8]{}{\color{darkblue} \tt{accuracy} }\;(\tt{cdf}_{2}\;{\tt{bins}_{10}}\;{\color{darkgreen} \texttt{1}}\;\Varid{networkTraffic})\;{\color{darkgreen} \texttt{0.05}}{}\<[E]%
\printlineend\\
\printlinebegin\>[5]{}\alpha\mathrel{=}{\color{darkgreen} \texttt{22}}{}\<[E]%
\printlineend\\[\blanklineskip]%
\printlinebegin\>[5]{}{\color{darkblue}  \texttt{>}}{}\<[8]%
\>[8]{}{\color{darkblue} \tt{accuracy} }\;(\tt{cdf}_{2}\;{\tt{bins}_{10}}\;{\color{darkgreen} \texttt{1}}\;\Varid{networkTraffic})\;{\color{darkgreen} \texttt{0.2}}{}\<[E]%
\printlineend\\
\printlinebegin\>[5]{}\alpha\mathrel{=}{\color{darkgreen} \texttt{20}}{}\<[E]%
\printlineend\ColumnHook
\end{hscode}\resethooks

\subsubsection{Exploring the privacy-accuracy trade-off}

\begin{figure}[t]
\vspace{-10pt}
\centering
\begin{tikzpicture}
  \begin{axis}[
    height=5cm,
    width=\selectformat{1\columnwidth}{0.7\textwidth},
    ylabel= $\alpha$,
    xlabel= Sub-queries,
    axis lines*=left,
    legend pos=north west,
    legend cell align={left},
    grid = major,
    grid style={dashed, gray!30},
    axis line style={->},
    every axis y label/.style={
      at={(ticklabel* cs:1.05)},
      anchor=south,},
    legend image post style={scale=0.5},
    ]
\addplot[red]
    table [x=bins, y=empiricalCDF1, col sep=comma]{./graphs/errComparison95.csv};
\addplot[blue]
    table [x=bins, y=theoreticalCDF1, col sep=comma]{./graphs/errComparison95.csv};
\addplot[dashed,red]
    table [x=bins, y=empiricalCDF2, col sep=comma]{./graphs/errComparison95.csv};
\addplot[dashed,blue]
    table [x=bins, y=theoreticalCDF2, col sep=comma]{./graphs/errComparison95.csv};
    \legend{\ensuremath{\tt{cdf}_{1}} Empiric, \ensuremath{\tt{cdf}_{1}} Theoretic
      , \ensuremath{\tt{cdf}_{2}} Empiric, \ensuremath{\tt{cdf}_{2}} Theoretic}
\end{axis}
\end{tikzpicture}
\caption{Error comparison (95\% confidence)\label{fig:comp}}
\end{figure}
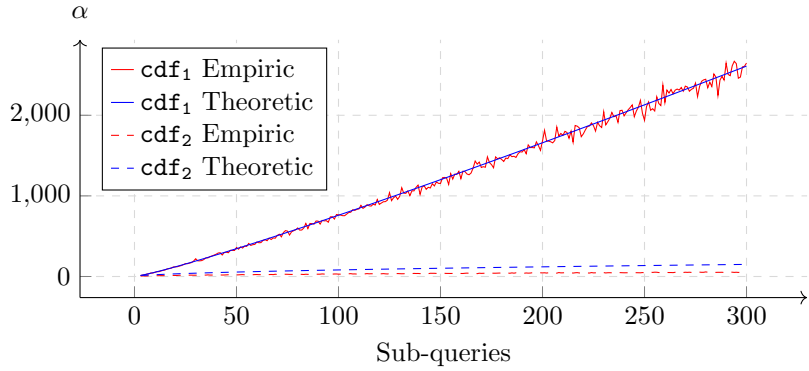

Let us assume that a data analyst is interested in running a CDF with
an error bounded with $90\%$ confidence, i.e., with $\beta = 0.1$,
having three bins (named \ensuremath{{\tt{bins}_3}}), and $\epsilon=1$.
With those assumptions in mind, which implementation should she use?
To answer that question, the data analyst can ask DPella:
%
\begin{hscode}\linenumsetup\printlinebegin\SaveRestoreHook
\column{B}{@{}>{\hspre}c<{\hspost}@{}}%
\column{BE}{@{}l@{}}%
\column{4}{@{}>{\hspre}l<{\hspost}@{}}%
\column{E}{@{}>{\hspre}l<{\hspost}@{}}%
\>[B]{}{\color{darkblue}  \texttt{>}}{}\<[BE]%
\>[4]{}{\color{darkblue} \tt{accuracy} }\;(\tt{cdf}_{1}\;{\tt{bins}_3}\;{\color{darkgreen} \texttt{1}}\;\Varid{networkTraffic})\;{\color{darkgreen} \texttt{0.1}}{}\<[E]%
\printlineend\\
\printlinebegin\>[4]{}\alpha\mathrel{=}{\color{darkgreen} \texttt{11}}{}\<[E]%
\printlineend\\
\printlinebegin\>[B]{}{\color{darkblue}  \texttt{>}}{}\<[BE]%
\>[4]{}{\color{darkblue} \tt{accuracy} }\;(\tt{cdf}_{2}\;{\tt{bins}_3}\;{\color{darkgreen} \texttt{1}}\;\Varid{networkTraffic})\;{\color{darkgreen} \texttt{0.1}}{}\<[E]%
\printlineend\\
\printlinebegin\>[4]{}\alpha\mathrel{=}{\color{darkgreen} \texttt{12}}{}\<[E]%
\printlineend\ColumnHook
\end{hscode}\resethooks
So, the analyst would know that using \ensuremath{\tt{cdf}_{1}} in this case would
give, likely, a lower error.
Suppose further that the data analyst realize that she prefers to have
a finer granularity and have 10 bins, instead of only 3.
Which implementation should she use?
Again, she can compute:
\begin{hscode}\linenumsetup\printlinebegin\SaveRestoreHook
\column{B}{@{}>{\hspre}c<{\hspost}@{}}%
\column{BE}{@{}l@{}}%
\column{4}{@{}>{\hspre}l<{\hspost}@{}}%
\column{E}{@{}>{\hspre}l<{\hspost}@{}}%
\>[B]{}{\color{darkblue}  \texttt{>}}{}\<[BE]%
\>[4]{}{\color{darkblue} \tt{accuracy} }\;(\tt{cdf}_{1}\;{\tt{bins}_{10}}\;{\color{darkgreen} \texttt{1}}\;\Varid{networkTraffic})\;{\color{darkgreen} \texttt{0.1}}{}\<[E]%
\printlineend\\
\printlinebegin\>[4]{}\alpha\mathrel{=}{\color{darkgreen} \texttt{46}}{}\<[E]%
\printlineend\\
\printlinebegin\>[B]{}{\color{darkblue}  \texttt{>}}{}\<[BE]%
\>[4]{}{\color{darkblue} \tt{accuracy} }\;(\tt{cdf}_{2}\;{\tt{bins}_{10}}\;{\color{darkgreen} \texttt{1}}\;\Varid{networkTraffic})\;{\color{darkgreen} \texttt{0.1}}{}\<[E]%
\printlineend\\
\printlinebegin\>[4]{}\alpha\mathrel{=}{\color{darkgreen} \texttt{20}}{}\<[E]%
\printlineend\ColumnHook
\end{hscode}\resethooks
So, the data analyst would know that using \ensuremath{\tt{cdf}_{2}} in this case would
give, likely, a lower error.
One can also use DPella to show a comparison between \ensuremath{\tt{cdf}_{1}} and \ensuremath{\tt{cdf}_{2}}
in terms of error when we keep the privacy parameter fixed and we
change the number of bins, where \ensuremath{\tt{cdf}_{2}} gives a better error when the
number of bins is large~\cite{Mcsherry2011network} as illustrated in
Figure~\ref{fig:comp}.
%
%
In the figure, we also show the empirical error to confirm that our estimate is
tight---the oscillations on the empirical \ensuremath{\tt{cdf}_{1}} are given by the relative small
(300) number of experimental runs we consider.

%
%
Now, what if the data analyst choose to use \ensuremath{\tt{cdf}_{2}} because of what we
discussed before but she realizes that she can afford an error
$\alpha \leq 50$;
%
what would be then the epsilon that gives such $\alpha$?
One of the feature of DPella is that the analyst \emph{can write a simple
  program that finds it by repetitively calling \ensuremath{{\color{darkblue} \tt{accuracy} }} with different
  epsilons}---this is one of the advantages of providing a programming
framework.
In this case, the answer is 0.42:
\begin{hscode}\linenumsetup\printlinebegin\SaveRestoreHook
\column{B}{@{}>{\hspre}c<{\hspost}@{}}%
\column{BE}{@{}l@{}}%
\column{4}{@{}>{\hspre}l<{\hspost}@{}}%
\column{E}{@{}>{\hspre}l<{\hspost}@{}}%
\>[B]{}{\color{darkblue}  \texttt{>}}{}\<[BE]%
\>[4]{}{\color{darkblue} \tt{accuracy} }\;(\tt{cdf}_{2}\;{\tt{bins}_{10}}\;{\color{darkgreen} \texttt{0.42}}\;\Varid{networkTraffic})\;{\color{darkgreen} \texttt{0.1}}{}\<[E]%
\printlineend\\
\printlinebegin\>[4]{}\alpha\mathrel{=}{\color{darkgreen} \texttt{49}}{}\<[E]%
\printlineend\ColumnHook
\end{hscode}\resethooks
%
These different use cases shows the flexibility of DPella for
different tasks in private data analysis.
The following sections will introduce the theoretical and technical
aspects of DPella, more concretely, Section~\ref{sec:pinq} presents
DPella's API and how does it enforce DP.
Section~\ref{sec:acc} describes the technicalities of the accuracy
calculations, followed by Section~\ref{sec:examples} where we expose
the versatility of the framework with case studies.
%
%
Lastly, we put our framework in context presenting some related work
in Section~\ref{sec:relWork} followed by the conclusions and future work
(Section~\ref{sec:conclusions}).

\section{Privacy}
\label{sec:pinq}



%
%
A query written in DPella can be thought as a sequence of operations responsible
to transform sensitive dataset for then obtaining aggregated data like the
amount of rows in a dataset.
%
%
%
%
DPella provides noised version of data aggregation operations by implementing a
\emph{Laplace mechanism}~as described in Theorem \ref{lap:mech}.
%
%
%
To ensure that data releases satisfy differential privacy despite
transformations, the noise injected is calibrated to the impact that such
transformations can have on the data---a concept known as \emph{stability}
\cite{PINQ09} (explained below).
Different than, e.g., PINQ \cite{PINQ09}, one novelty of DPella is that it
computes stability \emph{statically} using Haskell's type-system.
%

%

Queries written in DPella have an invariant enforced by construction: \emph{it
  is not possible to branch on results produced by aggregations}.
While it might seem restrictive, it enables to write counting queries, which are
the bread and butter of statistical analysis and have been the focus of the
majority of the work in DP.
Despite such restriction, DPella can be extended with data analyses branching on
noisy-values by simply incorporating them as black-box primitives, e.g.,
noisy-max is incorporated in this manner by DPella.

%

\paragraph{Terminology}
DPella have two kind of actors: \emph{data curators}, who decide the global
privacy budget and split that budget among data analysts, and \emph{data
  analysts}, who write queries to mine useful information from the data and
spend the budget they received.
DPella is designed to help data analysts to have an informed decision about how
to spend their budget based on exploring the trade-offs between privacy and
accuracy.

\subsection{Components of the API}
\begin{figure}
\centering
\numbersreset
\numbersoff

{\small
\selectformat{
\begin{hscode}\linenumsetup\printlinebegin\SaveRestoreHook
\column{B}{@{}>{\hspre}l<{\hspost}@{}}%
\column{10}{@{}>{\hspre}l<{\hspost}@{}}%
\column{14}{@{}>{\hspre}l<{\hspost}@{}}%
\column{15}{@{}>{\hspre}c<{\hspost}@{}}%
\column{15E}{@{}l@{}}%
\column{19}{@{}>{\hspre}l<{\hspost}@{}}%
\column{20}{@{}>{\hspre}l<{\hspost}@{}}%
\column{23}{@{}>{\hspre}c<{\hspost}@{}}%
\column{23E}{@{}l@{}}%
\column{26}{@{}>{\hspre}l<{\hspost}@{}}%
\column{27}{@{}>{\hspre}l<{\hspost}@{}}%
\column{30}{@{}>{\hspre}l<{\hspost}@{}}%
\column{34}{@{}>{\hspre}c<{\hspost}@{}}%
\column{34E}{@{}l@{}}%
\column{36}{@{}>{\hspre}l<{\hspost}@{}}%
\column{38}{@{}>{\hspre}l<{\hspost}@{}}%
\column{41}{@{}>{\hspre}l<{\hspost}@{}}%
\column{42}{@{}>{\hspre}l<{\hspost}@{}}%
\column{48}{@{}>{\hspre}l<{\hspost}@{}}%
\column{59}{@{}>{\hspre}l<{\hspost}@{}}%
\column{E}{@{}>{\hspre}l<{\hspost}@{}}%
\>[B]{}\mbox{\onelinecomment  Transformations (data analyst)}{}\<[E]%
\printlineend\\
\printlinebegin\>[B]{}{\color{darkred} \tt{dpWhere} }{}\<[15]%
\>[15]{}\mathbin{::}{}\<[15E]%
\>[19]{}({\color{darkblue} \tt{r} }\to \Conid{Bool})\to {\color{darkred} \tt{Data}}\;{\color{darkblue} \tt{s} }\;{\color{darkblue} \tt{r} }\to {\color{darkblue} \tt{Query}}\;({\color{darkred} \tt{Data}}\;{\color{darkblue} \tt{s} }\;{\color{darkblue} \tt{r} }){}\<[E]%
\printlineend\\[\blanklineskip]%
\printlinebegin\>[B]{}{\color{darkred} \tt{dpGroupBy} }{}\<[15]%
\>[15]{}\mathbin{::}{}\<[15E]%
\>[19]{}\Conid{Eq}\;\Varid{k}\Rightarrow ({\color{darkblue} \tt{r} }\to \Varid{k})\to {}\<[42]%
\>[42]{}{\color{darkred} \tt{Data}}\;{\color{darkblue} \tt{s} }\;{\color{darkblue} \tt{r} }{}\<[E]%
\printlineend\\
\printlinebegin\>[15]{}\to {}\<[15E]%
\>[20]{}{\color{darkblue} \tt{Query}}\;({\color{darkred} \tt{Data}}\;({\color{darkgreen} \texttt{2}}{\color{darkblue}  \texttt{*}}{\color{darkblue} \tt{s} })\;(\Varid{k},[\mskip1.5mu {\color{darkblue} \tt{r} }\mskip1.5mu])){}\<[E]%
\printlineend\\[\blanklineskip]%
\printlinebegin\>[B]{}{\color{darkred} \tt{dpIntersect} }{}\<[15]%
\>[15]{}\mathbin{::}{}\<[15E]%
\>[19]{}\Conid{Eq}\;{\color{darkblue} \tt{r} }{}\<[27]%
\>[27]{}\Rightarrow {\color{darkred} \tt{Data}}\;{\color{darkblue} \tt{s} }_{1}\;{\color{darkblue} \tt{r} }\to {}\<[48]%
\>[48]{}{\color{darkred} \tt{Data}}\;{\color{darkblue} \tt{s} }_{2}\;{\color{darkblue} \tt{r} }{}\<[E]%
\printlineend\\
\printlinebegin\>[15]{}\to {}\<[15E]%
\>[19]{}{\color{darkblue} \tt{Query}}\;({\color{darkred} \tt{Data}}\;({\color{darkblue} \tt{s} }_{1}{\color{darkblue}  \texttt{+}}{\color{darkblue} \tt{s} }_{2})\;{\color{darkblue} \tt{r} }){}\<[E]%
\printlineend\\[\blanklineskip]%
\printlinebegin\>[B]{}{\color{darkred} \tt{dpSelect} }{}\<[15]%
\>[15]{}\mathbin{::}{}\<[15E]%
\>[19]{}({\color{darkblue} \tt{r} }\to {\color{darkblue} \tt{r} }')\to {\color{darkred} \tt{Data}}\;{\color{darkblue} \tt{s} }\;{\color{darkblue} \tt{r} }\to {\color{darkblue} \tt{Query}}\;({\color{darkred} \tt{Data}}\;{\color{darkblue} \tt{s} }\;{\color{darkblue} \tt{r} }'){}\<[E]%
\printlineend\\[\blanklineskip]%
\printlinebegin\>[B]{}{\color{darkred} \tt{dpUnion} }{}\<[15]%
\>[15]{}\mathbin{::}{}\<[15E]%
\>[19]{}{\color{darkred} \tt{Data}}\;{\color{darkblue} \tt{s} }_{1}\;{\color{darkblue} \tt{r} }{}\<[34]%
\>[34]{}\to {}\<[34E]%
\>[38]{}{\color{darkred} \tt{Data}}\;{\color{darkblue} \tt{s} }_{2}\;{\color{darkblue} \tt{r} }{}\<[E]%
\printlineend\\
\printlinebegin\>[15]{}\to {}\<[15E]%
\>[19]{}{\color{darkblue} \tt{Query}}\;{}\<[26]%
\>[26]{}({\color{darkred} \tt{Data}}\;({\color{darkblue} \tt{s} }_{1}{\color{darkblue}  \texttt{+}}{\color{darkblue} \tt{s} }_{2})\;{\color{darkblue} \tt{r} }){}\<[E]%
\printlineend\\[\blanklineskip]%
\printlinebegin\>[B]{}{\color{darkred} \tt{dpPart} }{}\<[15]%
\>[15]{}\mathbin{::}{}\<[15E]%
\>[19]{}\Conid{Ord}\;\Varid{k}{}\<[26]%
\>[26]{}\Rightarrow {}\<[30]%
\>[30]{}({\color{darkblue} \tt{r} }\to \Varid{k})\to {\color{darkred} \tt{Data}}\;{\color{darkblue} \tt{s} }\;{\color{darkblue} \tt{r} }{}\<[E]%
\printlineend\\
\printlinebegin\>[15]{}\to {}\<[15E]%
\>[19]{}\Conid{Map}\;\Varid{k}\;({\color{darkred} \tt{Data}}\;{\color{darkblue} \tt{s} }\;{\color{darkblue} \tt{r} }){}\<[41]%
\>[41]{}\to {\color{darkblue} \tt{Query}}\;({\color{darkgreen} \tt{Value}}\;\Varid{a})){}\<[E]%
\printlineend\\
\printlinebegin\>[15]{}\to {}\<[15E]%
\>[19]{}{\color{darkblue} \tt{Query}}\;(\Conid{Map}\;\Varid{k}\;({\color{darkgreen} \tt{Value}}\;\Varid{a})){}\<[E]%
\printlineend\\[\blanklineskip]%
\printlinebegin\>[B]{}\mbox{\onelinecomment  Aggregations (data analyst)}{}\<[E]%
\printlineend\\
\printlinebegin\>[B]{}{\color{darkblue} \tt{dpCount} }{}\<[10]%
\>[10]{}\mathbin{::}{}\<[14]%
\>[14]{}\Conid{Stb}\;{\color{darkblue} \tt{s} }{}\<[23]%
\>[23]{}\Rightarrow {}\<[23E]%
\>[27]{}\epsilon{}\<[36]%
\>[36]{}\to {\color{darkred} \tt{Data}}\;{\color{darkblue} \tt{s} }\;{\color{darkblue} \tt{r} }\to {\color{darkblue} \tt{Query}}\;({\color{darkgreen} \tt{Value}}\;\Conid{Double}){}\<[E]%
\printlineend\\
\printlinebegin\>[B]{}{\color{darkblue} \tt{dpSum} }{}\<[10]%
\>[10]{}\mathbin{::}{}\<[14]%
\>[14]{}\Conid{Stb}\;{\color{darkblue} \tt{s} }{}\<[23]%
\>[23]{}\Rightarrow {}\<[23E]%
\>[27]{}\epsilon{}\<[36]%
\>[36]{}\to ({\color{darkblue} \tt{r} }\to \Conid{Double})\to {}\<[59]%
\>[59]{}{\color{darkred} \tt{Data}}\;{\color{darkblue} \tt{s} }\;{\color{darkblue} \tt{r} }{}\<[E]%
\printlineend\\
\printlinebegin\>[10]{}\to {}\<[14]%
\>[14]{}{\color{darkblue} \tt{Query}}\;({\color{darkgreen} \tt{Value}}\;\Conid{Double}){}\<[E]%
\printlineend\\
\printlinebegin\>[B]{}{\color{darkblue} \tt{dpAvg} }{}\<[10]%
\>[10]{}\mathbin{::}{}\<[14]%
\>[14]{}\Conid{Stb}\;{\color{darkblue} \tt{s} }{}\<[23]%
\>[23]{}\Rightarrow {}\<[23E]%
\>[27]{}\epsilon{}\<[36]%
\>[36]{}\to ({\color{darkblue} \tt{r} }\to \Conid{Double})\to {}\<[59]%
\>[59]{}{\color{darkred} \tt{Data}}\;{\color{darkblue} \tt{s} }\;{\color{darkblue} \tt{r} }{}\<[E]%
\printlineend\\
\printlinebegin\>[10]{}\to {}\<[14]%
\>[14]{}{\color{darkblue} \tt{Query}}\;({\color{darkgreen} \tt{Value}}\;\Conid{Double}){}\<[E]%
\printlineend\\
\printlinebegin\>[B]{}{\color{darkblue} \tt{dpMax} }{}\<[10]%
\>[10]{}\mathbin{::}{}\<[14]%
\>[14]{}\epsilon\to ({\color{darkblue} \tt{r} }\to \Conid{Double})\to {\color{darkred} \tt{Data}}\;{\color{darkgreen} \texttt{1}}\;{\color{darkblue} \tt{r} }{}\<[E]%
\printlineend\\
\printlinebegin\>[10]{}\to {\color{darkblue} \tt{Query}}\;({\color{darkgreen} \tt{Value}}\;\Conid{Integer}){}\<[E]%
\printlineend\\[\blanklineskip]%
\printlinebegin\>[B]{}\mbox{\onelinecomment  Budget}{}\<[E]%
\printlineend\\
\printlinebegin\>[B]{}{\color{darkblue} \tt{budget} }\mathbin{::}{\color{darkblue} \tt{Query}}\;\Varid{a}\to \epsilon{}\<[E]%
\printlineend\\
\printlinebegin\>[B]{}\mbox{\onelinecomment  Execution (data curator)}{}\<[E]%
\printlineend\\
\printlinebegin\>[B]{}{\color{darkred} \tt{dpEval} }\mathbin{::}({\color{darkred} \tt{Data}}\;{\color{darkgreen} \texttt{1}}\;{\color{darkblue} \tt{r} }\to {\color{darkblue} \tt{Query}}\;({\color{darkgreen} \tt{Value}}\;\Varid{a}))\to [\mskip1.5mu {\color{darkblue} \tt{r} }\mskip1.5mu]\to \epsilon\to \Conid{IO}\;\Varid{a}{}\<[E]%
\printlineend\ColumnHook
\end{hscode}\resethooks
}{
\begin{hscode}\linenumsetup\printlinebegin\SaveRestoreHook
\column{B}{@{}>{\hspre}l<{\hspost}@{}}%
\column{10}{@{}>{\hspre}c<{\hspost}@{}}%
\column{10E}{@{}l@{}}%
\column{14}{@{}>{\hspre}l<{\hspost}@{}}%
\column{15}{@{}>{\hspre}c<{\hspost}@{}}%
\column{15E}{@{}l@{}}%
\column{19}{@{}>{\hspre}l<{\hspost}@{}}%
\column{23}{@{}>{\hspre}c<{\hspost}@{}}%
\column{23E}{@{}l@{}}%
\column{27}{@{}>{\hspre}c<{\hspost}@{}}%
\column{27E}{@{}l@{}}%
\column{31}{@{}>{\hspre}l<{\hspost}@{}}%
\column{36}{@{}>{\hspre}l<{\hspost}@{}}%
\column{46}{@{}>{\hspre}l<{\hspost}@{}}%
\column{50}{@{}>{\hspre}l<{\hspost}@{}}%
\column{59}{@{}>{\hspre}l<{\hspost}@{}}%
\column{65}{@{}>{\hspre}c<{\hspost}@{}}%
\column{65E}{@{}l@{}}%
\column{69}{@{}>{\hspre}l<{\hspost}@{}}%
\column{76}{@{}>{\hspre}l<{\hspost}@{}}%
\column{94}{@{}>{\hspre}l<{\hspost}@{}}%
\column{E}{@{}>{\hspre}l<{\hspost}@{}}%
\>[B]{}\mbox{\onelinecomment  Transformations (data analyst)}{}\<[E]%
\printlineend\\
\printlinebegin\>[B]{}{\color{darkred} \tt{dpWhere} }{}\<[15]%
\>[15]{}\mathbin{::}{}\<[15E]%
\>[19]{}({\color{darkblue} \tt{r} }\to \Conid{Bool}){}\<[46]%
\>[46]{}\to {}\<[50]%
\>[50]{}{\color{darkred} \tt{Data}}\;{\color{darkblue} \tt{s} }\;{\color{darkblue} \tt{r} }{}\<[65]%
\>[65]{}\to {}\<[65E]%
\>[69]{}{\color{darkblue} \tt{Query}}\;({\color{darkred} \tt{Data}}\;{\color{darkblue} \tt{s} }\;{\color{darkblue} \tt{r} }){}\<[E]%
\printlineend\\
\printlinebegin\>[B]{}{\color{darkred} \tt{dpGroupBy} }{}\<[15]%
\>[15]{}\mathbin{::}{}\<[15E]%
\>[19]{}\Conid{Eq}\;\Varid{k}{}\<[27]%
\>[27]{}\Rightarrow {}\<[27E]%
\>[31]{}({\color{darkblue} \tt{r} }\to \Varid{k}){}\<[46]%
\>[46]{}\to {}\<[50]%
\>[50]{}{\color{darkred} \tt{Data}}\;{\color{darkblue} \tt{s} }\;{\color{darkblue} \tt{r} }{}\<[65]%
\>[65]{}\to {}\<[65E]%
\>[69]{}{\color{darkblue} \tt{Query}}\;({\color{darkred} \tt{Data}}\;({\color{darkgreen} \texttt{2}}{\color{darkblue}  \texttt{*}}{\color{darkblue} \tt{s} })\;(\Varid{k},[\mskip1.5mu {\color{darkblue} \tt{r} }\mskip1.5mu])){}\<[E]%
\printlineend\\
\printlinebegin\>[B]{}{\color{darkred} \tt{dpIntersect} }{}\<[15]%
\>[15]{}\mathbin{::}{}\<[15E]%
\>[19]{}\Conid{Eq}\;{\color{darkblue} \tt{r} }{}\<[27]%
\>[27]{}\Rightarrow {}\<[27E]%
\>[31]{}{\color{darkred} \tt{Data}}\;{\color{darkblue} \tt{s} }_{1}\;{\color{darkblue} \tt{r} }{}\<[46]%
\>[46]{}\to {}\<[50]%
\>[50]{}{\color{darkred} \tt{Data}}\;{\color{darkblue} \tt{s} }_{2}\;{\color{darkblue} \tt{r} }{}\<[65]%
\>[65]{}\to {}\<[65E]%
\>[69]{}{\color{darkblue} \tt{Query}}\;({\color{darkred} \tt{Data}}\;({\color{darkblue} \tt{s} }_{1}{\color{darkblue}  \texttt{+}}{\color{darkblue} \tt{s} }_{2})\;{\color{darkblue} \tt{r} }){}\<[E]%
\printlineend\\
\printlinebegin\>[B]{}{\color{darkred} \tt{dpSelect} }{}\<[15]%
\>[15]{}\mathbin{::}{}\<[15E]%
\>[19]{}({\color{darkblue} \tt{r} }\to {\color{darkblue} \tt{r} }'){}\<[46]%
\>[46]{}\to {}\<[50]%
\>[50]{}{\color{darkred} \tt{Data}}\;{\color{darkblue} \tt{s} }\;{\color{darkblue} \tt{r} }{}\<[65]%
\>[65]{}\to {}\<[65E]%
\>[69]{}{\color{darkblue} \tt{Query}}\;({\color{darkred} \tt{Data}}\;{\color{darkblue} \tt{s} }\;{\color{darkblue} \tt{r} }'){}\<[E]%
\printlineend\\
\printlinebegin\>[B]{}{\color{darkred} \tt{dpUnion} }{}\<[15]%
\>[15]{}\mathbin{::}{}\<[15E]%
\>[19]{}{\color{darkred} \tt{Data}}\;{\color{darkblue} \tt{s} }_{1}\;{\color{darkblue} \tt{r} }{}\<[46]%
\>[46]{}\to {}\<[50]%
\>[50]{}{\color{darkred} \tt{Data}}\;{\color{darkblue} \tt{s} }_{2}\;{\color{darkblue} \tt{r} }{}\<[65]%
\>[65]{}\to {}\<[65E]%
\>[69]{}{\color{darkblue} \tt{Query}}\;{}\<[76]%
\>[76]{}({\color{darkred} \tt{Data}}\;({\color{darkblue} \tt{s} }_{1}{\color{darkblue}  \texttt{+}}{\color{darkblue} \tt{s} }_{2})\;{\color{darkblue} \tt{r} }){}\<[E]%
\printlineend\\
\printlinebegin\>[B]{}{\color{darkred} \tt{dpPart} }{}\<[15]%
\>[15]{}\mathbin{::}{}\<[15E]%
\>[19]{}\Conid{Ord}\;\Varid{k}{}\<[27]%
\>[27]{}\Rightarrow {}\<[27E]%
\>[31]{}({\color{darkblue} \tt{r} }\to \Varid{k}){}\<[46]%
\>[46]{}\to {\color{darkred} \tt{Data}}\;{\color{darkblue} \tt{s} }\;{\color{darkblue} \tt{r} }{}\<[65]%
\>[65]{}\to {}\<[65E]%
\>[69]{}\Conid{Map}\;\Varid{k}\;({\color{darkred} \tt{Data}}\;{\color{darkblue} \tt{s} }\;{\color{darkblue} \tt{r} })\to {}\<[94]%
\>[94]{}{\color{darkblue} \tt{Query}}\;({\color{darkgreen} \tt{Value}}\;\Varid{a})){}\<[E]%
\printlineend\\
\printlinebegin\>[27]{}\to {}\<[27E]%
\>[31]{}{\color{darkblue} \tt{Query}}\;(\Conid{Map}\;\Varid{k}\;({\color{darkgreen} \tt{Value}}\;\Varid{a})){}\<[E]%
\printlineend\\[\blanklineskip]%
\printlinebegin\>[B]{}\mbox{\onelinecomment  Aggregations (data analyst)}{}\<[E]%
\printlineend\\
\printlinebegin\>[B]{}{\color{darkblue} \tt{dpCount} }{}\<[10]%
\>[10]{}\mathbin{::}{}\<[10E]%
\>[14]{}\Conid{Stb}\;{\color{darkblue} \tt{s} }{}\<[23]%
\>[23]{}\Rightarrow {}\<[23E]%
\>[27]{}\epsilon{}\<[27E]%
\>[36]{}\to {\color{darkred} \tt{Data}}\;{\color{darkblue} \tt{s} }\;{\color{darkblue} \tt{r} }\to {\color{darkblue} \tt{Query}}\;({\color{darkgreen} \tt{Value}}\;\Conid{Double}){}\<[E]%
\printlineend\\
\printlinebegin\>[B]{}{\color{darkblue} \tt{dpSum} }{}\<[10]%
\>[10]{}\mathbin{::}{}\<[10E]%
\>[14]{}\Conid{Stb}\;{\color{darkblue} \tt{s} }{}\<[23]%
\>[23]{}\Rightarrow {}\<[23E]%
\>[27]{}\epsilon{}\<[27E]%
\>[36]{}\to ({\color{darkblue} \tt{r} }\to \Conid{Double})\to {}\<[59]%
\>[59]{}{\color{darkred} \tt{Data}}\;{\color{darkblue} \tt{s} }\;{\color{darkblue} \tt{r} }\to {}\<[76]%
\>[76]{}{\color{darkblue} \tt{Query}}\;({\color{darkgreen} \tt{Value}}\;\Conid{Double}){}\<[E]%
\printlineend\\
\printlinebegin\>[B]{}{\color{darkblue} \tt{dpAvg} }{}\<[10]%
\>[10]{}\mathbin{::}{}\<[10E]%
\>[14]{}\Conid{Stb}\;{\color{darkblue} \tt{s} }{}\<[23]%
\>[23]{}\Rightarrow {}\<[23E]%
\>[27]{}\epsilon{}\<[27E]%
\>[36]{}\to ({\color{darkblue} \tt{r} }\to \Conid{Double})\to {}\<[59]%
\>[59]{}{\color{darkred} \tt{Data}}\;{\color{darkblue} \tt{s} }\;{\color{darkblue} \tt{r} }\to {}\<[76]%
\>[76]{}{\color{darkblue} \tt{Query}}\;({\color{darkgreen} \tt{Value}}\;\Conid{Double}){}\<[E]%
\printlineend\\
\printlinebegin\>[B]{}{\color{darkblue} \tt{dpMax} }{}\<[10]%
\>[10]{}\mathbin{::}{}\<[10E]%
\>[14]{}\epsilon\to ({\color{darkblue} \tt{r} }\to \Conid{Double})\to {\color{darkred} \tt{Data}}\;{\color{darkgreen} \texttt{1}}\;{\color{darkblue} \tt{r} }\to {\color{darkblue} \tt{Query}}\;({\color{darkgreen} \tt{Value}}\;\Conid{Integer}){}\<[E]%
\printlineend\\[\blanklineskip]%
\printlinebegin\>[B]{}\mbox{\onelinecomment  Budget}{}\<[E]%
\printlineend\\
\printlinebegin\>[B]{}{\color{darkblue} \tt{budget} }\mathbin{::}{\color{darkblue} \tt{Query}}\;\Varid{a}\to \epsilon{}\<[E]%
\printlineend\\
\printlinebegin\>[B]{}\mbox{\onelinecomment  Execution (data curator)}{}\<[E]%
\printlineend\\
\printlinebegin\>[B]{}{\color{darkred} \tt{dpEval} }\mathbin{::}({\color{darkred} \tt{Data}}\;{\color{darkgreen} \texttt{1}}\;{\color{darkblue} \tt{r} }\to {\color{darkblue} \tt{Query}}\;({\color{darkgreen} \tt{Value}}\;\Varid{a}))\to [\mskip1.5mu {\color{darkblue} \tt{r} }\mskip1.5mu]\to \epsilon\to \Conid{IO}\;\Varid{a}{}\<[E]%
\printlineend\ColumnHook
\end{hscode}\resethooks
}}

\vspace{-5pt}
\caption{DPella API: Part I\label{fig:dpella:dp}}
\vspace{-10pt}
\end{figure}
%
Figure \ref{fig:dpella:dp} shows part of DPella API.
DPella introduces two abstract data types to respectively denote datasets
and queries:
\numbersoff
\begin{hscode}\linenumsetup\printlinebegin\SaveRestoreHook
\column{B}{@{}>{\hspre}l<{\hspost}@{}}%
\column{5}{@{}>{\hspre}l<{\hspost}@{}}%
\column{25}{@{}>{\hspre}l<{\hspost}@{}}%
\column{E}{@{}>{\hspre}l<{\hspost}@{}}%
\>[5]{}\mathbf{data}\;{\color{darkred} \tt{Data}}\;{\color{darkblue} \tt{s} }\;{\color{darkblue} \tt{r} }{}\<[25]%
\>[25]{}\mbox{\onelinecomment  datasets}{}\<[E]%
\printlineend\\
\printlinebegin\>[5]{}\mathbf{data}\;{\color{darkblue} \tt{Query}}\;\Varid{a}{}\<[25]%
\>[25]{}\mbox{\onelinecomment  queries}{}\<[E]%
\printlineend\ColumnHook
\end{hscode}\resethooks
  The attentive reader might have observed that the API also introduces the data
  type \ensuremath{{\color{darkgreen} \tt{Value}}\;\Varid{a}}.
However, we defer its explanation for Section \ref{sec:acc} since it is only
used for accuracy calculations---for this section, readers can consider the
type \ensuremath{{\color{darkgreen} \tt{Value}}\;\Varid{a}} as isomorphic to the type \ensuremath{\Varid{a}}.
Values of type \ensuremath{{\color{darkred} \tt{Data}}\;{\color{darkblue} \tt{s} }\;{\color{darkblue} \tt{r} }} represent sensitive datasets with
\emph{accumulated stability} \ensuremath{{\color{darkblue} \tt{s} }}, where each row is of type \ensuremath{{\color{darkblue} \tt{r} }}.
%
%
%
Accumulated stability, on the other hand, is instantiated to type-level positive
natural numbers, i.e., \ensuremath{{\color{darkgreen} \texttt{1}}}, \ensuremath{{\color{darkgreen} \texttt{2}}}, etc.
Stability is a measure that captures the number of rows in the dataset that
could have been affected by transformations like selection or grouping of rows.
In DP research, stability is associated with dataset transformations rather than
with datasets themselves.
In order to simplify type signatures, DPella uses the type parameter \ensuremath{s} in
datasets to represent the accumulated stability of the transformations for which
datasets have gone through.
%
%

Values of type \ensuremath{{\color{darkblue} \tt{Query}}\;\Varid{a}} represent \emph{computations}, or queries, that yield
values of type \ensuremath{\Varid{a}}.
%
Type \ensuremath{{\color{darkblue} \tt{Query}}\;\Varid{a}} is a monad \cite{DBLP:journals/iandc/Moggi91}, and
%
because of this, computations of type \ensuremath{{\color{darkblue} \tt{Query}}\;\Varid{a}} are built by two fundamental
operations:
\begin{hscode}\linenumsetup\printlinebegin\SaveRestoreHook
\column{B}{@{}>{\hspre}l<{\hspost}@{}}%
\column{3}{@{}>{\hspre}l<{\hspost}@{}}%
\column{11}{@{}>{\hspre}c<{\hspost}@{}}%
\column{11E}{@{}l@{}}%
\column{15}{@{}>{\hspre}l<{\hspost}@{}}%
\column{E}{@{}>{\hspre}l<{\hspost}@{}}%
\>[3]{}\Varid{return}{}\<[11]%
\>[11]{}\mathbin{::}{}\<[11E]%
\>[15]{}\Varid{a}\to {\color{darkblue} \tt{Query}}\;\Varid{a}{}\<[E]%
\printlineend\\
\printlinebegin\>[3]{}(\bind ){}\<[11]%
\>[11]{}\mathbin{::}{}\<[11E]%
\>[15]{}{\color{darkblue} \tt{Query}}\;\Varid{a}\to (\Varid{a}\to {\color{darkblue} \tt{Query}}\;\Varid{b})\to {\color{darkblue} \tt{Query}}\;\Varid{b}{}\<[E]%
\printlineend\ColumnHook
\end{hscode}\resethooks
The operation \ensuremath{\Varid{return}\;\Varid{x}} returns a query that just produces the value \ensuremath{\Varid{x}}
without causing side-effects, i.e., without touching any dataset.
The function \ensuremath{(\bind )}---called \emph{bind}---is used to sequence queries and
their associated side-effects.
Specifically, \ensuremath{\Varid{qp}\bind \Varid{f}} executes the query \ensuremath{\Varid{qp}}, takes its \textit{result}, and
passes it to the function \ensuremath{\Varid{f}}, which then returns a second query to run.
Some languages, like Haskell, provide syntactic sugar for monadic computations
known as \ensuremath{\mathbf{do}}-notation.
For instance, the program \ensuremath{\tt{qp}_{1}\bind (\lambda \tt{x}_{1}\to \tt{qp}_{2}\bind (\lambda \tt{x}_{2}\to \Varid{return}\;(\tt{x}_{1},\tt{x}_{2})))},
which performs queries \ensuremath{\tt{qp}_{1}} and \ensuremath{\tt{qp}_{2}} and returns their results in a pair, can
be written as
\ensuremath{\mathbf{do}\;\tt{x}_{1}\leftarrow \tt{qp}_{1};\tt{x}_{2}\leftarrow \tt{qp}_{2};\Varid{return}\;(\tt{x}_{1},\tt{x}_{2})}
which gives a more ``imperative'' feeling to programs.
We split the API in four parts: transformations, aggregations, budget
prediction, and execution of queries.
(We defer to the next section the description of the API related to accuracy
calculations.)
The first three parts are intended to be used by data analysts, while the last
one is intended to be \emph{only} used by data curators\footnote{A separation
  that can be enforced via Haskell modules \cite{DBLP:conf/haskell/TereiMJM12}}.
%


\subsubsection{Transformations}
The primitive \ensuremath{{\color{darkred} \tt{dpWhere} }} filters rows in datasets based on a predicate functions (\ensuremath{{\color{darkblue} \tt{r} }\to \Conid{Bool}}).
The created query (of type \ensuremath{{\color{darkblue} \tt{Query}}\;({\color{darkred} \tt{Data}}\;{\color{darkblue} \tt{s} }\;{\color{darkblue} \tt{r} })}) produces a dataset with the
same row type \ensuremath{{\color{darkblue} \tt{r} }} and accumulated stability \ensuremath{{\color{darkblue} \tt{s} }} as the dataset given as
argument (\ensuremath{{\color{darkred} \tt{Data}}\;{\color{darkblue} \tt{s} }\;{\color{darkblue} \tt{r} }}).
Observe that if we consider two datasets which differ in \ensuremath{{\color{darkblue} \tt{s} }} rows in two given
executions, and we apply \ensuremath{{\color{darkred} \tt{dpWhere} }} to both of them, we will obtain datasets that
will still differ in \ensuremath{{\color{darkblue} \tt{s} }} rows---thus, the accumulated stability remains the
same.
The primitive \ensuremath{{\color{darkred} \tt{dpGroupBy} }} returns a dataset where rows with the same key are grouped
together.
The functional argument (of type \ensuremath{{\color{darkblue} \tt{r} }\to \Varid{k}}) maps rows to keys of type \ensuremath{\Varid{k}}.
The rows in the return dataset (\ensuremath{{\color{darkred} \tt{Data}}\;({\color{darkgreen} \texttt{2}}{\color{darkblue}  \texttt{*}}s)\;(\Varid{k},[\mskip1.5mu {\color{darkblue} \tt{r} }\mskip1.5mu])}) consist of key-rows
pairs of type \ensuremath{(\Varid{k},[\mskip1.5mu {\color{darkblue} \tt{r} }\mskip1.5mu])}---syntax \ensuremath{[\mskip1.5mu {\color{darkblue} \tt{r} }\mskip1.5mu]} denotes the type of lists of elements of
type \ensuremath{{\color{darkblue} \tt{r} }}.
What appears on the left-hand side of the symbol \ensuremath{\Rightarrow } are type constraints.
They can be seen as static demands for the types appearing on the right-hand
side of \ensuremath{\Rightarrow }.
Type constraint \ensuremath{\Conid{Eq}\;\Varid{k}} demands type \ensuremath{\Varid{k}}, denoting keys, to support equality;
otherwise grouping rows with the same keys is not possible.
The accumulated stability of the new dataset is multiplied by \ensuremath{{\color{darkgreen} \texttt{2}}} in accordance
with stability calculations for transformations
\cite{PINQ09,DBLP:journals/corr/EbadiS15}---observe that \ensuremath{{\color{darkgreen} \texttt{2}}{\color{darkblue}  \texttt{*}}{\color{darkblue} \tt{s} }} is a type-level
multiplication done by a type-level function (or type family
\cite{DBLP:conf/popl/EisenbergVJW14}) \ensuremath{{\color{darkblue}  \texttt{*}}}.
%
%
Our API also considers transformations similar to those found in SQL like
intersection (\ensuremath{{\color{darkred} \tt{dpIntersect} }}), union (\ensuremath{{\color{darkred} \tt{dpUnion} }}), and selection
(\ensuremath{{\color{darkred} \tt{dpSelect} }}) of datasets, where the accumulated stability is updated accordingly.
We do not provide joins as transformations since supporting them is knwon to be
challenging
\cite{PINQ09,DBLP:conf/osdi/NarayanH12,DBLP:conf/innovations/BlockiBDS13,DBLP:journals/pvldb/JohnsonNS18}.
The output of a join may contain duplicates of sensitive rows, which makes
difficult to bound the accumulated stability of datasets.
DPella therefore assumes that all the considered information is contained by the
rows of given datasets.

\subsubsection{Partition}
\label{sec:partition}
Primitive \ensuremath{{\color{darkred} \tt{dpPart} }} deserves special attention.
This primitive is a mixture of a transformation and aggregations since it
partitions the data (transformation) to subsequently apply aggregations on each
of them.
More specifically, this primitive partitions the given dataset (\ensuremath{{\color{darkred} \tt{Data}}\;{\color{darkblue} \tt{s} }\;{\color{darkblue} \tt{r} }})
based on a row-to-key mapping (\ensuremath{{\color{darkblue} \tt{r} }\to \Varid{k}}).
Then, it takes a partition for a given key \ensuremath{\Varid{k}} and applies it to the
corresponding function \ensuremath{{\color{darkred} \tt{Data}}\;{\color{darkblue} \tt{s} }\;{\color{darkblue} \tt{r} }\to {\color{darkblue} \tt{Query}}\;({\color{darkgreen} \tt{Value}}\;\Varid{a})}, which is given as an element
of a key-query mapping (\ensuremath{\Conid{Map}\;\Varid{k}\;(({\color{darkred} \tt{Data}}\;{\color{darkblue} \tt{s} }\;{\color{darkblue} \tt{r} })\to {\color{darkblue} \tt{Query}}\;({\color{darkgreen} \tt{Value}}\;\Varid{a}))}).
Subsequently, it returns the values produced at every partition as a key-value
mapping (\ensuremath{{\color{darkblue} \tt{Query}}\;(\Conid{Map}\;\Varid{k}\;({\color{darkgreen} \tt{Value}}\;\Varid{a}))}).
The primitive \ensuremath{{\color{darkred} \tt{dpPartRepeat} }} used by the examples in Section \ref{sec:teaser} is
implemented as a special case of \ensuremath{{\color{darkred} \tt{dpPart} }} and thus we do not discuss it
further.
%

\begin{figure}
\centering
\numbersreset
\numberson
{\small
\begin{hscode}\linenumsetup\printlinebegin\SaveRestoreHook
\column{B}{@{}>{\hspre}l<{\hspost}@{}}%
\column{4}{@{}>{\hspre}l<{\hspost}@{}}%
\column{5}{@{}>{\hspre}c<{\hspost}@{}}%
\column{5E}{@{}l@{}}%
\column{9}{@{}>{\hspre}l<{\hspost}@{}}%
\column{10}{@{}>{\hspre}l<{\hspost}@{}}%
\column{18}{@{}>{\hspre}l<{\hspost}@{}}%
\column{30}{@{}>{\hspre}l<{\hspost}@{}}%
\column{33}{@{}>{\hspre}l<{\hspost}@{}}%
\column{37}{@{}>{\hspre}l<{\hspost}@{}}%
\column{E}{@{}>{\hspre}l<{\hspost}@{}}%
\>[B]{}q{}\<[5]%
\>[5]{}\mathbin{::}{}\<[5E]%
\>[9]{}\epsilon\to [\mskip1.5mu \Conid{Color}\mskip1.5mu]\to {\color{darkred} \tt{Data}}\;{\color{darkgreen} \texttt{1}}\;\Conid{Double}{}\<[E]%
\printlineend\\
\printlinebegin\>[5]{}\to {}\<[5E]%
\>[9]{}{\color{darkblue} \tt{Query}}\;(\Conid{Map}\;\Conid{Color}\;\Conid{Double}){}\<[E]%
\printlineend\\
\printlinebegin\>[B]{}q\;\Varid{eps}\;\Varid{bins}\;\Varid{dataset}\mathrel{=}{\color{darkred} \tt{dpPart} }\;\Varid{id}\;\Varid{dataset}\;\Varid{dps}{}\<[E]%
\printlineend\\
\printlinebegin\>[B]{}\hsindent{4}{}\<[4]%
\>[4]{}\mathbf{where}\;\Varid{dps}\mathrel{=}{}\<[18]%
\>[18]{}\Varid{fromList}\;[\mskip1.5mu {}\<[30]%
\>[30]{}({}\<[33]%
\>[33]{}\Varid{c},{}\<[37]%
\>[37]{}\lambda \Varid{ds}\to {\color{darkblue} \tt{dpCount} }\;\Varid{eps}\;\Varid{dataset}){}\<[E]%
\printlineend\\
\printlinebegin\>[4]{}\hsindent{6}{}\<[10]%
\>[10]{}\mbox{\onelinecomment  \ensuremath{\Varid{dps}\mathrel{=}\Varid{fromList}\;[\mskip1.5mu (\Varid{c},\lambda \Varid{ds}\to {\color{darkblue} \tt{dpCount} }\;\Varid{eps}\;\Varid{ds}}}{}\<[E]%
\printlineend\\
\printlinebegin\>[10]{}\hsindent{20}{}\<[30]%
\>[30]{}\mid \Varid{c}\leftarrow \Varid{bins}\mskip1.5mu]{}\<[E]%
\printlineend\ColumnHook
\end{hscode}\resethooks
}
%
\caption{DP-histograms by using \ensuremath{{\color{darkred} \tt{dpPart} }}\label{fig:partition}}
\end{figure}

Partition is one of the most important operators to save privacy budget.
It allows to run the same query on a dataset's partitions but only paying for
one of them---recall Theorem~\ref{theo:parallel}.
The essential assumption that makes this possible is that every query runs on
\emph{disjoint} datasets.
Unfortunately, data analysts could ignore this assumption when writing
queries.
To illustrate this point, we present the code in Figure \ref{fig:partition}.
Query \ensuremath{q} produces a $\epsilon$-DP histogram of the colors found in the
argument \ensuremath{\Varid{dataset}}, which rows are of type \ensuremath{\Conid{Color}} and variable \ensuremath{\Varid{bins}}
enumerates all the possible values of such type.
The code partitions the dataset by using the function \ensuremath{\Varid{id}\mathbin{::}\Conid{Color}\to \Conid{Color}}
(line 2) and executes the aggregation counting query (\ensuremath{{\color{darkblue} \tt{dpCount} }}) in each
partition (line 3)---function \ensuremath{\Varid{fromList}} creates a map from a list of pairs.
The attentive reader could notice that \ensuremath{{\color{darkblue} \tt{dpCount} }} is applied to the original
\ensuremath{\Varid{dataset}} rather than the partitions.
This type of errors could lead to break privacy as well as inconsistencies when
estimating the required privacy budget.
A correct implementation consists on executing \ensuremath{{\color{darkblue} \tt{dpCount} }} on the corresponding
partition as shown in the commented line 4.
%

To catch possible coding errors as the one shown above, DPella deploys an static
information-flow control (IFC) analysis similar to that provided by MAC
\cite{Russo:2015}.
IFC ensures that queries run by \ensuremath{{\color{darkred} \tt{dpPart} }} do not perform queries on shared
datasets (as \ensuremath{\Varid{dataset}} in Figure \ref{fig:partition}).
For that, DPella attaches provenance labels to datasets \ensuremath{{\color{darkred} \tt{Data}}\;{\color{darkblue} \tt{s} }\;{\color{darkblue} \tt{r} }}
indicating to which part of the query they are associated with and propagates
that information accordingly.
For instance, \ensuremath{\Varid{dataset}} in \ensuremath{q} belongs to the top-level fragment of the query
rather than to sub-queries executed in each partition---and DPella will raise an
error at compile time.
Instead, if we comment line 3 and uncomment line 4, the query \ensuremath{\Varid{q}} is
successfully run by DPella (when there is enough privacy budget) since every
partition is only accessing their own partitioned data (denoted by variable
\ensuremath{\Varid{ds}}).
The implemented IFC mechanism is \emph{transparent} to data analysts and
curators, i.e., they do not need to understand how it works.
Analysts and curators only need to know that, when the IFC analysis raises an
alarm, is due to a possibly access to non-disjoint datasets when using
\ensuremath{{\color{darkred} \tt{dpPart} }}.

\subsubsection{Aggregations}
DPella presents primitives to count (\ensuremath{{\color{darkblue} \tt{dpCount} })}, sum (\ensuremath{{\color{darkblue} \tt{dpSum} }}), and average
(\ensuremath{{\color{darkblue} \tt{dpAvg} }}) rows in datasets.
These primitives take an argument \ensuremath{\Varid{eps}\mathbin{::}\epsilon}, a dataset, and build a
Laplace mechanism which is \ensuremath{\Varid{eps}}-differentially private from which a noisy
result gets return as a term of type \ensuremath{{\color{darkgreen} \tt{Value}}\;\Conid{Double}}.
The purpose of data type \ensuremath{{\color{darkgreen} \tt{Value}}\;\Varid{a}} is two fold: to encapsulate noisy values of
type \ensuremath{\Varid{a}} originating from aggregations of data, and to store information about
its accuracy---intuitively, how ``noisy'' the value is (explained in Section
\ref{sec:acc}).
The injected noise of these queries gets adjusted depending on three parameters:
the value of type \ensuremath{\epsilon}, the accumulated stability of the dataset \ensuremath{{\color{darkblue} \tt{s} }}, and the
sensitivity of the query (recall Definition~\ref{def:sensitivity}).
%
More specifically, the Laplace mechanism used by DPella uses accumulated stability $s$ to
scale the noise, i.e., it consider $b$ from Theorem~\ref{lap:mech} as
$b = s \cdot \frac{\Delta_Q}{\epsilon}$.
%
%
The sensitivity of DPella's aggregations are hard-coded into the
implementation---similar to what PINQ does.
The sensitivities of \ensuremath{{\color{darkblue} \tt{dpSum} }} and \ensuremath{{\color{darkblue} \tt{dpAvg} }} are set to \ensuremath{{\color{darkgreen} \texttt{1}}} and \ensuremath{{\color{darkgreen} \texttt{2}}},
respectively, by applying a clipping function (\ensuremath{{\color{darkblue} \tt{r} }\to \Conid{Double}}).
This function maps the values under scrutiny into the interval $[-1,1]$ before
executing the query.
%
The sensitivity of \ensuremath{{\color{darkblue} \tt{dpCount} }} and \ensuremath{{\color{darkblue} \tt{dpMax} }} is set to \ensuremath{{\color{darkgreen} \texttt{1}}}.
To implement the Laplace mechanism, the type constrain \ensuremath{\Conid{Stb}\;{\color{darkblue} \tt{s} }} in \ensuremath{{\color{darkblue} \tt{dpCount} }},
\ensuremath{{\color{darkblue} \tt{dpSum} }}, and \ensuremath{{\color{darkblue} \tt{dpAvg} }} demands the accumulated stability parameter \ensuremath{{\color{darkblue} \tt{s} }} to be a type-level
natural number in order to to obtain a term-level representation when injecting
noise.
Finally, primitive \ensuremath{{\color{darkblue} \tt{dpMax} }} implements noisy-max.
This query takes a scoring function (\ensuremath{{\color{darkblue} \tt{r} }\to \Conid{Double}}), applies it to every row,
adds a uniform noise to every score, and returns the \emph{index of the row}
with the highest noisy score.
This primitive becomes relevant to obtain the winner option in elections without
singling out any voter.
However, it requires that the stability of the dataset to be \ensuremath{{\color{darkgreen} \texttt{1}}} in order to be
sound \cite{DBLP:conf/icalp/BartheGGHS16}.
DPella guarantees such requirement by typing: the type of the given dataset as
argument is \ensuremath{{\color{darkred} \tt{Data}}\;{\color{darkgreen} \texttt{1}}\;{\color{darkblue} \tt{r} }}, i.e., its accumulated stability is set to \ensuremath{{\color{darkgreen} \texttt{1}}}.

\subsubsection{Privacy budget and execution of queries}

The primitive \ensuremath{{\color{darkblue} \tt{budget} }} statically computes how much privacy budget is required
to run a query.
%
It is worth notice that DPella returns an upper bound of the required privacy
budget rather than the exact one---an expected consequence of using a
type-system to compute it and provide early feedback to data analysts.
%
%
Finally, the primitive \ensuremath{{\color{darkred} \tt{dpEval} }} is used by data curators to run queries (\ensuremath{{\color{darkblue} \tt{Query}}\;\Varid{a}}) under given privacy budgets (\ensuremath{\epsilon}), where datasets are just lists of
rows (\ensuremath{[\mskip1.5mu {\color{darkblue} \tt{r} }\mskip1.5mu]}).
It assumes that the initial accumulated stability as \ensuremath{{\color{darkgreen} \texttt{1}}} (\ensuremath{{\color{darkred} \tt{Data}}\;{\color{darkgreen} \texttt{1}}\;{\color{darkblue} \tt{r} }}) since
the dataset has not yet gone through any transformation, and DPella will
automatically calculate the accumulated stability for datasets affected by
subsequent transformations via the Haskell's type system.
This primitive returns a computation of type \ensuremath{\Conid{IO}\;\Varid{a}}, which in Haskell are
computations responsible to perform side-effects---in this case, obtaining
randomness from the system in order to implement the Laplace mechanism.
%


\section{Accuracy}\label{sec:acc}



To deal with accuracy, DPella uses the data type \ensuremath{{\color{darkgreen} \tt{Value}}\;\Varid{a}} responsible to store
a result of type \ensuremath{\Varid{a}} as well as information about its accuracy.
For instance, a term of type \ensuremath{{\color{darkgreen} \tt{Value}}\;\Conid{Double}} stores a noisy number (e.g., coming
from executing \ensuremath{{\color{darkblue} \tt{dpCount} }}) together with its accuracy in terms of \emph{a bound
  on the noise introduced to protect privacy}.

DPella provides an static analysis capable to compute the accuracy of queries
via the following function
\numbersoff
\begin{hscode}\linenumsetup\printlinebegin\SaveRestoreHook
\column{B}{@{}>{\hspre}l<{\hspost}@{}}%
\column{E}{@{}>{\hspre}l<{\hspost}@{}}%
\>[B]{}{\color{darkblue} \tt{accuracy} }\mathbin{::}{\color{darkblue} \tt{Query}}\;({\color{darkgreen} \tt{Value}}\;\Varid{a})\to \beta\to \alpha{}\<[E]%
\printlineend\ColumnHook
\end{hscode}\resethooks
which takes as an argument a query and returns a function, called \emph{inverse
  Cumulative Distribution Function} (iCDF), capturing the theoretical error
\ensuremath{\alpha} for a given confidence \ensuremath{{\color{darkgreen} \texttt{1}}{\color{darkblue}  \texttt{-}}\beta}.
Function \ensuremath{{\color{darkblue} \tt{accuracy} }} does not execute queries but rather symbolically interpret
all of its components in order to compute the accuracy of the result based on
the sub-queries and how data gets aggregated.
%
%
%
%
DPella follows the principle of improving accuracy calculations by detecting
statistical independence.
For that, it implements taint analysis \cite{DBLP:conf/eurosp/SchoepeBPS16} in
order to track if values were drawn from statistically independent
distributions.

\subsection{Accuracy calculations}

\begin{figure}
\centering
{\small
  \begin{hscode}\linenumsetup\printlinebegin\SaveRestoreHook
\column{B}{@{}>{\hspre}l<{\hspost}@{}}%
\column{5}{@{}>{\hspre}l<{\hspost}@{}}%
\column{14}{@{}>{\hspre}c<{\hspost}@{}}%
\column{14E}{@{}l@{}}%
\column{15}{@{}>{\hspre}c<{\hspost}@{}}%
\column{15E}{@{}l@{}}%
\column{18}{@{}>{\hspre}l<{\hspost}@{}}%
\column{19}{@{}>{\hspre}l<{\hspost}@{}}%
\column{34}{@{}>{\hspre}c<{\hspost}@{}}%
\column{34E}{@{}l@{}}%
\column{36}{@{}>{\hspre}c<{\hspost}@{}}%
\column{36E}{@{}l@{}}%
\column{38}{@{}>{\hspre}l<{\hspost}@{}}%
\column{40}{@{}>{\hspre}l<{\hspost}@{}}%
\column{E}{@{}>{\hspre}l<{\hspost}@{}}%
\>[5]{}\mbox{\onelinecomment  Accuracy analysis (data analyst)}{}\<[E]%
\printlineend\\
\printlinebegin\>[5]{}{\color{darkblue} \tt{accuracy} }{}\<[15]%
\>[15]{}\mathbin{::}{}\<[15E]%
\>[19]{}{\color{darkblue} \tt{Query}}\;({\color{darkgreen} \tt{Value}}\;\Varid{a}){}\<[36]%
\>[36]{}\to {}\<[36E]%
\>[40]{}\beta\to \alpha{}\<[E]%
\printlineend\\
\printlinebegin\>[5]{}\mbox{\onelinecomment  Norms (data analyst)}{}\<[E]%
\printlineend\\
\printlinebegin\>[5]{}{\color{darkgreen} \tt{norm}_{\infty} }{}\<[14]%
\>[14]{}\mathbin{::}{}\<[14E]%
\>[18]{}[\mskip1.5mu {\color{darkgreen} \tt{Value}}\;\Conid{Double}\mskip1.5mu]{}\<[34]%
\>[34]{}\to {}\<[34E]%
\>[38]{}{\color{darkgreen} \tt{Value}}\;[\mskip1.5mu \Conid{Double}\mskip1.5mu]{}\<[E]%
\printlineend\\
\printlinebegin\>[5]{}{\color{darkgreen} \tt{norm}_{2} }{}\<[14]%
\>[14]{}\mathbin{::}{}\<[14E]%
\>[18]{}[\mskip1.5mu {\color{darkgreen} \tt{Value}}\;\Conid{Double}\mskip1.5mu]{}\<[34]%
\>[34]{}\to {}\<[34E]%
\>[38]{}{\color{darkgreen} \tt{Value}}\;[\mskip1.5mu \Conid{Double}\mskip1.5mu]{}\<[E]%
\printlineend\\
\printlinebegin\>[5]{}{\color{darkgreen} \tt{norm}_{1} }{}\<[14]%
\>[14]{}\mathbin{::}{}\<[14E]%
\>[18]{}[\mskip1.5mu {\color{darkgreen} \tt{Value}}\;\Conid{Double}\mskip1.5mu]{}\<[34]%
\>[34]{}\to {}\<[34E]%
\>[38]{}{\color{darkgreen} \tt{Value}}\;[\mskip1.5mu \Conid{Double}\mskip1.5mu]{}\<[E]%
\printlineend\\
\printlinebegin\>[5]{}{\color{darkgreen} \tt{rmsd} }{}\<[14]%
\>[14]{}\mathbin{::}{}\<[14E]%
\>[18]{}[\mskip1.5mu {\color{darkgreen} \tt{Value}}\;\Conid{Double}\mskip1.5mu]{}\<[34]%
\>[34]{}\to {}\<[34E]%
\>[38]{}{\color{darkgreen} \tt{Value}}\;[\mskip1.5mu \Conid{Double}\mskip1.5mu]{}\<[E]%
\printlineend\\
\printlinebegin\>[5]{}\mbox{\onelinecomment  Accuracy combinators (data analyst)}{}\<[E]%
\printlineend\\
\printlinebegin\>[5]{}{\color{darkgreen} \tt{add} }{}\<[14]%
\>[14]{}\mathbin{::}{}\<[14E]%
\>[18]{}[\mskip1.5mu {\color{darkgreen} \tt{Value}}\;\Conid{Double}\mskip1.5mu]{}\<[34]%
\>[34]{}\to {}\<[34E]%
\>[38]{}{\color{darkgreen} \tt{Value}}\;\Conid{Double}{}\<[E]%
\printlineend\\
\printlinebegin\>[5]{}{\color{darkgreen} \tt{neg} }{}\<[14]%
\>[14]{}\mathbin{::}{}\<[14E]%
\>[18]{}{\color{darkgreen} \tt{Value}}\;\Conid{Double}\to {\color{darkgreen} \tt{Value}}\;\Conid{Double}{}\<[E]%
\printlineend\ColumnHook
\end{hscode}\resethooks
}
\caption{DPella API: Part II \label{fig:dpella:acc}}
\end{figure}

DPella starts by generating iCDFs at the time of running aggregations based on
the following known result of the Laplace mechanism:
%
\begin{definition}[Accuracy for the Laplace mechanism]
  \label{acc:laplace}
  Given a randomized query $\tilde{Q}(\cdot) : db \to \mathbb{R}$ implemented
  with the Laplace mechanism as in Theorem~\ref{lap:mech}, where the scale is
  adjusted by stability $s$, we have that
  \begin{equation} \label{eq:laplace:acc}
    \Pr \Big[ |\tilde{Q}(D) - Q(D)| > \log(\sfrac{1}{\beta}) \cdot s \cdot \frac{\Delta_Q}{\epsilon} \Big] \leq \beta
  \end{equation}
\end{definition}
Consequently, DPella stores the iCDF
$\lambda \beta \to \log(\sfrac{1}{\beta}) \cdot s \cdot
\frac{\Delta_Q}{\epsilon}$
for the values of type \ensuremath{{\color{darkgreen} \tt{Value}}\;\Conid{Double}} returned by aggregation primitives like
\ensuremath{{\color{darkblue} \tt{dpCount} }}, \ensuremath{{\color{darkblue} \tt{dpSum} }}, and \ensuremath{{\color{darkblue} \tt{dpAvg} }}.
However, queries are often more complex than just calling aggregation
primitives---as shown by \ensuremath{\tt{CDF}_{2}} in Figure \ref{fig:cdf2}.
In this light, DPella provides combinators responsible to aggregate noisy values,
while computing its iCDFs based on the iCDFs of the arguments.
Figure \ref{fig:dpella:acc} shows DPella API when dealing with accuracy.

\subsubsection{Norms} DPella presents several primitives to aggregate the
magnitudes of several errors predictions into a \emph{single} measure---a useful
tool when dealing with vectors.
Primitives \ensuremath{{\color{darkgreen} \tt{norm}_{\infty} }}, \ensuremath{{\color{darkgreen} \tt{norm}_{2} }}, and \ensuremath{{\color{darkgreen} \tt{norm}_{1} }} take a list of values of type
\ensuremath{{\color{darkgreen} \tt{Value}}\;\Conid{Double}}, where each of them carries accuracy information, and produces a
\emph{single value} (or vector) that contains a list of elements (\ensuremath{{\color{darkgreen} \tt{Value}}\;[\mskip1.5mu \Conid{Double}\mskip1.5mu]}), which accuracy is set to be the well-known $\ell_\infty$-, $\ell_2$-,
$\ell_1$-norms, respectively.
Finally, primitive \ensuremath{{\color{darkgreen} \tt{rmsd} }} implements \emph{root-mean-square deviation} among the
elements given as arguments---i.e., the quadratic mean of the differences
between the noisy results and the original answer.
In our examples, we focus on using \ensuremath{{\color{darkgreen} \tt{norm}_{\infty} }}, but other norms are available for
the taste, and preference, of data analysts.

\subsubsection{Adding values}

\begin{figure}[t]
  \centering
\begin{tikzpicture}
  \begin{axis}[
            height=4cm,
            width=\selectformat{1\columnwidth}{0.7\textwidth},
            ylabel= $\alpha$,
            xlabel= Sub-queries,
            axis lines*=left,
            legend pos=north west,
            legend cell align={left},
            grid = major,
            grid style={dashed, gray!30},
            axis line style={->},
            every axis y label/.style={
              at={(ticklabel* cs:1.05)},
              anchor=south,},
            legend image post style={scale=0.5},
            ]
\addplot[red] table [x=nBuckets, y=union, col sep=comma]{./graphs/uniVScher.csv};
\addplot[blue] table [x=nBuckets, y=chernoff, col sep=comma]{./graphs/uniVScher.csv};

\addplot[mark=*,red] coordinates {(100,2303)} node[anchor=east]{$2303$};
\addplot[mark=*,blue] coordinates {(100,155)} node[anchor=south]{$155$};
\legend{Union,Chernoff}
\end{axis}
\end{tikzpicture}
\caption{Union vs. Chernoff bounds  \label{fig:bounds}}
\end{figure}
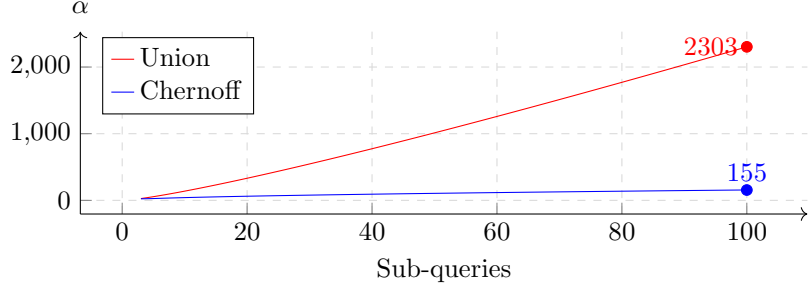
The primitive \ensuremath{{\color{darkgreen} \tt{add} }} aggregates values and, in order to compute accuracy of the
addition, it tries to apply the Chernoff bound if all the values are statistically
independent; otherwise, it applies the union bound.
More precisely, for the next definitions we assume that primitive \ensuremath{{\color{darkgreen} \tt{add} }}
receives $n$ terms \ensuremath{\tt{v}_{1}\mathbin{::}{\color{darkgreen} \tt{Value}}\;\Conid{Double}}, \ensuremath{\tt{v}_{2}\mathbin{::}{\color{darkgreen} \tt{Value}}\;\Conid{Double}}, ... ,
\ensuremath{\tt{v}_n\mathbin{::}{\color{darkgreen} \tt{Value}}\;\Conid{Double}}.
To support subtraction, DPella provides primitive \ensuremath{{\color{darkgreen} \tt{neg} }} responsible to change
the sign of a given value.
Importantly, since we are calculating the theoretical error, we should consider
random variables rather than specific numbers.
The next definition specifies how \ensuremath{{\color{darkgreen} \tt{add} }} behaves when applying union
bound.
\begin{definition}[\ensuremath{{\color{darkgreen} \tt{add} }} using union bound]
\label{def:union:bound}
  Given $n \geq 2$ random variables $V_j$ with their respective
  $\mathit{iCDF}_j$, where $j \in {1 \dots n}$, and
  $\alpha_j = iCDF_j (\frac{\beta}{n})$, then the addition
  $Z = \sum_{j=1}^n V_j$ has the following accuracy:
  \begin{equation}{\label{eq:unionBound}}
    \Pr[\textstyle \abs{Z} > \sum_{j=1}^n \alpha_j] \leq \beta
  \end{equation}
\end{definition}
Observe that to compute the $\mathit{iCDF}$ of $Z$, the formula uses the
$\mathit{iCDFs}$ from the operands applied to $\frac{\beta}{n}$, which is done
to obtain an error bound $Z$ with confidence $1 - \beta$.
Union bound makes no assumption about the distribution of the random variables
$V_j$.

In contrast, the Chernoff bound \emph{often} provides a tighter error estimation
than the commonly used union bound when adding several \emph{statistically
  independent} queries \emph{sampled from a Laplace distribution}.
To illustrate this point, Figure \ref{fig:bounds} shows that difference for the
\ensuremath{\tt{cdf}_{2}} function we presented in Section~\ref{sec:teaser} with $\epsilon=0.5$ (for each
DP sub-query) and $\beta=0.1$---the x-axis denotes the number of sub-queries
aggregated, while the y axis is the estimated $\alpha$.
Clearly, the Chernoff bound is asymptotically much better when estimating
accuracy, while the union bound works best with a reduced number of
sub-queries---observe how lines get crossed in Figure \ref{fig:bounds}.
In this light, and when possible, DPella computes both union bound and Chernoff
bound and selects the tighter error estimation.
However, to apply Chernoff bound, DPella needs to be certain that the events are
independent.
Before explaining how DPella detects that, we gives an specification of the
formula we use for Chernoff.

\begin{definition}[\ensuremath{{\color{darkgreen} \tt{add} }} using Chernoff bound~\cite{chan2011private}]
  \label{def:chernoff}
  Given $n \geq 2$ \emph{independent random variables} $V_j \sim Lap(0,b_j)$, where
  $j \in {1 \dots n}$, $b_M = \mathit{max}~ \{b_j\}_{j=1 \dots n}$, and
  $\nu = \mathit{max} \{ \sqrt{\sum_{j=1}^n b^2_j  }, b_M \sqrt{\ln
    {\frac{2}{\beta}}} \}$, then the addition $ Z = \sum_{j=1}^n V_j$ has the
  following accuracy:  \begin{equation}{\label{eq:chernoffBound}}
  \Pr[\textstyle \abs{Z} > (\nu + 0.00001) \sqrt{8 \ln{\frac{2}{\beta}}}] \leq \beta
 \end{equation}
\end{definition}
The presence of the number $0.00001$ in the formula is to obtain a number
strictly greater than $\nu$ in that part of the formula so that the formula
described above can be used---any other number that yields a number strictly
greater than $\nu$ works as well.
We next explain how DPella checks that values come from statistically
independent sampled variables.

\subsubsection{Detecting statistical independence}

\begin{figure}[t]
\centering
\numbersreset
\numberson
{\small
\begin{hscode}\linenumsetup\printlinebegin\SaveRestoreHook
\column{B}{@{}>{\hspre}l<{\hspost}@{}}%
\column{5}{@{}>{\hspre}l<{\hspost}@{}}%
\column{11}{@{}>{\hspre}c<{\hspost}@{}}%
\column{11E}{@{}l@{}}%
\column{15}{@{}>{\hspre}l<{\hspost}@{}}%
\column{24}{@{}>{\hspre}l<{\hspost}@{}}%
\column{27}{@{}>{\hspre}l<{\hspost}@{}}%
\column{30}{@{}>{\hspre}l<{\hspost}@{}}%
\column{E}{@{}>{\hspre}l<{\hspost}@{}}%
\>[B]{}\Varid{totalCount}\mathbin{::}{\color{darkblue} \tt{Query}}\;({\color{darkgreen} \tt{Value}}\;\Conid{Double}){}\<[E]%
\printlineend\\
\printlinebegin\>[B]{}\Varid{totalCount}\mathrel{=}\mathbf{do}{}\<[E]%
\printlineend\\
\printlinebegin\>[B]{}\hsindent{5}{}\<[5]%
\>[5]{}\tt{v}_{1}{}\<[11]%
\>[11]{}\leftarrow {}\<[11E]%
\>[15]{}{\color{darkblue} \tt{dpCount} }\;{}\<[24]%
\>[24]{}{\color{darkgreen} \texttt{0.3}}\;{}\<[30]%
\>[30]{}ds_{1}{}\<[E]%
\printlineend\\
\printlinebegin\>[B]{}\hsindent{5}{}\<[5]%
\>[5]{}\tt{v}_{2}{}\<[11]%
\>[11]{}\leftarrow {}\<[11E]%
\>[15]{}{\color{darkblue} \tt{dpCount} }\;{}\<[24]%
\>[24]{}{\color{darkgreen} \texttt{0.25}}\;{}\<[30]%
\>[30]{}ds_{2}{}\<[E]%
\printlineend\\
\printlinebegin\>[B]{}\hsindent{5}{}\<[5]%
\>[5]{}{\color{darkblue}  \texttt{...}}{}\<[E]%
\printlineend\\
\printlinebegin\>[B]{}\hsindent{5}{}\<[5]%
\>[5]{}v_{100}{}\<[11]%
\>[11]{}\leftarrow {}\<[11E]%
\>[15]{}{\color{darkblue} \tt{dpCount} }\;{}\<[24]%
\>[24]{}{\color{darkgreen} \texttt{0.5}}\;{}\<[30]%
\>[30]{}ds_{100}{}\<[E]%
\printlineend\\
\printlinebegin\>[B]{}\hsindent{5}{}\<[5]%
\>[5]{}\Varid{return}\;({\color{darkgreen} \tt{add} }\;[\mskip1.5mu \tt{v}_{1},\tt{v}_{2},{}\<[27]%
\>[27]{}\mathinner{\ldotp\ldotp},v_{100}\mskip1.5mu]){}\<[E]%
\printlineend\ColumnHook
\end{hscode}\resethooks
}
\caption{Combination of sub-queries results \label{fig:totalCount}}
\end{figure}
To detect statistical independence, we apply taint analysis when
considering terms of type \ensuremath{{\color{darkgreen} \tt{Value}}\;\Varid{a}}.
Specifically, every time a result of type \ensuremath{{\color{darkgreen} \tt{Value}}\;\Conid{Double}} gets generated by an
aggregation query in DPella's API (i.e., \ensuremath{{\color{darkblue} \tt{dpCount} }}, \ensuremath{{\color{darkblue} \tt{dpSum} }}, etc.), it gets
assigned a label indicating that it is \emph{untainted} and thus statistically
independent.
The label also carries information about the scale of the Laplace distribution
from which it was sampled---a useful information when applying Definition
\ref{def:chernoff}.
When the primitive \ensuremath{{\color{darkgreen} \tt{add} }} receives all untainted values as arguments,
the accuracy of the aggregation is determined by the best estimation
provided by either the union bound (Definition~\ref{def:union:bound})
or the Chernoff bound (Definition~\ref{def:chernoff}).
Importantly, values produced by \ensuremath{{\color{darkgreen} \tt{add} }} are considered \emph{tainted} since they
depend on other results.
When \ensuremath{{\color{darkgreen} \tt{add} }} receives any tainted argument, it proceeds to estimate the error of
the addition by just using union bound.

To illustrate how our taint analysis works, Figure \ref{fig:totalCount} presents
the query plan \ensuremath{\Varid{totalCount}} which adds the results of hundred \ensuremath{{\color{darkblue} \tt{dpCount} }} queries
over different datasets, namely \ensuremath{ds_{1}}, \ensuremath{ds_{2}}, $\dots$, \ensuremath{ds_{100}}.
(The \ensuremath{{\color{darkblue}  \texttt{...}}} denotes code intentionally left unspecified.)
The code calls the primitive \ensuremath{{\color{darkgreen} \tt{add} }} with the results of calling \ensuremath{{\color{darkblue} \tt{dpCount} }}.
(We use \ensuremath{[\mskip1.5mu x_{1},x_{2},x_{3}\mskip1.5mu]} to denote the list with elements \ensuremath{x_{1}}, \ensuremath{x_{2}}, and
\ensuremath{x_{3}}.)
What would it be then the theoretical error of \ensuremath{\Varid{totalCount}}?
The accuracy calculation depends on whether all the values are untainted in line 7.
When no dependencies are detected between \ensuremath{\tt{v}_{1}}, \ensuremath{\tt{v}_{2}}, $\dots$, \ensuremath{v_{100}},
namely all the values are untainted, DPella applies Chernoff bound in
order to give a tighter error estimation.
Instead, for instance, if \ensuremath{\tt{v}_{3}} were computed as an aggregation of \ensuremath{\tt{v}_{1}} and \ensuremath{\tt{v}_{2}},
e.g., \ensuremath{\mathbf{let}\;\tt{v}_{3}\mathrel{=}{\color{darkgreen} \tt{add} }\;[\mskip1.5mu \tt{v}_{1},\tt{v}_{2}\mskip1.5mu]},
%
 then line 7 applies union bound since \ensuremath{\tt{v}_{3}} is a tainted value.
 With taint analysis, DPella is capable to detect dependencies among terms of
 type \ensuremath{{\color{darkgreen} \tt{Value}}\;\Conid{Double}}, and leverages that information to apply different
 concentrations bounds.

%





\section{Case studies}
\label{sec:examples}

\begin{table}[t]
  \small
  \centering
\begin{tabular}{p{1.4cm} l p{2cm}}
  \hline
  Category & Application & Programs \\ \hline
  \multirow{9}{1.5cm}{PINQ-like}
  & CDFs~\cite{Mcsherry2011network}
  & \text{\tt cdf1}, \text{\tt cdf2}, \text{\tt cdfSmart} \\ \cline{2-3}

  & \multirow{3}{1.7cm}{Term frequency~\cite{PINQ09}}
  & \text{\tt queryFreq}, \text{\tt queriesFreq}, \text{\tt ip2location} \\ \cline{2-3}

  & \multirow{2}{1.7cm}{Network analysis~\cite{Mcsherry2011network}}
  & \text{\tt packetSize}, \text{\tt portSize} \\ \cline{2-3}

  & \multirow{3}{1.7cm}{Cumulative sums~\cite{BartheGAHKS14}}
  & \text{\tt cumulSum1} \text{\tt cumulSum2} \text{\tt cumulSumSmart} \\ \hline

  \multirow{3}{1.5cm}{Counting queries}
  & Range queries via Identity
  & \text{\tt i\char95{}n} \\

  & Range queries via  Histograms~\cite{HayRMS10Boosting}
  & \text{\tt h\char95{}n} \\

  & Range queries via Wavelet~\cite{XiaoWG11Privlet}
  & \text{\tt y\char95{}n} \\ \hline

\end{tabular}
\caption{Implemented literature examples \label{tab:testCases}}
\end{table}


In this section, we will discuss the advantages and limitations of our
programming framework.
Moreover, we will go in-depth into using DPella to analyze the interplay of
privacy and accuracy parameters in hierarchical histograms.

\subsection{DPella expressiveness}
First, we start by exploring the expressiveness of DPella.
For this, we have built several analyses found in the DP
literature---see Table~\ref{tab:testCases}---which we classify into two
categories, \emph{PINQ-like queries} and \emph{counting queries}. The
former class allows us to compare DPella expressivity with the one of
PINQ, while the latter allow us to compare DPella expressivity with
the one of APEx.
\paragraph*{PINQ-like queries}
We have implemented most of the examples that have been implemented in
PINQ~\cite{PINQ09,Mcsherry2011network}, such as,
different versions of CDFs (sequential, parallel, and hybrid) and
network tracing-like analyses.
Additionally, we considered examples of \emph{cumulative sums}
\cite{BartheGAHKS14}---which are queries that share some commonalities
with CDFs.
By construction, DPella's support these queries naturally, since the
expressiveness of DPella relies on its primitives that were designed
following PINQ's one very closely.
However, as stated in previous sections, our framework goes a step
further and exposes to data analysts the accuracy bound achieved by
the specific implementation.
This specific feature allows the data analyst to reason about accuracy
of the results---without actually executing the query---by varying
\begin{inparaenum}[i)]
\item the strategy of the implementation
\item the parameters of the query.
\end{inparaenum}
For instance, in Section~\ref{sec:teaser}, we have shown how the
analyst can inspect the error of a sequential and parallel strategy to
compute the CDF of packet lengths.
Furthermore, the data analyst can take advantage of DPella being an
embedded DSL and write a Haskell function that takes any of the
approaches (\text{\tt cdf1} or \text{\tt cdf2}) and varies epsilon aiming to certain
error tolerance (for a fixed confidence interval), or vice versa.
Such a function can be as simple as a brute force analysis or as
complex as an heuristic algorithm.


\begin{figure}
  \centering
  \begin{subfigure}[c]{0.15\columnwidth}
    \centering
    \begin{equation*}
      \begingroup 
      \setlength\arraycolsep{2pt}
       {\small
      \begin{bmatrix}
        1 & 0 & 0 & 0 \\
        1 & 1 & 0 & 0 \\
        1 & 1 & 1 & 0 \\
        1 & 1 & 1 & 1 \\
        0 & 1 & 0 & 0 \\
        0 & 1 & 1 & 0 \\
        0 & 1 & 1 & 1 \\
        0 & 0 & 1 & 0 \\
        0 & 0 & 1 & 1 \\
        0 & 0 & 0 & 1 \\
      \end{bmatrix}
      }
      \endgroup
    \end{equation*}
    \caption*{$\mathbf{W_{R_4}}$}
  \end{subfigure}
  ~
  \begin{subfigure}[c]{0.15\columnwidth}
    \centering
    \begin{equation*}
      \begingroup 
      \setlength\arraycolsep{2pt}
      {\small
      \begin{bmatrix}
        1 & 0 & 0 & 0\\
        0 & 1 & 0 & 0\\
        0 & 0 & 1 & 0\\
        0 & 0 & 0 & 1\\
      \end{bmatrix}
      }
      \endgroup
    \end{equation*}
    \caption*{$\mathbf{I_4}$}
  \end{subfigure}
  ~
  \begin{subfigure}[c]{0.15\columnwidth}
    \centering
    \begin{equation*}
      \begingroup 
      \setlength\arraycolsep{2pt}
      {\small
      \begin{bmatrix}
        1 & 1 & 1 & 1 \\
        1 & 1 & 0 & 0 \\
        0 & 0 & 1 & 1 \\
        1 & 0 & 0 & 0 \\
        0 & 1 & 0 & 0 \\
        0 & 0 & 1 & 0 \\
        0 & 0 & 0 & 1 \\
      \end{bmatrix}
      }
      \endgroup
    \end{equation*}
    \caption*{$\mathbf{H_4}$}
  \end{subfigure}
  ~
  \begin{subfigure}[c]{0.2\columnwidth}
    \centering
    \begin{equation*}
      \begingroup 
      \setlength\arraycolsep{2pt}
      {\small
      \begin{bmatrix}
        1 & \phantom{-}1 & \phantom{-}1 & \phantom{-}1 \\
        1 & \phantom{-}1 & -1           & -1           \\
        1 & -1           & \phantom{-}0 & \phantom{-}0 \\
        0 & \phantom{-}0 & \phantom{-}1 & -1           \\
      \end{bmatrix}
      }
      \endgroup
    \end{equation*}
    \caption*{$\mathbf{Y_4}$}
  \end{subfigure}
  \caption{\label{fig:strat} Workload of all range queries and query
    strategies for 4 ranges}
\vspace{-10pt}
\end{figure}

\paragraph*{Counting queries}
The second class of queries that we  considered are counting
queries. More specifically, we focused on range queries, a specific
subclass of counting queries. We considered this class to compare our
approach with the one implemented in the tool APEx~\cite{Apex}.

To answering counting queries, APEx uses the \emph{matrix
  mechanism}~\cite{Chao15MatMech}. This algorithm answers a set of
linear queries (called the \emph{workload}) by calibrating the noise
to specific properties of the workload while preserving differential privacy.
More in details, the matrix mechanism uses some \emph{query
  strategies} as an intermediate device to answer a workload. The
mechanism returns a DP version of the query strategies (obtained using
the Laplace or Gaussian mechanism), from which noisy answers of the
workload are derived.
By adding independent noise to the strategies instead of adding it to
the whole workload, the matrix mechanism exploits the correlation between the
queries in order to add less noise for the same level of privacy,
boosting in this way the accuracy of the results.


The matrix mechanism achieves an almost optimal error on counting
queries. In order to achieve this error, the algorithm uses several
non-trivial transformations which cannot be implemented easily in terms of other
components. APEx implements it as a black-box and we could do the same
in DPella.
Instead, in this section we show how DPella can be directly used to
answer sets of counting queries using some of the ideas behind the
design of the matrix mechanism, and how these answers improve with
respect to answering the queries naively, thanks to the use of
partition and the Chernoff bound.

To do this, we have
implemented several strategies to answer an specific workload
$\mathbf{W_{R}}$: the set of all range queries over a domain.
Figure~\ref{fig:strat} illustrates the workload that would be answer
for a frequency count of four ranges.
The identity strategy $\mathbf{I_4}$, represents 4 queries (number of
rows) computing the noisy count of each range (number of columns).
The hierarchical strategy $\mathbf{H_4}$ contains seven queries
representing a binary hierarchy of sums, for instance, the first row
represent the sum of all elements, the second one is the sum of half
of the ranges, and so on.
Finally, the wavelet strategy $\mathbf{Y_4}$ contains four queries
representing the Haar wavelet matrix.

To replicate the behavior of answering a workload $\mathbf{W_{R}}$ for
$n$ ranges using a strategy $A$ in DPella we follow three steps:

\begin{enumerate}[\hspace{1pt}1)]
\item Partition the table for each range and perform a noisy count in
  each partition obtaining
  $\tilde{x} = [\tilde{q_1}, \tilde{q_2}, \cdots, \tilde{q_n}]$,
  here each $\tilde{q_i}$ has type \ensuremath{{\color{darkgreen} \tt{Value}}\;\Varid{a}}

\item Compute the queries described by an strategy $A$ with $\tilde{x}$
\numbersoff
\begin{hscode}\linenumsetup\printlinebegin\SaveRestoreHook
\column{B}{@{}>{\hspre}l<{\hspost}@{}}%
\column{E}{@{}>{\hspre}l<{\hspost}@{}}%
\>[B]{}\Varid{strategy}\mathbin{::}[\mskip1.5mu [\mskip1.5mu \Conid{Int}\mskip1.5mu]\mskip1.5mu]\to [\mskip1.5mu {\color{darkgreen} \tt{Value}}\;\Varid{a}\mskip1.5mu]\to [\mskip1.5mu [\mskip1.5mu {\color{darkgreen} \tt{Value}}\;\Varid{a}\mskip1.5mu]\mskip1.5mu]{}\<[E]%
\printlineend\\
\printlinebegin\>[B]{}\Varid{strategy}\;\Varid{matA}\;\Varid{vectX}\mathrel{=}{\color{darkblue}  \texttt{...}}{}\<[E]%
\printlineend\ColumnHook
\end{hscode}\resethooks

Note that instead of adding each noisy count to obtain the value of a
row, we return a list of queries involved in the computation.
For instance, for strategy $\mathbf{H_4}$ each row will be represented as
follows:

{\small
\begin{equation*}
\mathbf{H_4}\tilde{x} =
\begingroup 
\setlength\arraycolsep{2pt}
\begin{bmatrix}
1 & 1 & 1 & 1 \\
1 & 1 & 0 & 0 \\
0 & 0 & 1 & 1 \\
1 & 0 & 0 & 0 \\
0 & 1 & 0 & 0 \\
0 & 0 & 1 & 0 \\
0 & 0 & 0 & 1
\end{bmatrix}
\endgroup
\begin{bmatrix}
\tilde{q_1}\\
\tilde{q_2}\\
\tilde{q_3}\\
\tilde{q_4}
\end{bmatrix}
=
\begin{bmatrix*}[l]
[ \tilde{q_1},\tilde{q_2},\tilde{q_3},\tilde{q_4} ] \\
[ \tilde{q_1},\tilde{q_2} ] \\
[ \tilde{q_3},\tilde{q_4} ] \\
[ \tilde{q_1} ] \\
[ \tilde{q_2} ] \\
[ \tilde{q_3} ] \\
[ \tilde{q_4} ]
\end{bmatrix*}
\begin{matrix}
(\vec{y}_1)\\
(\vec{y}_2)\\
(\vec{y}_3)\\
(\vec{y}_4)\\
(\vec{y}_5)\\
(\vec{y}_6)\\
(\vec{y}_7)\\
\end{matrix}
\end{equation*}
}

Which produces
$\vec{y} = [\vec{y}_1, \vec{y}_2, \vec{y}_3, \vec{y}_4, \vec{y}_5,
\vec{y}_6, \vec{y}_7]$, this is, \ensuremath{\Varid{strategy}\;\Varid{matH}\;\Varid{vectX}\mathrel{=}[\mskip1.5mu {\tt{y}_{1}},{\tt{y}_{2}},{\tt{y}_{3}},{\tt{y}_{4}},{\tt{y}_{5}},{\tt{y}_{6}},{\tt{y}_{7}}\mskip1.5mu]}

\item Construct the noisy answer of $\mathbf{W_{R}}$ using $\vec{y}$
  and adding each row using operator \ensuremath{{\color{darkgreen} \tt{add} }}.
  For instance, answering the 3rd range (3rd column) query of
  $\mathbf{W_{R4}}$ can be done by adding $\vec{y}_2$ and $\vec{y}_6$,
  this is \ensuremath{{\color{darkgreen} \tt{add} }\;({\tt{y}_{2}}\plus {\tt{y}_{6}})\mathrel{=}{\color{darkgreen} \tt{add} }\;[\mskip1.5mu {\tilde{\tt{q}}_{1}},{\tilde{\tt{q}}_{2}},{\tilde{\tt{q}}_{3}}\mskip1.5mu]}.
\end{enumerate}

Observe that there are several ways to combine $\vec{y}_i$ lists to
answer $\mathbf{{W}_{R4}}$, in particular, we could have used
only the identity matrix $\vec{y}_4 \cdots \vec{y}_7$ wich would
correspond to using $\mathbf{I_4}$ strategy.
Additionally, since we delayed the combination of noisy values
$\tilde{q}_i$ until the end, any possible combination will yield (at
least) the same error as using strategy $\mathbf{I_4}$.
We can conclude that by following the previous steps the more accurate
answer for $\mathbf{W_R}$ will be yield by the identity strategy.
This is not unexpected, since in order to use the other queries
strategies more efficiently we would need transformation similar to the
ones used in the matrix
mechanism.

%
\begin{figure}[t]
  \centering
  \includegraphics[scale=0.35]{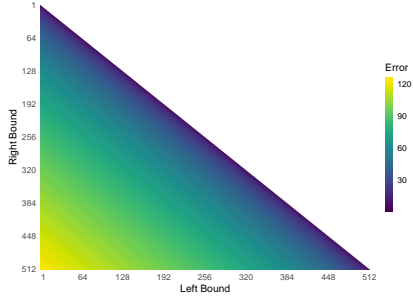}
  \caption{\label{fig:errorIn} Error of each range query in
    $\mathbf{W_R}$ using strategy $\mathbf{I_n}$ with
    $n=512, \epsilon = 1$, and $\beta = 0.05$}
  \vspace{-10pt}
\end{figure}
In order to understand how the accuracy of our method based on the identity strategy
compares to the one of the matrix mechanism we show in
Figure~\ref{fig:errorIn} the error of answering each range query
(i.e., each row) in $\mathbf{W_R}$ with strategy $\mathbf{I_n}$ and
$n=512$.
While we use the same kind of plot, this error cannot be directly compared with the one
shown in Figure 7 of~\cite{Chao15MatMech}, since we use a different error
metrics: ($\alpha$,$\beta$)-accuracy vs MSE.
Nonetheless, we share the tendency of having lower error on small
ranges and significant error on large ranges.
Now, since the noisy values that will be added (using the function \ensuremath{{\color{darkgreen} \tt{add} }}) are
statistically independent, we can use the Chernoff bound to show that the error
is approximately $\mathcal{O}(\sqrt{n})$ for each range query, and a maximum
error of $\mathcal{O}(\sqrt{n\log{n}})$ for answering any query in
$\mathbf{W_R}$.
If we compare our maximum error $\mathcal{O}(\sqrt{n\log{n}})$ with
the one of the matrix mechanism based on the identity strategy $\mathcal{O}(n / \epsilon^2)$, it
becomes evident how Chernoff bound is useful to provides tighter
accuracy bounds.
Unfortunately, as previously stated, the error of strategies
$\mathbf{H_n}$ and $\mathbf{Y_n}$ in DPella is not better than the one
of the strategy $\mathbf{I_n}$, so we cannot reach the same accuracy
the matrix mechanism achieves with these strategies (see Figure 7
of~\cite{Chao15MatMech}).
This limitation can be addressed by leveraging the fact that DPella is a
programming framerwork that could be \emph{extended} by adding the matrix
mechanism---and some other features---as black-box primitives.

\subsection{Privacy and accuracy trade-off analysis in
  DPella}\label{subsec:hierachHist}


We showcase how DPella helps the analyst to reason about
the privacy and accuracy trade-off while designing a query.
We study histograms with certain hierarchical structure (commonly seen in
Census Bureaus analyses) where different accuracy requirements are imposed per
level and where
%
varying one privacy or accuracy parameter can have a \emph{cascade impact} on
the privacy or accuracy of others.
We consider the scenario where we would like to generate histograms from the
Adult
database\footnote{\href{https://archive.ics.uci.edu/ml/datasets/adult}{https://archive.ics.uci.edu/ml/datasets/adult}}
in order to perform studies on gender balance.
The information that we need to mine is not only an histogram of the genders
(for simplicity, just male and female) but also how the gender distributes over
age, and within that, how age distributes over nationality---thus exposing a
hierarchical structure of three levels.
As a data analyst faced with this task, how should we proceed to implement the analysis?
%
%
%


\begin{figure}[t]
\numberson
\numbersreset


%

%
\begin{subfigure}[b]{0.9\columnwidth}
\centering
{\small
\begin{hscode}\linenumsetup\printlinebegin\SaveRestoreHook
\column{B}{@{}>{\hspre}l<{\hspost}@{}}%
\column{3}{@{}>{\hspre}l<{\hspost}@{}}%
\column{7}{@{}>{\hspre}c<{\hspost}@{}}%
\column{7E}{@{}l@{}}%
\column{11}{@{}>{\hspre}l<{\hspost}@{}}%
\column{24}{@{}>{\hspre}l<{\hspost}@{}}%
\column{29}{@{}>{\hspre}l<{\hspost}@{}}%
\column{E}{@{}>{\hspre}l<{\hspost}@{}}%
\>[B]{}{\tt{hierarchical}}_{1}\;[\mskip1.5mu {\tt{e}_1},{\tt{e}_2},{\tt{e}_3}\mskip1.5mu]\;\Varid{dat}\mathrel{=}\mathbf{do}{}\<[E]%
\printlineend\\
\printlinebegin\>[B]{}\mbox{\onelinecomment  \ensuremath{{\tt{h}_1}\mathbin{::}\Conid{Map}\;\Conid{Gen}\;({\color{darkgreen} \tt{Value}}\;\Conid{Double})}}{}\<[E]%
\printlineend\\
\printlinebegin\>[B]{}\mbox{\onelinecomment  \ensuremath{{\tt{h}_2}\mathbin{::}\Conid{Map}\;(\Conid{Gen},\Conid{Age})\;({\color{darkgreen} \tt{Value}}\;\Conid{Double})}}{}\<[E]%
\printlineend\\
\printlinebegin\>[B]{}\mbox{\onelinecomment  \ensuremath{{\tt{h}_3}\mathbin{::}\Conid{Map}\;(\Conid{Gen},\Conid{Age},\Conid{Nationality})\;({\color{darkgreen} \tt{Value}}\;\Conid{Double})}}{}\<[E]%
\printlineend\\[\blanklineskip]%
\printlinebegin\>[B]{}\hsindent{3}{}\<[3]%
\>[3]{}{\tt{h}_1}{}\<[7]%
\>[7]{}\leftarrow {}\<[7E]%
\>[11]{}\Varid{byGen}\;{}\<[24]%
\>[24]{}{\tt{e}_1}\;{}\<[29]%
\>[29]{}\Varid{dat}{}\<[E]%
\printlineend\\
\printlinebegin\>[B]{}\hsindent{3}{}\<[3]%
\>[3]{}{\tt{h}_2}{}\<[7]%
\>[7]{}\leftarrow {}\<[7E]%
\>[11]{}\Varid{byGenAge}\;{}\<[24]%
\>[24]{}{\tt{e}_2}\;{}\<[29]%
\>[29]{}\Varid{dat}{}\<[E]%
\printlineend\\
\printlinebegin\>[B]{}\hsindent{3}{}\<[3]%
\>[3]{}{\tt{h}_3}{}\<[7]%
\>[7]{}\leftarrow {}\<[7E]%
\>[11]{}\Varid{byGenAgeNat}\;{}\<[24]%
\>[24]{}{\tt{e}_3}\;{}\<[29]%
\>[29]{}\Varid{dat}{}\<[E]%
\printlineend\\
\printlinebegin\>[B]{}\hsindent{3}{}\<[3]%
\>[3]{}\Varid{return}\;({\tt{h}_1},{\tt{h}_2},{\tt{h}_3}){}\<[E]%
\printlineend\ColumnHook
\end{hscode}\resethooks
}
\caption{Hierarchical histogram I: distribute budget among the
  levels\label{fig:hist1}}
\end{subfigure}

\begin{subfigure}[b]{0.9\columnwidth}
\centering
{\small
\begin{hscode}\linenumsetup\printlinebegin\SaveRestoreHook
\column{B}{@{}>{\hspre}l<{\hspost}@{}}%
\column{3}{@{}>{\hspre}l<{\hspost}@{}}%
\column{7}{@{}>{\hspre}c<{\hspost}@{}}%
\column{7E}{@{}l@{}}%
\column{11}{@{}>{\hspre}l<{\hspost}@{}}%
\column{19}{@{}>{\hspre}l<{\hspost}@{}}%
\column{E}{@{}>{\hspre}l<{\hspost}@{}}%
\>[B]{}{\tt{hierarchical}}_{2}\;\Varid{e}\;\Varid{dat}\mathrel{=}\mathbf{do}{}\<[E]%
\printlineend\\
\printlinebegin\>[B]{}\hsindent{3}{}\<[3]%
\>[3]{}{\tt{h}_3}{}\<[7]%
\>[7]{}\leftarrow {}\<[7E]%
\>[11]{}\Varid{byGenAgeNat}\;\Varid{e}\;\Varid{dat}{}\<[E]%
\printlineend\\
\printlinebegin\>[B]{}\hsindent{3}{}\<[3]%
\>[3]{}{\tt{h}_2}{}\<[7]%
\>[7]{}\leftarrow {}\<[7E]%
\>[11]{}{\tt{level}}_{2}\;{}\<[19]%
\>[19]{}{\tt{h}_3}{}\<[E]%
\printlineend\\
\printlinebegin\>[B]{}\hsindent{3}{}\<[3]%
\>[3]{}{\tt{h}_1}{}\<[7]%
\>[7]{}\leftarrow {}\<[7E]%
\>[11]{}{\tt{level}}_{1}\;{}\<[19]%
\>[19]{}{\tt{h}_3}{}\<[E]%
\printlineend\\
\printlinebegin\>[B]{}\hsindent{3}{}\<[3]%
\>[3]{}\Varid{return}\;({\tt{h}_1},{\tt{h}_2},{\tt{h}_3}){}\<[E]%
\printlineend\ColumnHook
\end{hscode}\resethooks
}
\caption{Hierarchical histogram II: spend budget only on the most
  detailed histogram \label{fig:hist2}}
\end{subfigure}

\caption{Implementation of hierarchical histograms \label{fig:histograms}}
\vspace{-10pt}
\end{figure}

Our first approach is depicted in Figure \ref{fig:hist1}, where query
\ensuremath{{\tt{hierarchical}}_{1}} generates three histograms with different levels of
details.
This query puts together the results produced by queries \ensuremath{\Varid{byGen}},
\ensuremath{\Varid{byGenAge}}, and \ensuremath{\Varid{byGenAgeNationality}} where each query generates an
histogram of the specified set of attributes.
%
%
%
Observe that these sub-queries are called with potentially different
epsilons, namely \ensuremath{{\tt{e}_1}}, \ensuremath{{\tt{e}_2}}, and \ensuremath{{\tt{e}_3}}, then, we expect \ensuremath{{\tt{hierarchical}}_{1}}
to be \ensuremath{{\tt{e}_1}{\color{darkblue}  \texttt{+}}{\tt{e}_2}{\color{darkblue}  \texttt{+}}{\tt{e}_3}}-differentially private.


We need to proceed to explore the possibilities to tune the
privacy and accuracy parameters to our needs.
In this case, we want a confidence of $95\%$ for accuracy, i.e.,
$\beta = 0.05$, with a total budget of 3 ($\epsilon = 3$).
We could manually try to take the budget $\epsilon = 3$ and distribute it to the
different histograms in many different ways and analyze the implication for
accuracy by calling \ensuremath{{\color{darkblue} \tt{accuracy} }} on each sub-query.
Instead, we write a small (simple, brute force) optimizer in Haskell
that splits the budget uniformly among the queries, i.e., \ensuremath{{\tt{e}_1}\mathrel{=}{\color{darkgreen} \texttt{1}}},
\ensuremath{{\tt{e}_2}\mathrel{=}{\color{darkgreen} \texttt{1}}}, and \ensuremath{{\tt{e}_3}\mathrel{=}{\color{darkgreen} \texttt{1}}}, and tries to find the minimum epsilon that
meets the accuracy demands per histogram.
In other words, we are interested in minimizing the \emph{privacy loss} at each
level bounding the maximum accepted error.
The optimizer essentially adjusts the different epsilons and calls
\ensuremath{{\color{darkblue} \tt{accuracy} }} during the minimization process\showappendix{---see
Appendix~\ref{app:hierach:opt}}.
To ensure termination, the optimizer aborts after a fixed number of
iterations, or when the local budget \ensuremath{{\tt{e}_i}} is exhausted.

\begin{table}
\centering
{\small
\begin{tabular}{l c c l l}
  \hline
  Histogram     & $\alpha$ tolerance & Status & $\epsilon$ & $\alpha$ \\\hline
  \text{\tt byGen~~~~~~} & 100  & $\cmark$                    & 0.06  & 61.48 \\
  \text{\tt byGenAge~~~} & 100  & $\cmark$                    & 0.06  & 96.13 \\
  \text{\tt byGenAgeNat} & 100  & $\cmark$                    & 0.11  & 85.74 \\\hline

  \text{\tt byGen~~~~~~} & 10   & $\cmark$                    & 0.41  & 8.99  \\
  \text{\tt byGenAge~~~} & 50   & $\cmark$                    & 0.16  & 36.05 \\
  \text{\tt byGenAgeNat} & 5    & $\color{red}\times$ MaxBud  & 1     & 9.43  \\\hline

  \text{\tt byGen~~~~~~} & 5    & $\cmark$                    & 0.76  & 4.85  \\
  \text{\tt byGenAge~~~} & 5    & $\color{red}\times$ MaxBud  & 1     & 5.76  \\
  \text{\tt byGenAgeNat} & 10   & $\cmark$                    & 0.96  & 9.82  \\\hline
\end{tabular}
}
\vspace{5pt}
\caption{Budgeting with $\alpha$ tolerances, $\beta = 0.05$, \& total $\epsilon=3$ \label{tab:optOutput}}
\vspace{-10pt}
\end{table}

Table \ref{tab:optOutput} shows some of our findings.
The first row shows what happens when we impose an error of $100$ at every level
of detail, i.e., each bar in all the histograms could be at most $+/-100$ off.
Then, we only need to spend a little part of our budget---the optimizer finds
the minimum epsilons that adheres to the accuracy constrains.
Instead, the second row shows that if we ask to be \emph{gradually} more
accurate on more detailed histograms, then the optimizer aborted at the most
detailed one.
While it could fulfill the first two demands, it aborted on the most detailed
histogram (\ensuremath{\Varid{byGenAgeNat}}) since it could not find an epsilon that fulfills that
requirement---the best we can do is spending all the budget and obtain and error
bound of $9.43$.
Finally, the last row shows what happens if we want \emph{gradually} tighter
error bounds on the less detailed histograms.
In this case, the middle layer can be ``almost'' fulfilled by expending all the
budget and obtaining an error bound of $5.76$ instead of $5$.
While the results from Table \ref{tab:optOutput} could be acceptable for some
data analysts, they might not be for others.
We propose an alternative manner to implement the same query which
consists on spending privacy budget \emph{only} for the most detailed
histogram.
As shown in Figure~\ref{fig:hist2}, this new approach spends all the budget \ensuremath{\Varid{e}}
on calling \ensuremath{{\tt{h}_3}\leftarrow \Varid{byGenAgeNat}\;\Varid{e}\;\Varid{dat}}.
Subsequently, the query builds the other histograms based on the
information extracted from the most detailed one.
For that, we add the noisy values of \ensuremath{{\tt{h}_3}} (using helper functions
\ensuremath{{\tt{level}}_{2}} and \ensuremath{{\tt{level}}_{1}}) creating the rest of the histograms
representing the Cartesian products of gender and age, and gender,
respectively.
These methodology will use \ensuremath{{\color{darkgreen} \tt{add} }} and \ensuremath{{\color{darkgreen} \tt{norm}_{\infty} }} to compute the derived
histograms, and therefore will not consume more privacy budget.
Observe that the query proceeds in a bottom-up fashion, i.e., it starts with the
most detailed histogram and finishes with the less detailed one.
Now that we have two implementations, which one is better?
Which one yields the better trade-offs between privacy and accuracy?
For that, Figure \ref{fig:h1vsh2} shows the accuracy of the different level of
histograms, i.e., \ensuremath{{\tt{h}_1}}, \ensuremath{{\tt{h}_2}}, and \ensuremath{{\tt{h}_3}}, when fixing $\beta = 0.05$ and a global
budget of $\epsilon = 1$ (h1-$\epsilon$1, h2-$\epsilon$2, and h3-$\epsilon$3)
and $\epsilon = 3$ (h1-$\epsilon$3, h2-$\epsilon$3, and h3-$\epsilon$3)---we
obtained all this information by running repetitively the function \ensuremath{{\color{darkblue} \tt{accuracy} }}.
Form the graphics, we can infer that the splitting of the privacy
budget per level often gives rise to more accurate histograms.
However, observe the exception when $\epsilon=3$ for \ensuremath{{\tt{hierarchical}}_{2}}: in this
case, \ensuremath{{\tt{hierarchical}}_{1}} will use an $\epsilon=1$ in that histogram so it will
receive a more noisy count than using $\epsilon=3$.

\begin{figure}
\centering
\begin{tikzpicture}
  \begin{axis}[
    ybar=2pt,
    height=4cm,
    width=\selectformat{1\columnwidth}{0.7\textwidth},
    ylabel= $\alpha$,
    nodes near coords,
    point meta= round(y),
    symbolic x coords= {h1-$\epsilon$1, h2-$\epsilon$1, h3-$\epsilon$1,
                        h1-$\epsilon$3, h2-$\epsilon$3, h3-$\epsilon$3},
    xtick = data,
    axis lines*=left,
    grid = major,
    grid style={dashed, gray!30},
    legend cell align={left},
    enlarge y limits={0.1,upper},
    every axis y label/.style={
      at={(ticklabel* cs:1.05)},
      anchor=south,
},
    ]
    \addplot coordinates {
      (h1-$\epsilon$1, 11.178)
      (h2-$\epsilon$1, 17.48)
      (h3-$\epsilon$1, 28.581)
      (h1-$\epsilon$3, 3.689)
      (h2-$\epsilon$3, 5.768)
      (h3-$\epsilon$3, 9.432)};

    \addplot coordinates {
      (h1-$\epsilon$1,104.583)
      (h2-$\epsilon$1,44.9)
      (h3-$\epsilon$1,9.432)
      (h1-$\epsilon$3,34.861)
      (h2-$\epsilon$3,14.967)
      (h3-$\epsilon$3,3.144)};
    \legend{\ensuremath{{\tt{hierarchical}}_{1}},\ensuremath{{\tt{hierarchical}}_{2}}}
\end{axis}
\end{tikzpicture}
\vspace{1ex}
 {\raggedright \centering \ensuremath{{\tt{h}_1}\mathrel{=}\Varid{byGen},{\tt{h}_2}\mathrel{=}\Varid{byGenAge},{\tt{h}_3}\mathrel{=}\Varid{byGenAgeNat}} \par}
\caption{\ensuremath{{\tt{hierarchical}}_{1}} vs. \ensuremath{{\tt{hierarchical}}_{2}} \label{fig:h1vsh2}}
\end{figure}


\section{Related work}
\label{sec:relWork}

PINQ~\cite{PINQ09} is a programming framework for writing queries that
has inspired most of the subsequent works.
Queries can use basic aggregation mechanisms as well as transformations similar
to those supported by the DPella API.
PINQ uses dynamic tracking and sensitivity information to guarantee privacy of
computations.
Airavat~\cite{RoySKSW10} is a programming framework similar to PINQ but based on the
map-reduce model.
wPINQ~\cite{DBLP:journals/pvldb/ProserpioGM14} is an extension of the PINQ
framework that supports a more general form of join operators.
DJoin~\cite{DBLP:conf/osdi/NarayanH12} enforces differential privacy for
distributed databases, but for a restricted join operator.
Flex~\cite{DBLP:journals/pvldb/JohnsonNS18} enforces differential privacy for a
broader range of query plans, using the notion of elastic sensitivity.
None of these works keeps track of accuracy, nor static analysis for privacy or
accuracy.
We leave as future work to support join operations with accuracy---which is an
active area of research.

Fuzz~\cite{ReedP10} is a programming language which enforces (pure)
differential privacy of computations using a linear type system which
keeps track of program sensitivity.
DFuzz~\cite{GaboardiHHNP13} is a generalization of Fuzz based on linear
dependent types, a richer typing discipline which is able to accommodate
recursion.
Adaptive Fuzz~\cite{DBLP:journals/pacmpl/Winograd-CortHR17} is an adaptation of
Fuzz which supports adaptive data analysis using a combination of static and
dynamic techniques.
Finally, Fuzzi~\cite{zhang-2019-fuzzi} is a three layer system which
supports a more rich class of data analyses.
Ektelo~\cite{DBLP:conf/sigmod/ZhangMKHMM18} is a programming framework from
writing privacy-aware differentially private computations based on the matrix
mechanism and other components.
Ektelo allows one to write quite involved query strategies.
While several of Ektelo components are not supported in the current
implementation of DPella, these can be easily added as black-box components.
All these systems support reasoning about privacy, but not about accuracy.
In contrast, DPella supports accuracy but restricts the programming framework to
rule out certain analysis (e.g., adaptive ones) where supporting accuracy in a
compositional manner is still an open problem.

Hoare2~\cite{BartheGAHRS15} is a programming language which enforces (pure or
approximate) differential privacy using program verification.
Hoare2 combines a graded monad with a relational refinement type
system which keeps track of the relationship between two executions of
the program on adjacent inputs.
PrivInfer~\cite{BartheFGAGHS16} is an extension of Hoare2 that supports
differentially private Bayesian programming.
Both works are based on a relaxation of probabilistic couplings,
introduced in~\cite{BartheKOB12}.
In principle, both works allow programs that branch over query outputs.
However, their verification component is too weak to prove interesting
properties for such programs.
Subsquent work~\cite{BartheGGHS16} clarifies the relationship with probabilistic
couplings and provides stronger support to reason about programs that branch on
query outputs, such as the sparse vector technique.
Later, Albarghouthi and Hsu~\cite{DBLP:journals/pacmpl/AlbarghouthiH18} improve
the coupling-based approach and derive automated methods for proving privacy of
a broad range of computations.
Zhang et al.~\cite{ZhangK17} develop a similar automated approach but based on
the idea of aligning the randomness of two executions of a program.
More recently, this approach has been extended to deal with more involved
algorithms using an extra shadow execution~\cite{WangDWKZ19}.
Again. all these works focus on privacy for advanced data analyses while
neglecting accuracy.

In contrast, as we discussed before, several tools also support reasoning about
accuracy, but they restrict in general the kind of queries they support.
GUPT~\cite{MohanTSSC12} is a tool based on the sample-and-aggregate framework
for differential privacy~\cite{NissimRS07}.
GUPT allows analysts to specify the target accuracy of the output, and compute
privacy from it---or vice versa.
This approach has inspired several of the subsequent works and also our design.
The limitation of GUPT is that it supports only analyses that fit in the
sample-and-aggregate framework.
While this framework has several remarkable advantages, such as allowing
arbitrary queries to be run on the subsampled data, it only effectively supports
analyses that can be aggregated through some basic aggregation operations.
In contrast, DPella supports analyses of a more general class, such as the ones
we discussed in Section~\ref{sec:teaser} and Section~\ref{sec:examples}.
In principle, some of these analyses (e.g. CDF) could be seen as a
post-processed combination of analyses that also GUPT supports, the further step
that DPella takes is that it also allows to reason about the accuracy of the
combined form, rather that just about the individual queries.
PSI~\cite{DBLP:journals/corr/GaboardiHKNUV16} offers to the data analyst an
interface for selecting either the level of accuracy that she wants to reach, or
the level of privacy she wants to impose.
The error estimates that PSI provides are similar to the ones that are supported
in DPella.
However, similarly to GUPT, PSI supports only a limited set of transformations
and primitives, and in its current form it does not allow analysts to submit
their own (programmed) queries.
Once again, DPella can be seen as a programming environment that could be
combined with some of the analyses supported by tools similar to PSI or GUPT in
order to reason about the accuracy of the combined queries.

Ligett et al.~\cite{DBLP:journals/corr/LigettNRWW17} propose a framework for
developing differentially private algorithms under accuracy constraints.
This allows one to chose a given level of accuracy first, and then finding the
private algorithm meeting this accuracy.
This framework is so far limited to empirical risk minimization problems and it
is not supported by a system, yet.

APEx~\cite{Apex} is probably the most mature framework that has similar goals as
DPella since it supports reasoning about both accuracy and
privacy. Moreover, it allows data analysts to write queries as SQL-like statements.
However, the model that APEx uses is quite different from the one used by
DPella. First, APEx consider only three kind of queries:
WCQ (counting queries), ICQ (iceberg counting queries), and TCQ (top-k
counting queries).
To answer WCQ queries, as discussed in Section \ref{sec:examples}, APEx uses the
matrix mechanism.
While the matrix mechanism theoretical error is formulated in term of MSE
(see~\cite{Chao15MatMech}), APEx uses Monte Carlo simulations to achieve
accuracy bounds in terms of $\alpha$ and $\beta$, and to determine the least
privacy parameter ($\epsilon$) that fits those bounds.
We have shown how DPella con be used to answer queries based on the identity
strategies and the use of partition and concentration bounds.
To use effectively other query strategies we would need to extend DPella with
the matrix mechanism as a black-box.
This can be done easily but in this paper we want to keep the focus on the
programming support offered by DPella and we leave this for future work.
%

%
%
ICQ queries return the aggregate of bins greater than a threshold. To
answer these queries APEx applies novel data dependent accuracy bounds, so different
datasets might require different $\epsilon_s$ that fit the bounds---DPella
provides only data independent analyses.
TCQ queries are a generalization of report-noisy-max, this
generalization is not yet supported by DPella.
While APEx supports advanced queries, it does not provide means to reason about
combinations of analyses, e.g., it does not support reasoning about the accuracy
of a query using results from WCQs queries to perform TCQs ones.
DPella instead has been designed specifically to support the combination of
different queries. So, we can think about DPella as complimentary to APEx.

Focusing on core calculi, Barthe et al.~\cite{BartheGAHKS14} devise a method for
proving differential privacy using Hoare logic.
Their method uses accuracy bounds for the Laplace Mechanism for proving privacy
bounds of the Propose-Test-Release Mechanism, but cannot be used to prove
accuracy bounds of arbitrary computations.
Later, Barthe et al.~\cite{DBLP:conf/icalp/BartheGGHS16} develop a program logic
for proving accuracy bounds of differentially private computations based on the
Laplace Mechanism. Further, Barthe et al~\cite{BartheFGGHS16} use this logic, in
combination with a logic supporting reasoning by coupling, to verify
differentially private algorithms whose privacy guarantee depends on
the accuracy guarantee of some sub-component.
More recently, Smith et al.~\cite{DBLP:journals/pacmpl/SmithHA19} propose an
automated approach for computing accuracy bounds.
However, these methods use Union Bound and do not attempt to reason about
probabilistic independence to obtain tighter bounds.

%


\section{Conclusions}
\label{sec:conclusions}
DPella is an expressive programming framework for reasoning about privacy,
accuracy, and their trade-offs. DPella leverages features of Haskell type-system
to achieve expressiveness, and uses taint analysis to detect probabilistic
independence and derive tighter accuracy bounds using Chernoff bounds.

We believe that the principles behind DPella, i.e., the use of concentration
bounds guided by taint analysis, could also be used to support other mechanisms
(e.g., Gaussian) and thus providing support to $(\epsilon, \delta)$-DP, or
Renyi-DP~\cite{mironov2017renyi} as well as advanced forms of compositions.
We leave as future work to determine how to adapt the types and symbolic
interpreters of DPella to support such extensions.
Other future work also includes lifting the restriction that programs should not
branch on query outputs, and extending the scope of the language to other
notions of privacy, including concentrated differential
privacy~\cite{dwork2016concentrated}, zero concentrated differential
privacy~\cite{bun2016concentrated}, or truncated concentrated differential
privacy~\cite{bun2018composable}.

\selectformat
{\bibliographystyle{IEEEtranN}}%
{
\paragraph{Acknowledgments}
We would like to show our gratitude to Gilles Barthe (MPI-SP and IMDEA
Software Institute) who provided insight and expertise that greatly
assisted the development of DPella.

\bibliographystyle{unsrtnat}
}

\bibliography{local}

\begin{thebibliography}{53}
\providecommand{\natexlab}[1]{#1}
\providecommand{\url}[1]{\texttt{#1}}
\expandafter\ifx\csname urlstyle\endcsname\relax
  \providecommand{\doi}[1]{doi: #1}\else
  \providecommand{\doi}{doi: \begingroup \urlstyle{rm}\Url}\fi

\bibitem[Dwork et~al.(2006)Dwork, McSherry, Nissim, and Smith]{DMNS06}
Cynthia Dwork, Frank McSherry, Kobbi Nissim, and Adam Smith.
\newblock Calibrating noise to sensitivity in private data analysis.
\newblock In \emph{Proceedings of the Third Conference on Theory of
  Cryptography}, TCC'06, pages 265--284, 2006.
\newblock ISBN 3-540-32731-2, 978-3-540-32731-8.

\bibitem[McSherry(2009)]{PINQ09}
Frank~D. McSherry.
\newblock Privacy integrated queries: an extensible platform for
  privacy-preserving data analysis.
\newblock In \emph{SIGMOD}. ACM, 2009.

\bibitem[Roy et~al.(2010)Roy, Setty, Kilzer, Shmatikov, and Witchel]{RoySKSW10}
Indrajit Roy, Srinath T.~V. Setty, Ann Kilzer, Vitaly Shmatikov, and Emmett
  Witchel.
\newblock Airavat: Security and privacy for {MapReduce}.
\newblock In \emph{Proc. {USENIX} Symposium on Networked Systems Design and
  Implementation, {NSDI}}, 2010.

\bibitem[Reed and Pierce(2010)]{ReedP10}
Jason Reed and Benjamin~C. Pierce.
\newblock Distance makes the types grow stronger: a calculus for differential
  privacy.
\newblock In \emph{Proc. {ACM} {SIGPLAN} International Conference on Functional
  Programming}, 2010.

\bibitem[Haeberlen et~al.(2011)Haeberlen, Pierce, and Narayan]{HaeberlenPN11}
Andreas Haeberlen, Benjamin~C. Pierce, and Arjun Narayan.
\newblock Differential privacy under fire.
\newblock In \emph{Proc. of {USENIX} Security Symposium}, 2011.

\bibitem[Gaboardi et~al.(2013)Gaboardi, Haeberlen, Hsu, Narayan, and
  Pierce]{GaboardiHHNP13}
Marco Gaboardi, Andreas Haeberlen, Justin Hsu, Arjun Narayan, and Benjamin~C.
  Pierce.
\newblock Linear dependent types for differential privacy.
\newblock In \emph{Proc. {ACM} {SIGPLAN-SIGACT} Symposium on Principles of
  Programming Languages}, 2013.

\bibitem[Barthe et~al.(2015)Barthe, Gaboardi, Gallego~Arias, Hsu, Roth, and
  Strub]{BartheGAHRS15}
Gilles Barthe, Marco Gaboardi, Emilio~Jes{\'{u}}s Gallego~Arias, Justin Hsu,
  Aaron Roth, and Pierre-Yves Strub.
\newblock Higher-order approximate relational refinement types for mechanism
  design and differential privacy.
\newblock In \emph{POPL'15}. ACM, 2015.

\bibitem[Barthe et~al.(2016{\natexlab{a}})Barthe, Farina, Gaboardi, Arias,
  Gordon, Hsu, and Strub]{BartheFGAGHS16}
Gilles Barthe, Gian~Pietro Farina, Marco Gaboardi, Emilio Jes{\'{u}}s~Gallego
  Arias, Andy Gordon, Justin Hsu, and Pierre{-}Yves Strub.
\newblock Differentially private bayesian programming.
\newblock In \emph{Proc. {ACM} {SIGSAC} Conference on Computer and
  Communications Security}, 2016{\natexlab{a}}.

\bibitem[Zhang and Kifer(2017)]{ZhangK17}
Danfeng Zhang and Daniel Kifer.
\newblock {LightDP}: towards automating differential privacy proofs.
\newblock In \emph{Proc. {ACM} {SIGPLAN} Symp. on Principles of Programming
  Languages}, 2017.

\bibitem[Winograd{-}Cort et~al.(2017)Winograd{-}Cort, Haeberlen, Roth, and
  Pierce]{DBLP:journals/pacmpl/Winograd-CortHR17}
Daniel Winograd{-}Cort, Andreas Haeberlen, Aaron Roth, and Benjamin~C. Pierce.
\newblock A framework for adaptive differential privacy.
\newblock \emph{{PACMPL}}, 1\penalty0 ({ICFP}), 2017.

\bibitem[Johnson et~al.(2018{\natexlab{a}})Johnson, Near, and
  Song]{JohnsonNS18}
Noah~M. Johnson, Joseph~P. Near, and Dawn Song.
\newblock Towards practical differential privacy for {SQL} queries.
\newblock \emph{{PVLDB}}, 11\penalty0 (5), 2018{\natexlab{a}}.

\bibitem[Zhang et~al.(2018)Zhang, McKenna, Kotsogiannis, Hay, Machanavajjhala,
  and Miklau]{DBLP:conf/sigmod/ZhangMKHMM18}
Dan Zhang, Ryan McKenna, Ios Kotsogiannis, Michael Hay, Ashwin Machanavajjhala,
  and Gerome Miklau.
\newblock {EKTELO:} {A} framework for defining differentially-private
  computations.
\newblock In \emph{Proc. International Conference on Management of Data}, 2018.

\bibitem[Zhang et~al.(2019)Zhang, Roth, Haeberlen, Pierce, and
  Roth]{zhang-2019-fuzzi}
Hengchu Zhang, Edo Roth, Andreas Haeberlen, Benjamin~C. Pierce, and Aaron Roth.
\newblock Fuzzi: {A} three-level logic for differential privacy.
\newblock In \emph{Proc. {ACM SIGPLAN} International Conference on Functional
  Programming (ICFP'19)}, 2019.

\bibitem[Machanavajjhala et~al.(2008)Machanavajjhala, Kifer, Abowd, Gehrke, and
  Vilhuber]{MachanavajjhalaKAGV08}
Ashwin Machanavajjhala, Daniel Kifer, John~M. Abowd, Johannes Gehrke, and Lars
  Vilhuber.
\newblock Privacy: Theory meets practice on the map.
\newblock In \emph{Proc. International Conference on Data Engineering, {ICDE}},
  2008.

\bibitem[Mohan et~al.(2012)Mohan, Thakurta, Shi, Song, and Culler]{MohanTSSC12}
Prashanth Mohan, Abhradeep Thakurta, Elaine Shi, Dawn Song, and David~E.
  Culler.
\newblock {GUPT:} privacy preserving data analysis made easy.
\newblock In \emph{Proc. {ACM} {SIGMOD} International Conference on Management
  of Data, {SIGMOD}}, 2012.

\bibitem[Mir et~al.(2013)Mir, Isaacman, C{\'{a}}ceres, Martonosi, and
  Wright]{MirICMW13}
Darakhshan~J. Mir, Sibren Isaacman, Ram{\'{o}}n C{\'{a}}ceres, Margaret
  Martonosi, and Rebecca~N. Wright.
\newblock {DP-WHERE:} differentially private modeling of human mobility.
\newblock In \emph{Proc. {IEEE} International Conference on Big Data}, 2013.

\bibitem[Gaboardi et~al.(2016)Gaboardi, Honaker, King, Nissim, Ullman, and
  Vadhan]{DBLP:journals/corr/GaboardiHKNUV16}
Marco Gaboardi, James Honaker, Gary King, Kobbi Nissim, Jonathan Ullman, and
  Salil~P. Vadhan.
\newblock {PSI} ({\(\Psi\)}): a private data sharing interface.
\newblock \emph{CoRR}, abs/1609.04340, 2016.

\bibitem[Ge et~al.(2019)Ge, He, Ilyas, and Machanavajjhala]{Apex}
Chang Ge, Xi~He, Ihab~F. Ilyas, and Ashwin Machanavajjhala.
\newblock {APEx}: Accuracy-aware differentially private data exploration.
\newblock In \emph{Proc. International Conference on Management of Data (to
  appear)}, 2019.

\bibitem[Dubhashi and Panconesi(2009)]{dubhashi2009concentration}
Devdatt~P Dubhashi and Alessandro Panconesi.
\newblock \emph{Concentration of measure for the analysis of randomized
  algorithms}.
\newblock Cambridge University Press, 2009.

\bibitem[Dwork and Roth(2014)]{DworkR14}
Cynthia Dwork and Aaron Roth.
\newblock The algorithmic foundations of differential privacy.
\newblock \emph{Foundations and Trends in Theoretical Computer Science},
  9\penalty0 (3-4):\penalty0 211--407, 2014.

\bibitem[Dwork et~al.(2010)Dwork, Rothblum, and Vadhan]{DworkRV10}
Cynthia Dwork, Guy~N. Rothblum, and Salil~P. Vadhan.
\newblock Boosting and differential privacy.
\newblock In \emph{51th Annual {IEEE} Symposium on Foundations of Computer
  Science, {FOCS} 2010, October 23-26, 2010, Las Vegas, Nevada, {USA}}, pages
  51--60, 2010.
\newblock \doi{10.1109/FOCS.2010.12}.
\newblock URL \url{https://doi.org/10.1109/FOCS.2010.12}.

\bibitem[Sabelfeld and Myers(2003)]{Sabelfeld:Myers:JSAC}
A.~Sabelfeld and A.~C. Myers.
\newblock {Language-Based Information-Flow Security}.
\newblock \emph{IEEE J. Selected Areas in Communications}, 21\penalty0
  (1):\penalty0 5--19, January 2003.

\bibitem[Schoepe et~al.(2016)Schoepe, Balliu, Pierce, and
  Sabelfeld]{DBLP:conf/eurosp/SchoepeBPS16}
Daniel Schoepe, Musard Balliu, Benjamin~C. Pierce, and Andrei Sabelfeld.
\newblock Explicit secrecy: {A} policy for taint tracking.
\newblock In \emph{{IEEE} European Symposium on Security and Privacy}, pages
  15--30, 2016.

\bibitem[Li and Zdancewic(2010)]{Li+:2010:arrows}
P.~Li and S.~Zdancewic.
\newblock {Arrows for secure information flow}.
\newblock \emph{Theoretical Computer Science}, 411\penalty0 (19):\penalty0
  1974--1994, 2010.

\bibitem[Russo et~al.(2008)Russo, Claessen, and Hughes]{Russo+:Haskell08}
A.~Russo, K.~Claessen, and J.~Hughes.
\newblock {A library for light-weight information-flow security in {H}askell}.
\newblock In \emph{{Proc. ACM {SIGPLAN} Symp. on {H}askell}}. ACM Press, 2008.

\bibitem[Smith et~al.(2019)Smith, Hsu, and
  Albarghouthi]{DBLP:journals/pacmpl/SmithHA19}
Calvin Smith, Justin Hsu, and Aws Albarghouthi.
\newblock Trace abstraction modulo probability.
\newblock \emph{{PACMPL}}, 3\penalty0 ({POPL}), 2019.

\bibitem[McSherry and Mahajan(2011)]{Mcsherry2011network}
Frank McSherry and Ratul Mahajan.
\newblock Differentially-private network trace analysis.
\newblock \emph{ACM SIGCOMM Computer Communication Review}, 41\penalty0
  (4):\penalty0 123--134, 2011.

\bibitem[Barthe et~al.(2014)Barthe, Gaboardi, Arias, Hsu, Kunz, and
  Strub]{BartheGAHKS14}
Gilles Barthe, Marco Gaboardi, Emilio Jes{\'{u}}s~Gallego Arias, Justin Hsu,
  C{\'{e}}sar Kunz, and Pierre{-}Yves Strub.
\newblock Proving differential privacy in {H}oare logic.
\newblock In \emph{Proc. {IEEE} Computer Security Foundations Symposium}, 2014.

\bibitem[Li et~al.(2015)Li, Miklau, Hay, McGregor, and Rastogi]{Chao15MatMech}
Chao Li, Gerome Miklau, Michael Hay, Andrew McGregor, and Vibhor Rastogi.
\newblock The matrix mechanism: optimizing linear counting queries under
  differential privacy.
\newblock \emph{{VLDB} J.}, 24\penalty0 (6), 2015.

\bibitem[Hay et~al.(2010)Hay, Rastogi, Miklau, and Suciu]{HayRMS10Boosting}
Michael Hay, Vibhor Rastogi, Gerome Miklau, and Dan Suciu.
\newblock Boosting the accuracy of differentially private histograms through
  consistency.
\newblock \emph{{PVLDB}}, 3\penalty0 (1), 2010.

\bibitem[Xiao et~al.(2011)Xiao, Wang, and Gehrke]{XiaoWG11Privlet}
Xiaokui Xiao, Guozhang Wang, and Johannes Gehrke.
\newblock Differential privacy via wavelet transforms.
\newblock \emph{{IEEE} Trans. Knowl. Data Eng.}, 23\penalty0 (8), 2011.

\bibitem[Moggi(1991)]{DBLP:journals/iandc/Moggi91}
Eugenio Moggi.
\newblock Notions of computation and monads.
\newblock \emph{Inf. Comput.}, 93\penalty0 (1):\penalty0 55--92, 1991.

\bibitem[Terei et~al.(2012)Terei, Marlow, {Peyton Jones}, and
  Mazi{\`{e}}res]{DBLP:conf/haskell/TereiMJM12}
David Terei, Simon Marlow, Simon~L. {Peyton Jones}, and David Mazi{\`{e}}res.
\newblock Safe {H}askell.
\newblock In \emph{Proceedings of the 5th {ACM} {SIGPLAN} Symposium on Haskell,
  Haskell 2012, Copenhagen, Denmark, 13 September 2012}, pages 137--148, 2012.

\bibitem[Ebadi and Sands(2017)]{DBLP:journals/corr/EbadiS15}
Hamid Ebadi and David Sands.
\newblock Featherweight {PINQ}.
\newblock \emph{Privacy and Confidentiality}, 7\penalty0 (2), 2017.

\bibitem[Eisenberg et~al.(2014)Eisenberg, Vytiniotis, {Peyton Jones}, and
  Weirich]{DBLP:conf/popl/EisenbergVJW14}
Richard~A. Eisenberg, Dimitrios Vytiniotis, Simon~L. {Peyton Jones}, and
  Stephanie Weirich.
\newblock Closed type families with overlapping equations.
\newblock In \emph{The {ACM} {SIGPLAN-SIGACT} Symposium on Principles of
  Programming Languages}, 2014.

\bibitem[Narayan and Haeberlen(2012)]{DBLP:conf/osdi/NarayanH12}
Arjun Narayan and Andreas Haeberlen.
\newblock Djoin: Differentially private join queries over distributed
  databases.
\newblock In Chandu Thekkath and Amin Vahdat, editors, \emph{10th {USENIX}
  Symposium on Operating Systems Design and Implementation, {OSDI}}. {USENIX}
  Association, 2012.

\bibitem[Blocki et~al.(2013)Blocki, Blum, Datta, and
  Sheffet]{DBLP:conf/innovations/BlockiBDS13}
Jeremiah Blocki, Avrim Blum, Anupam Datta, and Or~Sheffet.
\newblock Differentially private data analysis of social networks via
  restricted sensitivity.
\newblock In \emph{Innovations in Theoretical Computer Science, {ITCS}}, 2013.

\bibitem[Johnson et~al.(2018{\natexlab{b}})Johnson, Near, and
  Song]{DBLP:journals/pvldb/JohnsonNS18}
Noah~M. Johnson, Joseph~P. Near, and Dawn Song.
\newblock Towards practical differential privacy for {SQL} queries.
\newblock \emph{{PVLDB}}, 11\penalty0 (5), 2018{\natexlab{b}}.

\bibitem[Russo(2015)]{Russo:2015}
Alejandro Russo.
\newblock {Functional Pearl: Two Can Keep a Secret, if One of Them Uses
  Haskell}.
\newblock In \emph{Proc. of the {ACM SIGPLAN} {I}nternational Conference on
  Functional Programming}. ACM, 2015.

\bibitem[Barthe et~al.(2016{\natexlab{b}})Barthe, Gaboardi, Gr{\'{e}}goire,
  Hsu, and Strub]{DBLP:conf/icalp/BartheGGHS16}
Gilles Barthe, Marco Gaboardi, Benjamin Gr{\'{e}}goire, Justin Hsu, and
  Pierre{-}Yves Strub.
\newblock A program logic for union bounds.
\newblock In Ioannis Chatzigiannakis, Michael Mitzenmacher, Yuval Rabani, and
  Davide Sangiorgi, editors, \emph{International Colloquium on Automata,
  Languages, and Programming, {ICALP}}, volume~55 of \emph{LIPIcs}. Schloss
  Dagstuhl - Leibniz-Zentrum fuer Informatik, 2016{\natexlab{b}}.

\bibitem[Chan et~al.(2011)Chan, Shi, and Song]{chan2011private}
T-H~Hubert Chan, Elaine Shi, and Dawn Song.
\newblock Private and continual release of statistics.
\newblock \emph{ACM Transactions on Information and System Security (TISSEC)},
  14\penalty0 (3):\penalty0 26, 2011.

\bibitem[Proserpio et~al.(2014)Proserpio, Goldberg, and
  McSherry]{DBLP:journals/pvldb/ProserpioGM14}
Davide Proserpio, Sharon Goldberg, and Frank McSherry.
\newblock Calibrating data to sensitivity in private data analysis.
\newblock \emph{{PVLDB}}, 7\penalty0 (8), 2014.

\bibitem[Barthe et~al.(2012)Barthe, K{\"{o}}pf, Olmedo, and
  B{\'{e}}guelin]{BartheKOB12}
Gilles Barthe, Boris K{\"{o}}pf, Federico Olmedo, and Santiago~Zanella
  B{\'{e}}guelin.
\newblock Probabilistic relational reasoning for differential privacy.
\newblock In \emph{Proc. {ACM} {SIGPLAN-SIGACT} Symposium on Principles of
  Programming Languages}, 2012.

\bibitem[Barthe et~al.(2016{\natexlab{c}})Barthe, Gaboardi, Gr{\'{e}}goire,
  Hsu, and Strub]{BartheGGHS16}
Gilles Barthe, Marco Gaboardi, Benjamin Gr{\'{e}}goire, Justin Hsu, and
  Pierre{-}Yves Strub.
\newblock Proving differential privacy via probabilistic couplings.
\newblock In \emph{Proc. {ACM/IEEE} Symposium on Logic in Computer Science},
  2016{\natexlab{c}}.

\bibitem[Albarghouthi and Hsu(2018)]{DBLP:journals/pacmpl/AlbarghouthiH18}
Aws Albarghouthi and Justin Hsu.
\newblock Synthesizing coupling proofs of differential privacy.
\newblock \emph{{PACMPL}}, 2\penalty0 ({POPL}), 2018.

\bibitem[Wang et~al.(2019)Wang, Ding, Wang, Kifer, and Zhang]{WangDWKZ19}
Yuxin Wang, Zeyu Ding, Guanhong Wang, Daniel Kifer, and Danfeng Zhang.
\newblock Proving differential privacy with shadow execution.
\newblock In \emph{Proc. {ACM} {SIGPLAN} Conference on Programming Language
  Design and Implementation}, 2019.

\bibitem[Nissim et~al.(2007)Nissim, Raskhodnikova, and Smith]{NissimRS07}
Kobbi Nissim, Sofya Raskhodnikova, and Adam~D. Smith.
\newblock Smooth sensitivity and sampling in private data analysis.
\newblock In \emph{Proc. Annual {ACM} Symposium on Theory of Computing}, 2007.

\bibitem[Ligett et~al.(2017)Ligett, Neel, Roth, Waggoner, and
  Wu]{DBLP:journals/corr/LigettNRWW17}
Katrina Ligett, Seth Neel, Aaron Roth, Bo~Waggoner, and Zhiwei~Steven Wu.
\newblock Accuracy first: Selecting a differential privacy level for
  accuracy-constrained {ERM}.
\newblock \emph{CoRR}, abs/1705.10829, 2017.

\bibitem[Barthe et~al.(2016{\natexlab{d}})Barthe, Fong, Gaboardi,
  Gr{\'{e}}goire, Hsu, and Strub]{BartheFGGHS16}
Gilles Barthe, No{\'{e}}mie Fong, Marco Gaboardi, Benjamin Gr{\'{e}}goire,
  Justin Hsu, and Pierre{-}Yves Strub.
\newblock Advanced probabilistic couplings for differential privacy.
\newblock In \emph{Proc. {ACM} {SIGSAC} Conference on Computer and
  Communications Security}, 2016{\natexlab{d}}.

\bibitem[Mironov(2017)]{mironov2017renyi}
Ilya Mironov.
\newblock R{\'e}nyi differential privacy.
\newblock In \emph{2017 IEEE 30th Computer Security Foundations Symposium
  (CSF)}. IEEE, 2017.

\bibitem[Dwork and Rothblum(2016)]{dwork2016concentrated}
Cynthia Dwork and Guy~N Rothblum.
\newblock Concentrated differential privacy.
\newblock \emph{arXiv preprint arXiv:1603.01887}, 2016.

\bibitem[Bun and Steinke(2016)]{bun2016concentrated}
Mark Bun and Thomas Steinke.
\newblock Concentrated differential privacy: Simplifications, extensions, and
  lower bounds.
\newblock In \emph{Theory of Cryptography Conference}. Springer, 2016.

\bibitem[Bun et~al.(2018)Bun, Dwork, Rothblum, and Steinke]{bun2018composable}
Mark Bun, Cynthia Dwork, Guy~N Rothblum, and Thomas Steinke.
\newblock Composable and versatile privacy via truncated cdp.
\newblock In \emph{Proceedings of the 50th Annual ACM SIGACT Symposium on
  Theory of Computing}, pages 74--86. ACM, 2018.

\end{thebibliography}

\showappendix{
\appendix
\section{Appendix}{\label{sec:appendix}}

\subsection{Accuracy vs Privacy trade-offs}
\label{app:hierach:opt}

\begin{figure}[!h]
{\small
\numbersreset
\begin{hscode}\linenumsetup\printlinebegin\SaveRestoreHook
\column{B}{@{}>{\hspre}l<{\hspost}@{}}%
\column{5}{@{}>{\hspre}l<{\hspost}@{}}%
\column{9}{@{}>{\hspre}l<{\hspost}@{}}%
\column{10}{@{}>{\hspre}l<{\hspost}@{}}%
\column{11}{@{}>{\hspre}l<{\hspost}@{}}%
\column{14}{@{}>{\hspre}l<{\hspost}@{}}%
\column{15}{@{}>{\hspre}c<{\hspost}@{}}%
\column{15E}{@{}l@{}}%
\column{17}{@{}>{\hspre}l<{\hspost}@{}}%
\column{18}{@{}>{\hspre}c<{\hspost}@{}}%
\column{18E}{@{}l@{}}%
\column{19}{@{}>{\hspre}l<{\hspost}@{}}%
\column{20}{@{}>{\hspre}l<{\hspost}@{}}%
\column{21}{@{}>{\hspre}l<{\hspost}@{}}%
\column{23}{@{}>{\hspre}l<{\hspost}@{}}%
\column{24}{@{}>{\hspre}l<{\hspost}@{}}%
\column{28}{@{}>{\hspre}l<{\hspost}@{}}%
\column{29}{@{}>{\hspre}l<{\hspost}@{}}%
\column{31}{@{}>{\hspre}c<{\hspost}@{}}%
\column{31E}{@{}l@{}}%
\column{34}{@{}>{\hspre}l<{\hspost}@{}}%
\column{35}{@{}>{\hspre}l<{\hspost}@{}}%
\column{52}{@{}>{\hspre}l<{\hspost}@{}}%
\column{E}{@{}>{\hspre}l<{\hspost}@{}}%
\>[B]{}\mbox{\onelinecomment  Cases for which the optimizer might unsuccessfully stop}{}\<[E]%
\printlineend\\
\printlinebegin\>[B]{}\mathbf{data}\;\Conid{ErrorInfo}{}\<[17]%
\>[17]{}\mathrel{=}{}\<[20]%
\>[20]{}\mbox{\onelinecomment  Maximum iterations}{}\<[E]%
\printlineend\\
\printlinebegin\>[20]{}\Conid{MaxIteration}\;{}\<[34]%
\>[34]{}(\epsilon,\alpha){}\<[E]%
\printlineend\\
\printlinebegin\>[20]{}\mbox{\onelinecomment  Maximum budget}{}\<[E]%
\printlineend\\
\printlinebegin\>[17]{}\mid {}\<[20]%
\>[20]{}\Conid{MaxBudget}\;{}\<[34]%
\>[34]{}(\epsilon,\alpha){}\<[E]%
\printlineend\\[\blanklineskip]%
\printlinebegin\>[B]{}\mbox{\onelinecomment  Optimizator parameters}{}\<[E]%
\printlineend\\
\printlinebegin\>[B]{}\mathbf{data}\;\Conid{Input}\mathrel{=}\Conid{In}\;{}\<[18]%
\>[18]{}\{\mskip1.5mu {}\<[18E]%
\>[21]{}\Varid{budTotal}{}\<[31]%
\>[31]{}\mathbin{::}{}\<[31E]%
\>[35]{}\epsilon{}\<[E]%
\printlineend\\
\printlinebegin\>[18]{},{}\<[18E]%
\>[21]{}\Varid{minEps}{}\<[31]%
\>[31]{}\mathbin{::}{}\<[31E]%
\>[35]{}\epsilon{}\<[E]%
\printlineend\\
\printlinebegin\>[18]{},{}\<[18E]%
\>[21]{}\Delta{}\<[31]%
\>[31]{}\mathbin{::}{}\<[31E]%
\>[35]{}\epsilon{}\<[E]%
\printlineend\\
\printlinebegin\>[18]{},{}\<[18E]%
\>[21]{}\Varid{beta}{}\<[31]%
\>[31]{}\mathbin{::}{}\<[31E]%
\>[35]{}\beta{}\<[E]%
\printlineend\\
\printlinebegin\>[18]{},{}\<[18E]%
\>[21]{}\Varid{errorTol}{}\<[31]%
\>[31]{}\mathbin{::}{}\<[31E]%
\>[35]{}\alpha{}\<[E]%
\printlineend\\
\printlinebegin\>[18]{},{}\<[18E]%
\>[21]{}\Varid{iter}{}\<[31]%
\>[31]{}\mathbin{::}{}\<[31E]%
\>[35]{}\Conid{Int}\mskip1.5mu\}{}\<[E]%
\printlineend\\[\blanklineskip]%
\printlinebegin\>[B]{}\mbox{\onelinecomment  Inspect one program}{}\<[E]%
\printlineend\\
\printlinebegin\>[B]{}\Varid{iterateError}{}\<[15]%
\>[15]{}\mathbin{::}{}\<[15E]%
\>[19]{}(\epsilon\to {\color{darkblue} \tt{Query}}\;({\color{darkgreen} \tt{Value}}\;\Varid{a}))\to {}\<[52]%
\>[52]{}\Conid{Input}{}\<[E]%
\printlineend\\
\printlinebegin\>[15]{}\to {}\<[15E]%
\>[19]{}{\tt{Option}}\;\Conid{ErrorInfo}\;(\epsilon,\alpha){}\<[E]%
\printlineend\\
\printlinebegin\>[B]{}\Varid{iterateError}\;\Varid{prog}\;(\Conid{In}\;\Varid{bud}\;\Varid{eInit}\;\Delta\;\Varid{beta}\;\Varid{errorTol}\;\Varid{n})\mathrel{=}{}\<[E]%
\printlineend\\
\printlinebegin\>[B]{}\hsindent{5}{}\<[5]%
\>[5]{}\mathbf{if}\;\Varid{currentErr}{\color{darkblue}  \texttt{>}}\Varid{errorTol}{}\<[E]%
\printlineend\\
\printlinebegin\>[B]{}\hsindent{5}{}\<[5]%
\>[5]{}\mathbf{then}\;{}\<[11]%
\>[11]{}\mathbf{if}\;\Varid{n}{\color{darkblue}  \texttt{>}}{\color{darkgreen} \texttt{0}}{}\<[E]%
\printlineend\\
\printlinebegin\>[11]{}\mathbf{then}\;{}\<[17]%
\>[17]{}\mathbf{if}\;(\Varid{eInit}{\color{darkblue}  \texttt{+}}\Delta)\leq \Varid{bud}{}\<[E]%
\printlineend\\
\printlinebegin\>[17]{}\mathbf{then}\;\Varid{reduce}\;(\Varid{eInit}{\color{darkblue}  \texttt{+}}\Delta){}\<[E]%
\printlineend\\
\printlinebegin\>[17]{}\mathbf{else}\;{}\<[23]%
\>[23]{}\mathbf{if}\;\Varid{eInit}{\color{darkblue}  \texttt{<}}\Varid{bud}{}\<[E]%
\printlineend\\
\printlinebegin\>[23]{}\mathbf{then}\;\Varid{reduce}\;\Varid{bud}{}\<[E]%
\printlineend\\
\printlinebegin\>[23]{}\mathbf{else}\;{}\<[29]%
\>[29]{}\Conid{Left}{\color{darkblue} \tt{~\$~}}{}\<[E]%
\printlineend\\
\printlinebegin\>[29]{}\Conid{MaxBudget}\;(\Varid{eInit},\Varid{currentErr}){}\<[E]%
\printlineend\\
\printlinebegin\>[11]{}\mathbf{else}\;\Conid{Left}{\color{darkblue} \tt{~\$~}}\Conid{MaxIteration}\;(\Varid{eInit},\Varid{currentErr}){}\<[E]%
\printlineend\\
\printlinebegin\>[B]{}\hsindent{5}{}\<[5]%
\>[5]{}\mathbf{else}\;\Conid{Right}\;(\Varid{eInit},\Varid{currentErr}){}\<[E]%
\printlineend\\
\printlinebegin\>[B]{}\mathbf{where}\;{}\<[9]%
\>[9]{}\Varid{currentErr}{}\<[24]%
\>[24]{}\mathrel{=}({\color{darkblue} \tt{accuracy} }\mathbin{\$}\Varid{prog}\;\Varid{eInit})\;\Varid{beta}{}\<[E]%
\printlineend\\
\printlinebegin\>[9]{}\Varid{reduce}\;\Varid{eInit'}{}\<[24]%
\>[24]{}\mathrel{=}{}\<[E]%
\printlineend\\
\printlinebegin\>[9]{}\hsindent{5}{}\<[14]%
\>[14]{}\Varid{iterateError}\;{}\<[29]%
\>[29]{}\Varid{prog}\;{}\<[E]%
\printlineend\\
\printlinebegin\>[29]{}(\Conid{In}\;\Varid{bud}\;\Varid{eInit'}\;\Delta\;\Varid{beta}{}\<[E]%
\printlineend\\
\printlinebegin\>[29]{}\Varid{errorTol}\;(\Varid{n}{\color{darkblue}  \texttt{-}}{\color{darkgreen} \texttt{1}})){}\<[E]%
\printlineend\\[\blanklineskip]%
\printlinebegin\>[B]{}\mbox{\onelinecomment  Optimization for each histogram}{}\<[E]%
\printlineend\\
\printlinebegin\>[B]{}\Varid{chooseEps}\mathbin{::}\Conid{Input}\to [\mskip1.5mu \alpha\mskip1.5mu]{}\<[E]%
\printlineend\\
\printlinebegin\>[B]{}\hsindent{11}{}\<[11]%
\>[11]{}\to [\mskip1.5mu {\tt{Option}}\;\Conid{ErrorInfo}\;(\epsilon,\alpha)\mskip1.5mu]{}\<[E]%
\printlineend\\
\printlinebegin\>[B]{}\Varid{chooseEps}\;\Varid{args}\mathord{@}\Conid{In}\;\{\mskip1.5mu \Varid{budTotal}\mathrel{=}\Varid{epsTotal}\mskip1.5mu\}\;\Varid{tolerances}\mathrel{=}{}\<[E]%
\printlineend\\
\printlinebegin\>[B]{}\hsindent{5}{}\<[5]%
\>[5]{}\mathbf{let}\;{}\<[10]%
\>[10]{}\Varid{localEpsMax}\mathrel{=}\Varid{epsTotal}\mathbin{/}{\color{darkgreen} \texttt{3}}{}\<[E]%
\printlineend\\
\printlinebegin\>[10]{}\Varid{buildProg}\;\Varid{f}\;\Varid{e}\mathrel{=}\Varid{f}\;\Varid{e}\;\Varid{tabTest}{}\<[E]%
\printlineend\\
\printlinebegin\>[10]{}\Varid{programs}\mathrel{=}\Varid{map}\;\Varid{buildProg}\;[\mskip1.5mu \Varid{byGen},\Varid{byAge},\Varid{byNat}\mskip1.5mu]{}\<[E]%
\printlineend\\
\printlinebegin\>[10]{}\Varid{input}\;\Varid{tol}\mathrel{=}\Varid{args}\;{}\<[28]%
\>[28]{}\{\mskip1.5mu \Varid{budTotal}\mathrel{=}\Varid{localEpsMax}{}\<[E]%
\printlineend\\
\printlinebegin\>[28]{},\Varid{errorTol}\mathrel{=}\Varid{tol}\mskip1.5mu\}{}\<[E]%
\printlineend\\
\printlinebegin\>[10]{}\Varid{opt}\;(\Varid{f},\Varid{tol})\mathrel{=}\Varid{iterateError}\;\Varid{f}\;(\Varid{input}\;\Varid{tol}){}\<[E]%
\printlineend\\
\printlinebegin\>[B]{}\hsindent{5}{}\<[5]%
\>[5]{}\mathbf{in}\;\Varid{map}\;\Varid{opt}\;(\Varid{zip}\;\Varid{programs}\;\Varid{tolerances}){}\<[E]%
\printlineend\ColumnHook
\end{hscode}\resethooks
}
\vspace{-10pt}
\caption{Brute force optimizer for \ensuremath{{\tt{hierarchical}}_{1}} \label{fig:optH1}}
\end{figure}

Figure~\ref{fig:optH1} describes the implementation of a simple
optimizer for the \ensuremath{{\tt{hierarchical}}_{1}} query presented in
Section~\ref{sec:examples}. In a nutshell, \ensuremath{\Varid{iterateError}} takes a
function from \ensuremath{\epsilon} to a \ensuremath{{\color{darkblue} \tt{Query}}\;({\color{darkgreen} \tt{Value}}\;\Varid{a})} and iterativelly calls such
function with different values of \ensuremath{\epsilon} (starting with \ensuremath{\Varid{minEps}}
until \ensuremath{\Varid{budTotal}} with steps of \ensuremath{\Delta}) aiming to reach certain error
tolerance called \ensuremath{\Varid{errorTol}}.
The program finishes when one of the following events happened:

\begin{itemize}
\item Error \ensuremath{\Conid{Left}{\color{darkblue} \tt{~\$~}}\Conid{MaxIteration}\;(\epsilon,\alpha)} :
  It has reach the maximum number of iterations \ensuremath{\Varid{iter}} without finding
  an $\epsilon$ that satisfy the condition.

\item Error \ensuremath{\Conid{Left}{\color{darkblue} \tt{~\$~}}\Conid{MaxBudget}\;(\epsilon,\alpha)} :
  It has reach the maximum budget \ensuremath{\Varid{budTotal}} without finding an
  $\epsilon$ that satisfy the condition.

\item Success \ensuremath{\Conid{Rigth}\;(\epsilon,\alpha)} :
  The goal has been fullfilled.
\end{itemize}

Finally, function \ensuremath{\Varid{chooseEps}} will call \ensuremath{\Varid{iterError}} for each histogram
with their respective error tolerances \ensuremath{\Varid{errorTols}}.




\subsection{Implementation}
\label{sec:design}



\paragraph{Basic structures}
We start by describing the implementation of datasets and values.
Intuitively, datasets are simply encoded as Haskell lists:
%
\numbersoff
{\small
\begin{hscode}\linenumsetup\printlinebegin\SaveRestoreHook
\column{B}{@{}>{\hspre}l<{\hspost}@{}}%
\column{3}{@{}>{\hspre}l<{\hspost}@{}}%
\column{39}{@{}>{\hspre}c<{\hspost}@{}}%
\column{39E}{@{}l@{}}%
\column{42}{@{}>{\hspre}l<{\hspost}@{}}%
\column{51}{@{}>{\hspre}c<{\hspost}@{}}%
\column{51E}{@{}l@{}}%
\column{55}{@{}>{\hspre}l<{\hspost}@{}}%
\column{E}{@{}>{\hspre}l<{\hspost}@{}}%
\>[3]{}\mathbf{data}\;{\color{darkred} \tt{Data}}\;({\color{darkblue} \tt{s} }\mathbin{::}\Conid{Nat})\;{\color{darkblue} \tt{r} }\mathrel{=}\Conid{MkData}\;\{\mskip1.5mu {}\<[42]%
\>[42]{}\Varid{owner}{}\<[51]%
\>[51]{}\mathbin{::}{}\<[51E]%
\>[55]{}\Conid{ThreadId},{}\<[E]%
\printlineend\\
\printlinebegin\>[42]{}\Varid{content}{}\<[51]%
\>[51]{}\mathbin{::}{}\<[51E]%
\>[55]{}[\mskip1.5mu {\color{darkblue} \tt{r} }\mskip1.5mu]{}\<[E]%
\printlineend\\
\printlinebegin\>[3]{}\hsindent{36}{}\<[39]%
\>[39]{}\mskip1.5mu\}{}\<[39E]%
\printlineend\ColumnHook
\end{hscode}\resethooks
}
This definition enforces the accumulated stability of the dataset to be a
type-level natural number by requiring \ensuremath{{\color{darkblue} \tt{s} }} to have kind \ensuremath{\Conid{Nat}} (\ensuremath{{\color{darkblue} \tt{s} }\mathbin{::}\Conid{Nat}}).
(The symbol \ensuremath{\mathbin{::}} is overloaded to also talk about kinds at the type-level.)
The field \ensuremath{\Varid{owner}} is used to keep track of the provenance of the dataset in
order to safely apply \ensuremath{{\color{darkred} \tt{dpPart} }}---recall Section \ref{sec:partition}.
The record \ensuremath{\Varid{content}} stores the elements of the dataset in a list of type \ensuremath{[\mskip1.5mu {\color{darkblue} \tt{r} }\mskip1.5mu]}.
Values, on the other hand, are implemented as follows:

\numbersoff
{\small
\begin{hscode}\linenumsetup\printlinebegin\SaveRestoreHook
\column{B}{@{}>{\hspre}l<{\hspost}@{}}%
\column{3}{@{}>{\hspre}l<{\hspost}@{}}%
\column{27}{@{}>{\hspre}c<{\hspost}@{}}%
\column{27E}{@{}l@{}}%
\column{30}{@{}>{\hspre}l<{\hspost}@{}}%
\column{37}{@{}>{\hspre}l<{\hspost}@{}}%
\column{E}{@{}>{\hspre}l<{\hspost}@{}}%
\>[3]{}\mathbf{data}\;{\color{darkgreen} \tt{Value}}\;\Varid{a}\mathrel{=}\Conid{MkValue}\;{}\<[27]%
\>[27]{}\{\mskip1.5mu {}\<[27E]%
\>[30]{}\Varid{value}{}\<[37]%
\>[37]{}\mathbin{::}\Varid{a}{}\<[E]%
\printlineend\\
\printlinebegin\>[27]{},{}\<[27E]%
\>[30]{}\Varid{iCDF}{}\<[37]%
\>[37]{}\mathbin{::}\beta\to \alpha{}\<[E]%
\printlineend\\
\printlinebegin\>[27]{},{}\<[27E]%
\>[30]{}\Varid{scale}{}\<[37]%
\>[37]{}\mathbin{::}\Conid{Maybe}\;{\tt{Double}}{}\<[E]%
\printlineend\\
\printlinebegin\>[27]{},{}\<[27E]%
\>[30]{}\Varid{label}{}\<[37]%
\>[37]{}\mathbin{::}[\mskip1.5mu \Conid{Int}\mskip1.5mu]\mskip1.5mu\}{}\<[E]%
\printlineend\ColumnHook
\end{hscode}\resethooks
}
The field \ensuremath{\Varid{value}} stores the concrete value of type \ensuremath{\Varid{a}}, while \ensuremath{\Varid{iCDF}} captures
its accuracy estimation.
Field \ensuremath{\Varid{scale}} is a floating point number tracking the scale of the Laplace
distribution used to sample such value.
More specifically, we use the type \ensuremath{\Conid{Maybe}\;{\tt{Double}}} which denotes an optional
floating point number, where constructor \ensuremath{\Conid{Nothing}} indicates that the value is
tainted and constructor \ensuremath{\Conid{Just}\;\Varid{scale}} stores the \ensuremath{\Varid{scale}} of the Laplace
distribution used to independently sample the content of the field \ensuremath{\Varid{value}},
i.e., when the value is considered untainted---information that is required when
applying Chernoff bound (recall Definition \ref{def:chernoff}).
The field \ensuremath{\Varid{label}} attaches a provenance label to detect when untainted values
come from different ``noisy sources''---otherwise, union bound must be applied
when adding them.

\paragraph{Queries}
Queries are implemented using a deep-embedded DSL.
The characteristic of deep embedding is that they use an abstract syntax tree
(AST) to represent the domain---in this case, the queries.
By doing so, we are capable to provide \emph{multiple interpretations for queries},
namely the ones given by functions \ensuremath{{\color{darkblue} \tt{budget} }}, \ensuremath{{\color{darkblue} \tt{accuracy} }}, and \ensuremath{{\color{darkred} \tt{dpEval} }}.
Two of these interpretations are symbolic, i.e., \ensuremath{{\color{darkblue} \tt{budget} }} and \ensuremath{{\color{darkblue} \tt{accuracy} }}.

We present the (partial) definition of the data type encoding queries:
{\small
\begin{hscode}\linenumsetup\printlinebegin\SaveRestoreHook
\column{B}{@{}>{\hspre}l<{\hspost}@{}}%
\column{3}{@{}>{\hspre}l<{\hspost}@{}}%
\column{12}{@{}>{\hspre}l<{\hspost}@{}}%
\column{13}{@{}>{\hspre}l<{\hspost}@{}}%
\column{22}{@{}>{\hspre}l<{\hspost}@{}}%
\column{26}{@{}>{\hspre}l<{\hspost}@{}}%
\column{E}{@{}>{\hspre}l<{\hspost}@{}}%
\>[B]{}\mathbf{data}\;{\color{darkblue} \tt{Query}}\;\Varid{a}\;\mathbf{where}{}\<[E]%
\printlineend\\
\printlinebegin\>[B]{}\hsindent{3}{}\<[3]%
\>[3]{}\mbox{\onelinecomment  Transformations}{}\<[E]%
\printlineend\\
\printlinebegin\>[B]{}\hsindent{3}{}\<[3]%
\>[3]{}\Conid{Where}\mathbin{::}({\color{darkblue} \tt{r} }\to \Conid{Bool})\to {\color{darkred} \tt{Data}}\;{\color{darkblue} \tt{s} }\;{\color{darkblue} \tt{r} }\to {\color{darkblue} \tt{Query}}\;({\color{darkred} \tt{Data}}\;{\color{darkblue} \tt{s} }\;{\color{darkblue} \tt{r} }){}\<[E]%
\printlineend\\
\printlinebegin\>[B]{}\hsindent{3}{}\<[3]%
\>[3]{}\Conid{Part}\mathbin{::}{}\<[12]%
\>[12]{}\Conid{Ord}\;\Varid{k}\Rightarrow {}\<[22]%
\>[22]{}({\color{darkblue} \tt{r} }\to \Varid{k}){}\<[E]%
\printlineend\\
\printlinebegin\>[22]{}\to \Varid{z}\;{\color{darkred} \tt{Data}}\;{\color{darkblue} \tt{s} }\;{\color{darkblue} \tt{r} }{}\<[E]%
\printlineend\\
\printlinebegin\>[22]{}\to {}\<[26]%
\>[26]{}\Conid{Map}\;\Varid{k}\;({\color{darkred} \tt{Data}}\;{\color{darkblue} \tt{s} }\;{\color{darkblue} \tt{r} }\to {\color{darkblue} \tt{Query}}\;({\color{darkgreen} \tt{Value}}\;\Varid{a})){}\<[E]%
\printlineend\\
\printlinebegin\>[22]{}\to {}\<[26]%
\>[26]{}{\color{darkblue} \tt{Query}}\;(\Conid{Map}\;\Varid{k}\;({\color{darkgreen} \tt{Value}}\;\Varid{a})){}\<[E]%
\printlineend\\
\printlinebegin\>[B]{}\hsindent{3}{}\<[3]%
\>[3]{}{\color{darkblue}  \texttt{...}}{}\<[E]%
\printlineend\\
\printlinebegin\>[B]{}\hsindent{3}{}\<[3]%
\>[3]{}\mbox{\onelinecomment  Aggregations}{}\<[E]%
\printlineend\\
\printlinebegin\>[B]{}\hsindent{3}{}\<[3]%
\>[3]{}\Conid{Count}\mathbin{::}{}\<[13]%
\>[13]{}\Conid{Stb}\;{\color{darkblue} \tt{s} }\Rightarrow \epsilon\to {\color{darkred} \tt{Data}}\;{\color{darkblue} \tt{s} }\;{\color{darkblue} \tt{r} }\to {\color{darkblue} \tt{Query}}\;({\color{darkgreen} \tt{Value}}\;\Conid{Double}){}\<[E]%
\printlineend\\
\printlinebegin\>[B]{}\hsindent{3}{}\<[3]%
\>[3]{}{\color{darkblue}  \texttt{...}}{}\<[E]%
\printlineend\ColumnHook
\end{hscode}\resethooks
}%
Queries are described by a constructor for every primitive shown in Figure
\ref{fig:dpella:dp} except for \ensuremath{{\color{darkred} \tt{dpEval} }}.
For instance, constructors \ensuremath{\Conid{Where}} and \ensuremath{\Conid{Part}} describe the transformations provided by
\ensuremath{{\color{darkred} \tt{dpWhere} }} and \ensuremath{{\color{darkred} \tt{dpPart} }}, respectively---note that the type signatures of the
constructors are the same as their corresponding primitives.
Similarly, constructor \ensuremath{\Conid{Count}} describes the data aggregation provided by
\ensuremath{{\color{darkblue} \tt{dpCount} }}.

\subsection{Symbolic interpretation for privacy}
\label{app:budget:st}

\begin{figure}
\numbersreset
\small
\begin{hscode}\linenumsetup\printlinebegin\SaveRestoreHook
\column{B}{@{}>{\hspre}l<{\hspost}@{}}%
\column{3}{@{}>{\hspre}l<{\hspost}@{}}%
\column{5}{@{}>{\hspre}l<{\hspost}@{}}%
\column{9}{@{}>{\hspre}l<{\hspost}@{}}%
\column{11}{@{}>{\hspre}l<{\hspost}@{}}%
\column{19}{@{}>{\hspre}l<{\hspost}@{}}%
\column{22}{@{}>{\hspre}l<{\hspost}@{}}%
\column{23}{@{}>{\hspre}l<{\hspost}@{}}%
\column{30}{@{}>{\hspre}l<{\hspost}@{}}%
\column{33}{@{}>{\hspre}l<{\hspost}@{}}%
\column{45}{@{}>{\hspre}l<{\hspost}@{}}%
\column{52}{@{}>{\hspre}c<{\hspost}@{}}%
\column{52E}{@{}l@{}}%
\column{55}{@{}>{\hspre}l<{\hspost}@{}}%
\column{66}{@{}>{\hspre}c<{\hspost}@{}}%
\column{66E}{@{}l@{}}%
\column{E}{@{}>{\hspre}l<{\hspost}@{}}%
\>[B]{}\mbox{\onelinecomment  Interface}{}\<[E]%
\printlineend\\
\printlinebegin\>[B]{}{\color{darkblue} \tt{budget} }\mathbin{::}{\color{darkblue} \tt{Query}}\;\Varid{a}\to \epsilon{}\<[E]%
\printlineend\\
\printlinebegin\>[B]{}{\color{darkblue} \tt{budget} }\;\Varid{comp}\mathrel{=}\Varid{snd}\;(\Varid{runState}\;(\Varid{symInterBud}\;\Varid{comp})\;{\color{darkgreen} \texttt{0}}){}\<[E]%
\printlineend\\[\blanklineskip]%
\printlinebegin\>[B]{}\mbox{\onelinecomment  Symbolic interpreter}{}\<[E]%
\printlineend\\
\printlinebegin\>[B]{}\Varid{symInterPriv}\mathbin{::}{\color{darkblue} \tt{Query}}\;\Varid{a}\to \Conid{State}\;\epsilon\;\Varid{a}{}\<[E]%
\printlineend\\[\blanklineskip]%
\printlinebegin\>[B]{}\mbox{\onelinecomment  Aggregations}{}\<[E]%
\printlineend\\
\printlinebegin\>[B]{}\Varid{symInterPriv}\;(\Conid{Count}\;\Varid{eps}\;\Varid{ds}){}\<[30]%
\>[30]{}\mathrel{=}\mathbf{do}{}\<[E]%
\printlineend\\
\printlinebegin\>[B]{}\hsindent{3}{}\<[3]%
\>[3]{}\Varid{modify}\;({\color{darkblue}  \texttt{+}}\Varid{eps}){}\<[E]%
\printlineend\\
\printlinebegin\>[B]{}\hsindent{3}{}\<[3]%
\>[3]{}\Varid{return}\;\Varid{symValue}{}\<[E]%
\printlineend\\
\printlinebegin\>[B]{}\mbox{\onelinecomment  Transformations}{}\<[E]%
\printlineend\\
\printlinebegin\>[B]{}\Varid{symInterPriv}\;\Conid{Where}\;\{\mskip1.5mu \mskip1.5mu\}\mathrel{=}\Varid{return}\;\tt{symDS}{}\<[E]%
\printlineend\\[\blanklineskip]%
\printlinebegin\>[B]{}\Varid{symInterPriv}\;(\Conid{Part}\;\Varid{conts}\;\anonymous \;\anonymous )\mathrel{=}\mathbf{do}{}\<[E]%
\printlineend\\
\printlinebegin\>[B]{}\hsindent{3}{}\<[3]%
\>[3]{}\mathbf{let}\;\Varid{ks}\mathrel{=}\Varid{\Conid{Map}.keys}\;\Varid{conts}{}\<[E]%
\printlineend\\
\printlinebegin\>[B]{}\hsindent{3}{}\<[3]%
\>[3]{}\Varid{bud}\leftarrow \Varid{get}{}\<[E]%
\printlineend\\
\printlinebegin\>[B]{}\hsindent{3}{}\<[3]%
\>[3]{}\Varid{consumed}\leftarrow {}\<[E]%
\printlineend\\
\printlinebegin\>[3]{}\hsindent{2}{}\<[5]%
\>[5]{}\Varid{seqMap}\;(\Varid{\Conid{Map}.map}\;{}\<[22]%
\>[22]{}(\Varid{bracket}\mathbin{\circ}\Varid{symInterPriv}\mathbin{\circ}(\mathbin{\$}\tt{symDS}))\;{}\<[E]%
\printlineend\\
\printlinebegin\>[22]{}\Varid{conts}){}\<[E]%
\printlineend\\
\printlinebegin\>[B]{}\hsindent{3}{}\<[3]%
\>[3]{}\Varid{put}\;(\Varid{bud}{\color{darkblue}  \texttt{+}}\Varid{maximum}\;(\Varid{\Conid{Map}.elems}\;\Varid{consumed})){}\<[E]%
\printlineend\\
\printlinebegin\>[B]{}\hsindent{3}{}\<[3]%
\>[3]{}\Varid{return}\;(\Varid{symMap}\;\Varid{ks}){}\<[E]%
\printlineend\\
\printlinebegin\>[B]{}\hsindent{3}{}\<[3]%
\>[3]{}\mathbf{where}{}\<[E]%
\printlineend\\
\printlinebegin\>[3]{}\hsindent{6}{}\<[9]%
\>[9]{}\Varid{bracket}\;\Varid{m}\mathrel{=}\mathbf{do}\;\Varid{put}\;{\color{darkgreen} \texttt{0}};\Varid{m};\Varid{get}{}\<[E]%
\printlineend\\
\printlinebegin\>[3]{}\hsindent{6}{}\<[9]%
\>[9]{}\Varid{symMap}\;\Varid{ks}\mathrel{=}\Varid{\Conid{Map}.fromList}\;(\Varid{zip}\;\Varid{ks}\;(\Varid{repeat}\;\Varid{symValue})){}\<[E]%
\printlineend\\
\printlinebegin\>[3]{}\hsindent{6}{}\<[9]%
\>[9]{}\Varid{seqMap}\mathbin{::}(\Conid{MonadTrans}\;\Varid{m},\Conid{Monad}\;(\Varid{m}\;s),\Conid{Ord}\;\Varid{k})\Rightarrow {}\<[E]%
\printlineend\\
\printlinebegin\>[9]{}\hsindent{10}{}\<[19]%
\>[19]{}\Conid{\Conid{Map}.Map}\;\Varid{k}\;(\Varid{m}\;s\;\Varid{a})\to \Varid{m}\;s\;(\Conid{\Conid{Map}.Map}\;\Varid{k}\;\Varid{a}){}\<[E]%
\printlineend\\
\printlinebegin\>[3]{}\hsindent{6}{}\<[9]%
\>[9]{}\Varid{seqMap}\;\Varid{ms}\mathrel{=}\mathbf{do}{}\<[E]%
\printlineend\\
\printlinebegin\>[9]{}\hsindent{2}{}\<[11]%
\>[11]{}\Varid{as}\leftarrow \Varid{sequence}\;(\Varid{\Conid{Map}.elems}\;\Varid{ms}){}\<[E]%
\printlineend\\
\printlinebegin\>[9]{}\hsindent{2}{}\<[11]%
\>[11]{}\Varid{return}\;(\Varid{\Conid{Map}.fromList}\;(\Varid{zip}\;(\Varid{\Conid{Map}.keys}\;\Varid{ms})\;\Varid{as})){}\<[E]%
\printlineend\\[\blanklineskip]%
\printlinebegin\>[B]{}\mbox{\onelinecomment  Symbolic dataset}{}\<[E]%
\printlineend\\
\printlinebegin\>[B]{}\tt{symDS}\mathrel{=}\Conid{MkData}\;\{\mskip1.5mu \Varid{owner}\mathrel{=}\bot ,\Varid{content}\mathrel{=}\bot \mskip1.5mu\}{}\<[E]%
\printlineend\\
\printlinebegin\>[B]{}\mbox{\onelinecomment  Symbolic value}{}\<[E]%
\printlineend\\
\printlinebegin\>[B]{}\Varid{symValue}\mathrel{=}\Conid{MkValue}\;\{\mskip1.5mu {}\<[23]%
\>[23]{}\Varid{value}{}\<[30]%
\>[30]{}\mathrel{=}{}\<[33]%
\>[33]{}\bot ,{}\<[45]%
\>[45]{}\Varid{icdf}{}\<[52]%
\>[52]{}\mathrel{=}{}\<[52E]%
\>[55]{}\bot ,{}\<[E]%
\printlineend\\
\printlinebegin\>[23]{}\Varid{scale}{}\<[30]%
\>[30]{}\mathrel{=}{}\<[33]%
\>[33]{}\bot ,{}\<[45]%
\>[45]{}\Varid{label}{}\<[52]%
\>[52]{}\mathrel{=}{}\<[52E]%
\>[55]{}\bot {}\<[66]%
\>[66]{}\mskip1.5mu\}{}\<[66E]%
\printlineend\ColumnHook
\end{hscode}\resethooks
\caption{Symbolic interpretation of queries to obtain privacy
  budgets\label{bud:priv}}
\end{figure}

Figure \ref{bud:priv} presents the interesting parts when symbolically
interpreting queries to obtain the required (upper bound on the) privacy
budgets.
Function \ensuremath{{\color{darkblue} \tt{budget} }} is the interface for the data analysts but the work is done by
function \ensuremath{\Varid{symInterPriv}}.
In essence, \ensuremath{\Varid{symInterPriv}} symbolically interprets the query in the state monad,
where the state stores the privacy budget (\ensuremath{\Conid{State}\;\epsilon\;\Varid{a}}).
The definition traverses the query and essentially increments the privacy budget
when finding an aggregation.
For instance, when finding the constructor \ensuremath{\Conid{Count}} (line 7), it increases the
privacy budget by \ensuremath{\Varid{eps}} (\ensuremath{\Varid{modify}\;({\color{darkblue}  \texttt{+}}\Varid{eps})}).
The interpreter returns symbolic datasets and values, e.g., see line 9.
In our case, a symbolic dataset (value) is just a dataset (value) which content
is set to \ensuremath{\bot }---see line 29 (31).
(In Haskell, it is possible to write unspecified terms, written \ensuremath{\bot \mathbin{::}\Varid{a}},
to indicate a piece of code not yet defined---if \ensuremath{\bot } gets ever
evaluated, it throws an exception.)
Since transformations do not consume privacy budget, they return just a
symbolic dataset---see line 11.

The interpretation of \ensuremath{{\color{darkred} \tt{dpPart} }} is, as always, the most interesting case for the
analysis.
It essentially provides a symbolic dataset to the computations to run in each
partition
(\ensuremath{(\mathbin{\$}\tt{symDS})}) and computes the budget by calling \ensuremath{\Varid{symInterPriv}} while
assuming the initial required privacy budget is $0$---see the definition of
\ensuremath{\Varid{bracket}} where it does \ensuremath{\mathbf{do}\;\Varid{put}\;{\color{darkgreen} \texttt{0}};\Varid{m};\Varid{get}}.
All the privacy budget estimations computed in each partition are stored in
variable \ensuremath{\Varid{consume}} as a mapping key-to-budget---there is some very technical
detail about what \ensuremath{\Varid{seqMap}} does but the reader unfamiliar with advanced
funcitonal programming do not need to care about that.
Finally, the budget requirement for a partition is set up to be the requirement
we had \emph{before} running it (variable \ensuremath{\Varid{bud}} in line 14) and the maximum requirement
imposed by a partition (\ensuremath{\Varid{maximum}\;(\Varid{\Conid{Map}.elems}\;\Varid{consumed})})---see line 15.
To finish, the interpreter returns a symbolic mapping, where each value in the
mapping is symbolic---that is the purpose of function \ensuremath{\Varid{symMap}} (line
22).

\subsection{Symbolic interpretation for accuracy}

\begin{figure}
\numbersreset
\numberson
{\small
\begin{hscode}\linenumsetup\printlinebegin\SaveRestoreHook
\column{B}{@{}>{\hspre}l<{\hspost}@{}}%
\column{3}{@{}>{\hspre}l<{\hspost}@{}}%
\column{4}{@{}>{\hspre}l<{\hspost}@{}}%
\column{5}{@{}>{\hspre}l<{\hspost}@{}}%
\column{12}{@{}>{\hspre}l<{\hspost}@{}}%
\column{21}{@{}>{\hspre}c<{\hspost}@{}}%
\column{21E}{@{}l@{}}%
\column{24}{@{}>{\hspre}c<{\hspost}@{}}%
\column{24E}{@{}l@{}}%
\column{27}{@{}>{\hspre}l<{\hspost}@{}}%
\column{30}{@{}>{\hspre}c<{\hspost}@{}}%
\column{30E}{@{}l@{}}%
\column{31}{@{}>{\hspre}l<{\hspost}@{}}%
\column{34}{@{}>{\hspre}l<{\hspost}@{}}%
\column{37}{@{}>{\hspre}l<{\hspost}@{}}%
\column{E}{@{}>{\hspre}l<{\hspost}@{}}%
\>[B]{}\mbox{\onelinecomment  Interface}{}\<[E]%
\printlineend\\
\printlinebegin\>[B]{}{\color{darkblue} \tt{accuracy} }\mathbin{::}{\color{darkblue} \tt{Query}}\;({\color{darkgreen} \tt{Value}}\;\Varid{a})\to (\beta\to \alpha){}\<[E]%
\printlineend\\
\printlinebegin\>[B]{}{\color{darkblue} \tt{accuracy} }\mathrel{=}\Varid{iCDF}\mathbin{\circ}\Varid{symInterAcc}{}\<[E]%
\printlineend\\[\blanklineskip]%
\printlinebegin\>[B]{}\mbox{\onelinecomment  Symbolic interpreter for accuracy}{}\<[E]%
\printlineend\\
\printlinebegin\>[B]{}\Varid{symInterAcc}\mathbin{::}{\color{darkblue} \tt{Query}}\;({\color{darkgreen} \tt{Value}}\;\Varid{a})\to \Conid{State}\;\Conid{Int}\;({\color{darkgreen} \tt{Value}}\;\Varid{a}){}\<[E]%
\printlineend\\
\printlinebegin\>[B]{}\mbox{\onelinecomment  Aggregations}{}\<[E]%
\printlineend\\
\printlinebegin\>[B]{}\Varid{symInterAcc}\;(\Conid{Count}\;\Varid{eps}\;\Varid{ds})\mathrel{=}\mathbf{do}{}\<[E]%
\printlineend\\
\printlinebegin\>[B]{}\hsindent{3}{}\<[3]%
\>[3]{}\Varid{l}\leftarrow \Varid{get}{}\<[E]%
\printlineend\\
\printlinebegin\>[B]{}\hsindent{3}{}\<[3]%
\>[3]{}\Varid{modify}\;({\color{darkblue}  \texttt{+}}{\color{darkgreen} \texttt{1}}){}\<[E]%
\printlineend\\
\printlinebegin\>[B]{}\hsindent{3}{}\<[3]%
\>[3]{}\Varid{return}\;(\Conid{MkValue}\;{}\<[21]%
\>[21]{}\{\mskip1.5mu {}\<[21E]%
\>[24]{}\Varid{value}{}\<[24E]%
\>[31]{}\mathrel{=}{}\<[34]%
\>[34]{}\bot {}\<[E]%
\printlineend\\
\printlinebegin\>[21]{},{}\<[21E]%
\>[24]{}\Varid{iCDF}{}\<[24E]%
\>[31]{}\mathrel{=}{}\<[34]%
\>[34]{}\Varid{theoError}{}\<[E]%
\printlineend\\
\printlinebegin\>[21]{},{}\<[21E]%
\>[24]{}\Varid{scale}{}\<[24E]%
\>[31]{}\mathrel{=}{}\<[34]%
\>[34]{}\Conid{Just}\;\Varid{lapScale}{}\<[E]%
\printlineend\\
\printlinebegin\>[21]{},{}\<[21E]%
\>[24]{}\Varid{label}{}\<[24E]%
\>[31]{}\mathrel{=}{}\<[34]%
\>[34]{}[\mskip1.5mu \Varid{l}\mskip1.5mu]\mskip1.5mu\}){}\<[E]%
\printlineend\\
\printlinebegin\>[3]{}\hsindent{2}{}\<[5]%
\>[5]{}\mathbf{where}\;{}\<[12]%
\>[12]{}{\color{darkblue} \tt{s} }{}\<[24]%
\>[24]{}\mathrel{=}{}\<[24E]%
\>[27]{}\Varid{stability}\;\Varid{ds}{}\<[E]%
\printlineend\\
\printlinebegin\>[12]{}\Delta_Q{}\<[24]%
\>[24]{}\mathrel{=}{}\<[24E]%
\>[27]{}{\color{darkgreen} \texttt{1}}{}\<[E]%
\printlineend\\
\printlinebegin\>[12]{}\Varid{lapScale}{}\<[24]%
\>[24]{}\mathrel{=}{}\<[24E]%
\>[27]{}{\color{darkblue} \tt{s} }{\color{darkblue}  \texttt{*}}(\Delta_Q\mathbin{/}\Varid{eps}){}\<[E]%
\printlineend\\
\printlinebegin\>[12]{}\Varid{theoError}{}\<[24]%
\>[24]{}\mathrel{=}{}\<[24E]%
\>[27]{}\lambda \Varid{b}\to \Varid{log}\;({\color{darkgreen} \texttt{1}}\mathbin{/}\Varid{b}){\color{darkblue}  \texttt{*}}\Varid{lapScale}{}\<[E]%
\printlineend\\
\printlinebegin\>[B]{}\mbox{\onelinecomment  Transformations}{}\<[E]%
\printlineend\\
\printlinebegin\>[B]{}\Varid{symInterAcc}\;(\Conid{Where}\;\anonymous \;\anonymous :\!\bind \!:{}\<[31]%
\>[31]{}\Varid{f})\mathrel{=}\Varid{symInterAcc}\;(\Varid{f}\;\Varid{symDS}){}\<[E]%
\printlineend\\
\printlinebegin\>[B]{}\Varid{symInterAcc}\;(\Conid{Part}\;\Varid{conts}\;\anonymous \;\anonymous {}\<[30]%
\>[30]{}:\!\bind \!:{}\<[30E]%
\>[37]{}\Varid{f})\mathrel{=}{}\<[E]%
\printlineend\\
\printlinebegin\>[B]{}\hsindent{4}{}\<[4]%
\>[4]{}\Varid{symInterAcc}\;(\Varid{f}\;(\Varid{\Conid{Map}.map}\;(\Varid{symInterAcc}\mathbin{\circ}(\mathbin{\$}\Varid{symDS}))\;\Varid{conts})){}\<[E]%
\printlineend\\
\printlinebegin\>[B]{}{\color{darkblue}  \texttt{...}}{}\<[E]%
\printlineend\ColumnHook
\end{hscode}\resethooks
}
\vspace{-20pt}
\caption{Symbolic execution for accuracy \label{fig:interp:utility}}
\end{figure}

\begin{figure}
\numbersreset
\numberson
{\small
\begin{hscode}\linenumsetup\printlinebegin\SaveRestoreHook
\column{B}{@{}>{\hspre}l<{\hspost}@{}}%
\column{4}{@{}>{\hspre}l<{\hspost}@{}}%
\column{6}{@{}>{\hspre}l<{\hspost}@{}}%
\column{11}{@{}>{\hspre}l<{\hspost}@{}}%
\column{12}{@{}>{\hspre}l<{\hspost}@{}}%
\column{13}{@{}>{\hspre}c<{\hspost}@{}}%
\column{13E}{@{}l@{}}%
\column{16}{@{}>{\hspre}l<{\hspost}@{}}%
\column{23}{@{}>{\hspre}c<{\hspost}@{}}%
\column{23E}{@{}l@{}}%
\column{26}{@{}>{\hspre}l<{\hspost}@{}}%
\column{32}{@{}>{\hspre}c<{\hspost}@{}}%
\column{32E}{@{}l@{}}%
\column{35}{@{}>{\hspre}l<{\hspost}@{}}%
\column{E}{@{}>{\hspre}l<{\hspost}@{}}%
\>[B]{}{\color{darkgreen} \tt{add} }\mathbin{::}\Conid{Num}\;\Varid{a}\Rightarrow [\mskip1.5mu {\color{darkgreen} \tt{Value}}\;\Varid{a}\mskip1.5mu]\to {\color{darkgreen} \tt{Value}}\;\Varid{a}{}\<[E]%
\printlineend\\
\printlinebegin\>[B]{}{\color{darkgreen} \tt{add} }\;{}\<[6]%
\>[6]{}[\mskip1.5mu \mskip1.5mu]{}\<[11]%
\>[11]{}\mathrel{=}\bot {}\<[E]%
\printlineend\\
\printlinebegin\>[B]{}{\color{darkgreen} \tt{add} }\;{}\<[6]%
\>[6]{}[\mskip1.5mu \tt{v}\mskip1.5mu]{}\<[11]%
\>[11]{}\mathrel{=}\tt{v}{}\<[E]%
\printlineend\\
\printlinebegin\>[B]{}{\color{darkgreen} \tt{add} }\;{}\<[6]%
\>[6]{}\Varid{vs}{}\<[11]%
\>[11]{}\mathrel{=}{}\<[E]%
\printlineend\\
\printlinebegin\>[B]{}\hsindent{4}{}\<[4]%
\>[4]{}\Conid{MkValue}\;{}\<[13]%
\>[13]{}\{\mskip1.5mu {}\<[13E]%
\>[16]{}\Varid{value}{}\<[23]%
\>[23]{}\mathrel{=}{}\<[23E]%
\>[26]{}\Varid{sum}\;(\Varid{map}\;\Varid{value}\;\Varid{vs}){}\<[E]%
\printlineend\\
\printlinebegin\>[13]{},{}\<[13E]%
\>[16]{}\Varid{iCDF}{}\<[23]%
\>[23]{}\mathrel{=}{}\<[23E]%
\>[26]{}\Varid{selectBound}\;\Varid{vs}{}\<[E]%
\printlineend\\
\printlinebegin\>[13]{},{}\<[13E]%
\>[16]{}\Varid{scale}{}\<[23]%
\>[23]{}\mathrel{=}{}\<[23E]%
\>[26]{}\Conid{Nothing}{}\<[E]%
\printlineend\\
\printlinebegin\>[13]{},{}\<[13E]%
\>[16]{}\Varid{label}{}\<[23]%
\>[23]{}\mathrel{=}{}\<[23E]%
\>[26]{}\Varid{nub}\;(\Varid{concatMap}\;\Varid{label}\;\Varid{vs})\mskip1.5mu\}{}\<[E]%
\printlineend\\
\printlinebegin\>[B]{}\hsindent{4}{}\<[4]%
\>[4]{}\mathbf{where}\;{}\<[11]%
\>[11]{}\Varid{selectBound}\;\Varid{vs}{}\<[E]%
\printlineend\\
\printlinebegin\>[11]{}\hsindent{1}{}\<[12]%
\>[12]{}\mid \Varid{all}\;\Varid{isIndependent}\;\Varid{vs}\mathrel{=}{}\<[E]%
\printlineend\\
\printlinebegin\>[12]{}\hsindent{4}{}\<[16]%
\>[16]{}\lambda \Varid{b}\to \Varid{min}\;(\Varid{chernoff}\;\Varid{scales}\;\Varid{b})\;(\Varid{union}\;\Varid{b}){}\<[E]%
\printlineend\\
\printlinebegin\>[11]{}\hsindent{1}{}\<[12]%
\>[12]{}\mid \Varid{otherwise}\mathrel{=}\Varid{union}{}\<[E]%
\printlineend\\
\printlinebegin\>[12]{}\hsindent{4}{}\<[16]%
\>[16]{}\mathbf{where}\;{}\<[23]%
\>[23]{}\Varid{scales}{}\<[23E]%
\>[32]{}\mathrel{=}{}\<[32E]%
\>[35]{}\Varid{map}\;(\Varid{fromJust}\mathbin{\circ}\Varid{scale})\;\Varid{vs}{}\<[E]%
\printlineend\\
\printlinebegin\>[23]{}\Varid{iCDFs}{}\<[23E]%
\>[32]{}\mathrel{=}{}\<[32E]%
\>[35]{}\Varid{map}\;\Varid{iCDF}\;\Varid{vs}{}\<[E]%
\printlineend\\
\printlinebegin\>[23]{}\Varid{union}{}\<[23E]%
\>[32]{}\mathrel{=}{}\<[32E]%
\>[35]{}\Varid{unionBound}\;\Varid{iCDFs}{}\<[E]%
\printlineend\ColumnHook
\end{hscode}\resethooks
}
\vspace{-20pt}
\caption{Accuracy interpretation for \ensuremath{{\color{darkgreen} \tt{add} }} \label{fig:acc:add}}
\end{figure}

%
The interpretation for accuracy estimation consists on taking a query, which
returns a result of type \ensuremath{{\color{darkgreen} \tt{Value}}\;\Varid{a}}, and generating a function \ensuremath{\beta\to \alpha}---thus capturing the theoretical error bound for such a value.
However, function \ensuremath{{\color{darkblue} \tt{accuracy} }} is just the interface for data analysts and the
actual work is done by function \ensuremath{\Varid{symInterAcc}}---see Figure
\ref{fig:interp:utility}.

Function \ensuremath{\Varid{symInterAcc}} takes a query producing a \ensuremath{{\color{darkgreen} \tt{Value}}\;\Varid{a}} and symbolically
executed to produce a \ensuremath{{\color{darkgreen} \tt{Value}}\;\Varid{a}}, which \emph{\ensuremath{\Varid{iCDF}} field is calculated
  according to the structure of the query}.
Because the function only takes queries that produce a result of type \ensuremath{{\color{darkgreen} \tt{Value}}\;\Varid{a}},
the only way to build them is essentially by performing data aggregation queries
(e.g., \ensuremath{{\color{darkblue} \tt{dpCount} }}) preceded by a (possibly empty) sequence of other primitives
like data transformations.
%
In this light, the definition of \ensuremath{\Varid{symInterAcc}} gets split into two cases: data
aggregation queries (as the base case and denoted by constructors like \ensuremath{\Conid{Count}},
lines 7--13) and primitives (or sequences of them as the inductive case and
denoted by constructor \ensuremath{(:\!\bind \!:)}, line 19--21).

The insight to understand why \ensuremath{\Varid{symInterAcc}} works is that the symbolic execution
leaves \emph{unspecified} (\ensuremath{\bot }) calculations related to the record
\ensuremath{\Varid{value}} and only uses symbolic datasets---after all, the execution is going to
produce terms of type \ensuremath{{\color{darkgreen} \tt{Value}}\;\Varid{a}} with the accuracy information.
When interpreting an aggregation (line 7), \ensuremath{\Varid{symInterAcc}} creates a record of type
\ensuremath{{\color{darkgreen} \tt{Value}}\;\Conid{Double}} where the field \ensuremath{\Varid{value}} is left unspecified (line 10).
Since \ensuremath{\Varid{symInterAcc}} never inspects that field, there is no risk to raise an
exception.
However, it populates the field \ensuremath{\Varid{iCDF}} with the corresponding error calculations
for Laplace as presented in (\ref{eq:laplace:acc}).
Function \ensuremath{\Varid{stability}\;\Varid{ds}} synthesizes term-level numbers from the type-level information
found in the data set (\ensuremath{\Varid{ds}}), e.g., \ensuremath{\Varid{stability}\;(\Varid{ds}\mathbin{::}{\color{darkred} \tt{Data}}\;{\color{darkgreen} \texttt{5}}\;\Conid{Int})} generates
the number \ensuremath{{\color{darkgreen} \texttt{5}}}.
To interpret a sequence of instructions, \ensuremath{\Varid{symInterAcc}} traverses it interpreting
each primitive individually.
Often, function \ensuremath{\Varid{symInterAcc}} ignores primitives related to dataset
operations---after all, they do not generate terms of type \ensuremath{{\color{darkgreen} \tt{Value}}\;\Varid{a}} and
therefore require no accuracy estimations---in such cases, we use the symbolic
dataset \ensuremath{\Varid{symDS}}.
To illustrate this point, line 19 shows the execution of \ensuremath{\Conid{Where}} (corresponding
to primitive \ensuremath{{\color{darkred} \tt{dpWhere} }} in Figure \ref{fig:dpella:dp}).
In this case, the continuation \ensuremath{\Varid{f}} is waiting for a term of type \ensuremath{{\color{darkred} \tt{Data}}\;{\color{darkblue} \tt{s} }\;{\color{darkblue} \tt{r} }}
as an argument.
Since such term is not relevant for accuracy estimation, \ensuremath{\Varid{symInterAcc}} simply
uses the symbolic value \ensuremath{\Varid{symDS}} when passing it to the continuation, i.e., \ensuremath{\Varid{f}\;\Varid{symDS}}, and subsequently proceeds to symbolically execute the rest of the yield
instructions (i.e., \ensuremath{\Varid{symInterAcc}\;(\Varid{f}\;\Varid{symDS})}).
Symbolically executing most of the transformation primitives are interpreted in
this way except for \ensuremath{{\color{darkred} \tt{dpPart} }}.

The execution of \ensuremath{{\color{darkred} \tt{dpPart} }}, associated with constructor \ensuremath{\Conid{Part}} (line 20),
deserves special attention.
Constructor \ensuremath{\Conid{Part}} takes as an argument a mapping (\ensuremath{\Varid{conts}\mathbin{::}\Conid{Map}\;\Varid{k}\;({\color{darkred} \tt{Data}}\;{\color{darkblue} \tt{s} }\;{\color{darkblue} \tt{r} }\to {\color{darkblue} \tt{Query}}\;({\color{darkgreen} \tt{Value}}\;\Varid{a}))}) that describes the queries to execute once given the
corresponding bins.
Since these queries produce values (\ensuremath{{\color{darkgreen} \tt{Value}}\;\Varid{a}}), we need to symbolically execute
each of them in order to obtain their accuracy estimations.
To achieve that, \ensuremath{\Varid{symInterAcc}} gets applied to each of those queries when given a
symbolic dataset
(\ensuremath{\mathbin{\$}\Varid{symDS}}), which results in a key-value mapping---see code \ensuremath{\Varid{\Conid{Map}.map}\;(\Varid{symInterAcc}\mathbin{\circ}(\mathbin{\$}\Varid{symDS}))\;\Varid{conts}\mathbin{::}\Conid{Map}\;\Varid{k}\;({\color{darkgreen} \tt{Value}}\;\Varid{a})} in line 21.
The continuation \ensuremath{\Varid{f}} is then applied to such term and \ensuremath{\Varid{symInterAcc}} continues
executing the rest of the yield instructions.
%
Observe that using \ensuremath{\Varid{f}\;\Varid{symDS}} as we did before is incorrect in this case
since we would have been leaving unspecified parts of our query that need
accuracy estimations, namely the values produced by the queries run at each
partition.

\subsection{Taint analysis}
The DPella taint analysis is carried out when executing the primitives
aggregating terms of type \ensuremath{{\color{darkgreen} \tt{Value}}\;\Varid{a}}---which also results in a term of that type.
In our case, we provide the function \ensuremath{{\color{darkgreen} \tt{add} }} to perform addition of
values---shown in Figure \ref{fig:acc:add}.
The interesting case is when receiving a list of values with two
or more elements (line 4).
In such a case, the record \ensuremath{\Varid{value}} stores the addition of the elements in the
list \ensuremath{\Varid{vs}} (line 5).
The value return by \ensuremath{{\color{darkgreen} \tt{add} }} is always tainted (line 7)---this indicates that the
produced term \ensuremath{{\color{darkgreen} \tt{Value}}\;\Varid{a}} depends on others.
The resulting value has as a label the the union of arguments' labels (see line
8).
Finally, the field \ensuremath{\Varid{iCDF}} consists of two cases: if all the values were
independently generated (\ensuremath{\Varid{all}\;\Varid{isIndependent}\;\Varid{vs}}), then the error estimation
produces the minimal error (line 11) between applying Chernoff
(\ref{eq:chernoffBound}) and union bound (\ref{eq:unionBound}); otherwise, it
applies union bound (line 12).
%


}

\end{document}